\documentclass[12pt,letter]{article}

\usepackage{graphicx, epsfig, color}
\usepackage{amsmath,amssymb,hyperref}
\usepackage{subcaption}

\def\eslt{\not\!\!{E_T}}
\def\to{\rightarrow}

\def\bi{\begin{itemize}}
\def\ei{\end{itemize}}
\def\te{\tilde e}
\def\ta{\tilde a}
\def\tG{\tilde G}
\def\tchi{\tilde\chi}

\def\tu{\tilde u}
\def\sps1ap{SPS1a$^\prime$}
\def\c1p{C1$^\prime$}

\def\td{\tilde d}

\def\tst{\tilde t}
\def\ttau{\tilde \tau}
\def\tmu{\tilde \mu}
\def\tg{\tilde g}

\def\tell{\tilde\ell}
\def\tq{\tilde q}
\def\tw{\widetilde W}

\def\alt{\lesssim}
\def\agt{\gtrsim}
\def\be{\begin{equation}}  
\def\ee{\end{equation}}  
\def\bea{\begin{eqnarray}}  
\def\eea{\end{eqnarray}}  
\def\beas{\begin{eqnarray*}}  
\def\eeas{\end{eqnarray*}}

\begin{document}
\begin{titlepage}
\begin{flushright}
OUHEP-250130
\end{flushright}

\begin{center}
  {\Large \bf Prospects for supersymmetry at high luminosity LHC}\\
\vspace{0.3cm} \renewcommand{\thefootnote}{\fnsymbol{footnote}}
{\large Howard Baer$^1$\footnote[1]{Email: baer@nhn.ou.edu }, 
Vernon Barger$^2$\footnote[2]{Email: barger@pheno.wisc.edu },
Jessica Bolich$^1$\footnote[3]{Email: Jessica.R.Bolich-1@ou.edu},\\
Juhi Dutta$^1,^3$\footnote[4]{Email: juhi.dutta@ou.edu, juhidutta@imsc.res.in},
Dakotah Martinez$^1$\footnote[5]{Email: dakotah.martinez@ou.edu},
Shadman Salam$^4$\footnote[6]{Email: ext.shadman.salam@bracu.ac.bd},\\
Dibyashree Sengupta$^{5,6}$\footnote[7]{Email: Dibyashree.Sengupta@roma1.infn.it}
and Kairui Zhang$^1$\footnote[8]{Email: kzhang25@ou.edu}
}\\ 
\vspace{0.2cm} \renewcommand{\thefootnote}{\arabic{footnote}}
{\it 
$^1$Department of Physics and Astronomy,
University of Oklahoma, Norman, OK 73019, USA \\
}
{\it 
$^2$Department of Physics,
University of Wisconsin, Madison, WI 53706, USA \\
}
{\it 
$^3$ The Institute of Mathematical Sciences, 
4th Cross St, CIT Campus, Tharamani, Chennai, Tamil Nadu, India, 600113 \\[3pt]
}
{\it 
$^4$Department of Mathematics and Natural Sciences,
Brac University, Dhaka 1212, Bangladesh \\[3pt]
}
{\it
$^5$ INFN, Laboratori Nazionali di Frascati,
Via E. Fermi 54, 00044 Frascati (RM), Italy} \\[3pt]
{\it
$^6$ INFN, Sezione di Roma, c/o Dipartimento di Fisica, Sapienza Università di Roma, Piazzale Aldo Moro 2, I-00185 Rome, Italy}
%
\end{center}

\begin{abstract}
\noindent 
Weak scale supersymmetry (SUSY) is highly motivated in that it
provides a 't Hooft technically natural solution to the gauge hierarchy problem.
However, recent strong limits from superparticle searches at LHC Run 2
may exacerbate a so-called Little Hierarchy problem (LHP) 
which is a matter of practical naturalness: why is $m_{weak}\ll m_{soft}$?
We review recent LHC and WIMP dark matter search bounds as well as their
impact on a variety of proposed SUSY models:
gravity-, gauge-, anomaly-, mirage- and gaugino-mediation
along with some dark matter proposals such as well-tempered neutralinos.
We address the naturalness question.
We also address the emergence of the string landscape at the beginning of
the $21^{st}$ century and its impact on expectations for SUSY.
Rather generally, the string landscape statistically
prefers large soft SUSY breaking terms but subject to the
anthropic requirement that the derived value of the weak scale for
each pocket universe (PU) within the greater
multiverse lies with the ABDS window of values.
This {\it stringy natural} (SN) approach implies $m_h\sim 125$ GeV more often than not with
sparticles beyond or well-beyond present LHC search limits.
We review detailed reach calculations of the high-lumi LHC (HL-LHC) for 
non-universal Higgs mass models which present perhaps the most plausible
realization of SUSY from the string landscape. 
In contrast to conventional wisdom, from a stringy naturalness point of view, the search for SUSY at LHC has only just begun to explore the interesting regimes of parameter space.
We comment on how non-universal Higgs models could be differentiated from
other expressions of natural SUSY such as natural anomaly-mediation and natural mirage mediation at HL-LHC.


\end{abstract}

\end{titlepage}

\tableofcontents

\pagebreak

\section{Introduction}
\label{sec:intro}

The Standard Model (SM) is based on the gauge symmetries
$SU(3)_C\times SU(2)_L\times U(1)_Y$ with three generations of quarks
and leptons, and where the fermions and weak bosons gain mass when
electroweak symmetry is spontaneously broken via a Higgs field vacuum
expectation value\cite{Barger:1987nn,Quigg:2013ufa,Langacker:2010zza}.
With the discovery of a very SM-like Higgs boson with mass $m_h\simeq 125$ GeV,
the CERN Large Hadron Collider (LHC) has completed its
primary mission\cite{ATLAS:2012yve,CMS:2012qbp}, and all predicted
SM matter states have been accounted for.

Yet, it is highly implausible that the SM is the final word in the
description of fundamental physics at the TeV-scale and beyond.
This is due in part to three finetuning problems inherent in the theory:
\begin{enumerate}
\item the cosmological constant (CC) problem\cite{Weinberg:1988cp,Polchinski:2006gy},
\item  the strong CP problem\cite{Kim:2008hd} and
\item the gauge hierarchy problem (GHP)\cite{Susskind:1982mw,Hook:2023yzd}.
\end{enumerate}

These theory problems are accompanied by
several problems arising from cosmological data, namely that the SM has no
candidate for the required cold dark matter (CDM) in the universe, no way to
account for the matter-antimatter asymmetry and no way to account for the
apparent accelerated expansion of the universe.
After decades of focused research, the most plausible solutions to the 
three finetuning problems are the following.

\bi
\item  Weinberg's anthropic selection of a tiny cosmological constant\cite{Weinberg:1987dv}
  from all the possible available values in an eternally inflating
  multiverse\cite{Linde:2015edk}.
  Weinberg argued that the CC should exist at a non-zero value, but if it is too big, then the universe will expand too rapidly and large scale structure
  (galaxy formation) will fail to occur; in such a universe, observers are
  unlikely to form. A value of the CC in rough accord with Weinberg's
  expectations was observed in 1998\cite{SupernovaCosmologyProject:1998vns,SupernovaSearchTeam:1998fmf} and confirmed later in other
  channels\cite{Planck:2018vyg}.
  This explanation gained footing with the discovery of an immense number of
  string vacuum solutions within the context of string flux
  compactifications\cite{Bousso:2000xa,Susskind:2003kw,Ashok:2003gk,Douglas:2006es}. 
\item The Peccei-Quinn\cite{Peccei:1977hh} solution to the strong CP problem which relies on
  introducing a global $U(1)_{PQ}$ symmetry which is spontaneously broken
  at a scale $f_a\sim 10^{11}$ GeV gives rise to a pseudo-Goldstone boson,
  the {\it axion}\cite{Weinberg:1977ma,Wilczek:1977pj,Kim:1979if,Shifman:1979if,Dine:1981rt,Zhitnitsky:1980tq}.
  The axion can also serve as a good candidate for CDM in the
  universe\cite{Dine:1982ah,Abbott:1982af,Preskill:1982cy}. 
\item The GHP arises from a plethora of quadratically divergent loop
  corrections to the Higgs boson mass, necessitating highly unnatural
  finetuning of parameters in order to avoid its blow-up to the highest
  mass scales allowed in the model\cite{Georgi:1974yf}.
  A loop-by-loop just-right cancellation
  of quadratic divergences\cite{Witten:1981nf,Kaul:1981wp,Drees:1996ca}
  is gained by extending the SM to a supersymmetric (SUSY) SM --
  the Minimal Supersymmetric Standard Model or MSSM\cite{Dimopoulos:1981zb} --
  with soft SUSY breaking terms at scale $m_{soft}\sim 1$ TeV
  which allow for SUSY breaking whilst
  maintaining cancellation of quadratic divergences.
  While SUSY yields a 't Hooft technical naturalness\cite{tHooft:1979rat}
  solution to stabilizing the weak scale against blow-up to much higher values
  (in that the symmetry of the theory is increased by taking the
  soft SUSY breaking terms to zero), LHC SUSY search results have engendered a
  Little Hierarchy Problem (LHP) -- why is $m_{weak}\simeq m_{W,Z,h}\sim 100$ GeV
  much smaller than $m_{soft}\agt 1$ TeV -- based on practical
  naturalness\cite{Baer:2015rja,Baer:2023cvi}:
  the notion that all {\it independent}
  contributions to an observable should be comparable to or less than the
  observable in question. Under so-called $R$-parity conservation, SUSY
  theories also give rise to a CDM candidate\cite{Goldberg:1983nd,Ellis:1983ew}, the lightest SUSY particle(LSP),
  a weakly interacting massive particle or WIMP, and also provide
  several viable mechanisms for generating the matter-antimatter
  asymmetry\cite{Dine:2003ax,Bae:2015efa}.
\ei

This review article will focus mainly on the status of the
{\it weak scale supersymmetry} (WSS) solution to the
GHP\cite{Chung:2003fi,Drees:2004jm,Baer:2006rs,Dreiner:2023yus}.
One of the main consequences of SUSY as a solution to the
GHP is the expectation that supersymmetric matter -- the so-called superpartners -- should exist with mass values not too far from the weak scale
$m_{weak}\sim m_{W,Z,h}\sim 100$ GeV.
While highly motivated theoretically, WSS is also supported indirectly
via four different quantum loop effects.
\bi
\item The measured values of the three SM gauge couplings all (nearly)
  unify at the scale $m_{GUT}\simeq 2\times 10^{16}$ GeV under
  renormalization group evolution using the MSSM beta functions\cite{Dimopoulos:1981yj,Amaldi:1991cn,Ellis:1990wk,Langacker:1991an}.
  This does not occur in the SM or many of its other extensions.
\item In supersymmetric models, the electroweak gauge symmetry is broken radiatively due to the large top-quark Yukawa coupling, requiring a top quark
  in the rather heavy range of 100-200 GeV\cite{Ibanez:1982fr,Alvarez-Gaume:1983drc}.
  The present measured value of the top quark is $m_t=172.6\pm 0.3$ GeV\cite{ParticleDataGroup:2024cfk}.
\item The light Higgs boson mass in the MSSM is bounded at tree-level to be
  $m_h<m_Z$, but by including radiative correction, especially from the top-stop sector, this bound is raised to the vicinity of $m_h\alt 130$ GeV\cite{Carena:2002es}.
  The measured value $m_h=125.11\pm 0.11$ GeV\cite{ParticleDataGroup:2024cfk}.
\item Electroweak radiative corrections to $m_W$ prefer heavy SUSY over the
  SM\cite{Heinemeyer:2006px}.
\ei

Owing to its strong theoretical motivation and its support from virtual
quantum effects, WSS was widely expected to emerge in the early stages of
LHC running. The fact that it hasn't (so far) has exacerbated the LHP --
why is $m_{weak}\ll m_{sparticle}$ -- which in turn calls into question the
so-called {\it naturalness principle}. In fact, Weinberg \cite{Weinberg:2015exp} declares
in his book {\it To Explain the World: The Discovery of Modern Science} that
\begin{quotation}
  The appearance of finetuning in a scientific theory is like a cry of distress from nature, complaining that something needs to be better explained.
\end{quotation}
Furthermore, Arkani-Hamed {\it et al.}\cite{Arkani-Hamed:2015vfh} exclaim that
\begin{quotation}
  settling the ultimate fate of naturalness is perhaps the most profound
  theoretical question of our time $\ldots$
  Given the magnitude of the stakes involved, it is vital to get a clear
  verdict on naturalness from experiment.
\end{quotation}

Whereas we are in close accord with the above sentiments, we take this
opportunity to collect some recent results from LHC Run 2
and from ton-scale noble liquid WIMP search experiments in
Sec. \ref{sec:status} and to confront these with present day
theoretical understanding of WSS.
In Sec. \ref{sec:nat}, we present a critical assessment of the naturalness
question. We find early naturalness measures to be flawed, and advocate
instead for the more conservative, model independent measure $\Delta_{EW}$.
In Sec. \ref{sec:string}, we review the present state of string theory,
namely the string landscape, but in this case apply it to gain statistical
predictions for superparticle masses, in a manner akin to Weinberg's
solution to the CC problem.
Rather generally, the landscape favors large values of soft SUSY breaking
terms subject to the constraint that the derived value of the weak scale is
within a factor of a few of its measured value in our universe.
This methodology, dubbed by Douglas as {\it stringy naturalness}\cite{Douglas:2004zg},
has a close connection with the $\Delta_{EW}$ naturalness measure, and predicts\cite{Baer:2017uvn}
$m_h\sim 125$ GeV with sparticles somewhat or well-beyond present LHC
search limits. In this sense, modern string theory predicts exactly what
LHC is seeing.
In Sec. \ref{sec:DM}, we revisit the SUSY dark matter situation and its
relation to naturalness. Here, it is possible that both R-parity and Peccei-Quinn (PQ) symmetry can arise from a more fundamental discrete $R$-symmetry which
could emerge as a remnant from string compactification.
An example is given which solves the SUSY $\mu$ problem via the Kim-Nilles
mechanism, which solves the axion quality problem and generates naturally a
PQ scale $f_a\sim 10^{11}$ GeV, in the cosmological sweet spot. As a
consequence, DM is expected as an axion/higgsino-like WIMP admixture and
numerical results suggest it is mainly axions with a smattering of WIMPs.
The cosmological moduli problem (CMP) is also addressed in the case
where a light stringy modulus field $\phi$ persists down to the weak scale.
In Sec. \ref{sec:models}, we assess which popular SUSY models seem
ruled out or are highly implausible based on present LHC data, and
which models remain viable or even favored.
Our main results are collected in Sec. \ref{sec:hllhc}, where we survey
the capacity of high-luminosity LHC to discover natural supersymmetry.
We review HL-LHC prospects to discover
\bi
\item higgsino pair production,
\item gluino pair production,
\item gaugino (mainly wino) pair production,
\item top-squark pair production,
\item light tau-slepton pair production and
  \item heavy Higgs boson $H$, $A$, $H^\pm$ production.
\ei
In the higgsino pair production channel, both ATLAS and CMS have at present
of order $2\sigma$ excesses and it will be exciting to see whether these
build up in the accumulating Run 3 data sample or whether the excesses turn out
to be statistical flukes. Some SUSY models, such as natural anomaly-mediation,
seem completely testable at HL-LHC while others like non-universal Higgs models
(as expected from gravity-mediation) and mirage-mediation seem to have natural
regions of parameter space that can lie beyond the HL-LHC purview.
In Sec. \ref{sec:conclude}, we pull all the pieces together to give an
updated assessment of the status of supersymmetric theories of particle physics.

\section{Status of SUSY after LHC Run 2 and ton-scale noble liquid WIMP searches}\label{sec:status}

SUSY theories have been subject to two sets of strong tests in recent years:
1. the direct search for sparticle and Higgs boson production by the
ATLAS\cite{ATLAS:2024lda} and CMS\cite{Canepa:2019hph} experiments and 2. the direct and indirect search for
SUSY WIMPs by various dark matter search experiments.

\subsection{Post Run 2 status of LHC SUSY searches}

\begin{figure}[tbp]
\begin{center}
  \includegraphics[height=0.5\textheight]{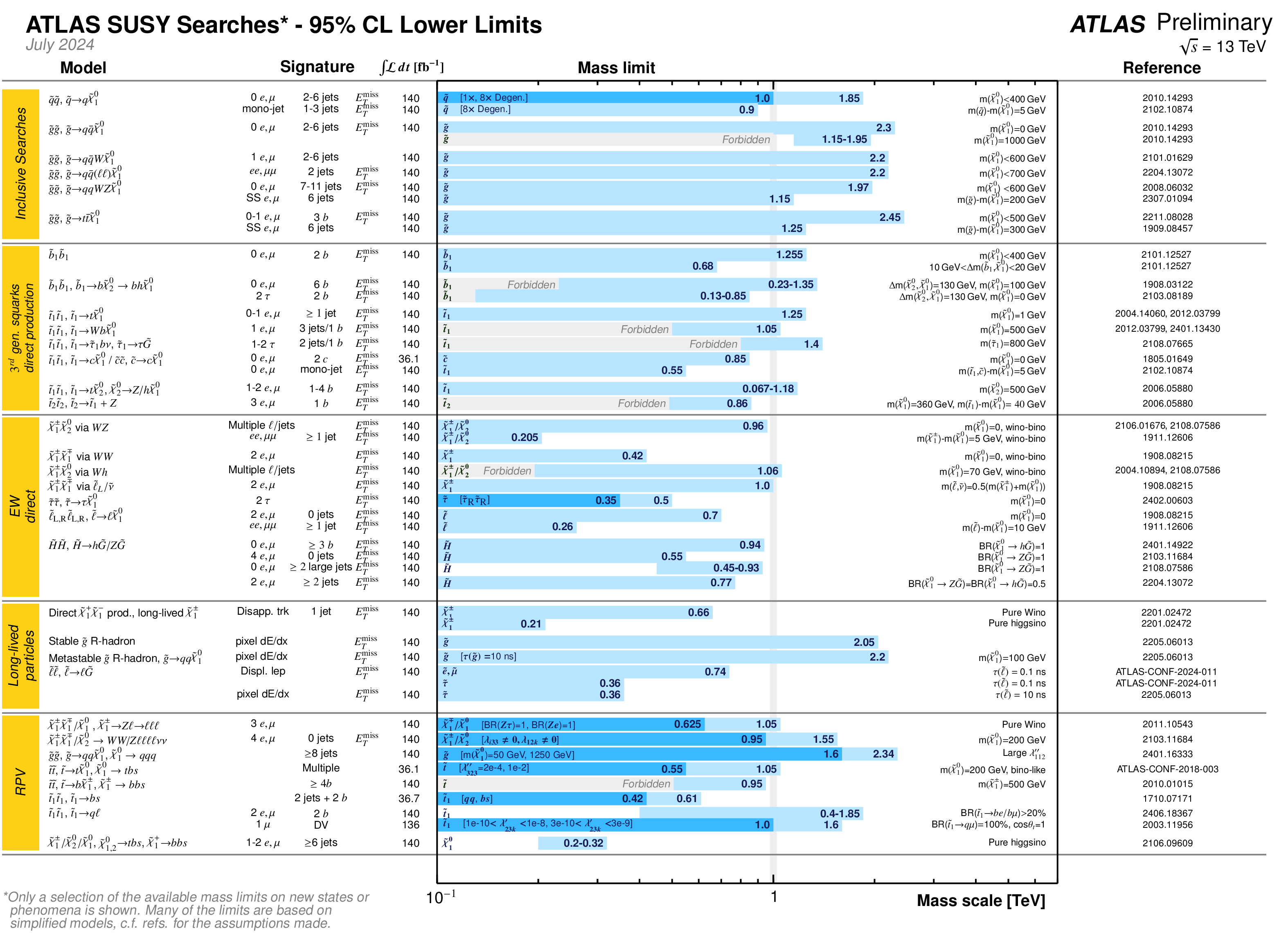}
\caption{Summary of some of the ATLAS run 2 SUSY search limits
  in context of various simplified models.
    \label{fig:atlas}}
\end{center}
\end{figure}

The bulk of LHC SUSY searches have taken place in recent years within the
context of simplified models\cite{LHCNewPhysicsWorkingGroup:2011mji}.
Simplified models usually assume a specific $2\to 2$ sparticle
(or $2\to 1$ Higgs) production mode along with 100\% branching fraction
of the new matter state into a particular decay mode. As such,
simplified models do not correspond to actual SUSY models where
typically hundreds of production modes may be simultaneously
occuring whilst the sparticles may undergo rather complex cascade
decays\cite{Baer:1986au}.\footnote{Another unfortunate outcome of the advance of
  simplified models may be that they have deterred communication
  between experimentalists and their theoretical colleagues.}
A consequence of the rise of simplified models has been a concommitant
rise in a mini-industry of computer codes designed to bridge the gap
between simplified models and models\cite{Drees:2013wra,Kraml:2013mwa,GAMBIT:2017yxo}.
Another consequence is that simplified models allow for a very large number
of different experimental SUSY analyses, and these can be motivated by the
strategy of ``leaving no rock unturned'' or
``searching beneath each lamp post''.
The bevy of SUSY searches, as illustrated partially in Fig. \ref{fig:atlas},
gives the impression of an exhaustive effort despite the case that most
analyses are performed in rather implausible settings. This gives the
general impression that LHC experiments have searched over the bulk of SUSY
parameter space whereas as we argue later, in some contexts -- for instance,
in the context of our current understanding of string theory -- the searches
have only begun to explore the lower edge of plausible parameter
space.\footnote{It may be that the bulk of expected 20th century parameter
space has been explored.}

\subsubsection{Gluino pair production}

\begin{figure}[tbp]
\begin{center}
  \includegraphics[height=0.3\textheight]{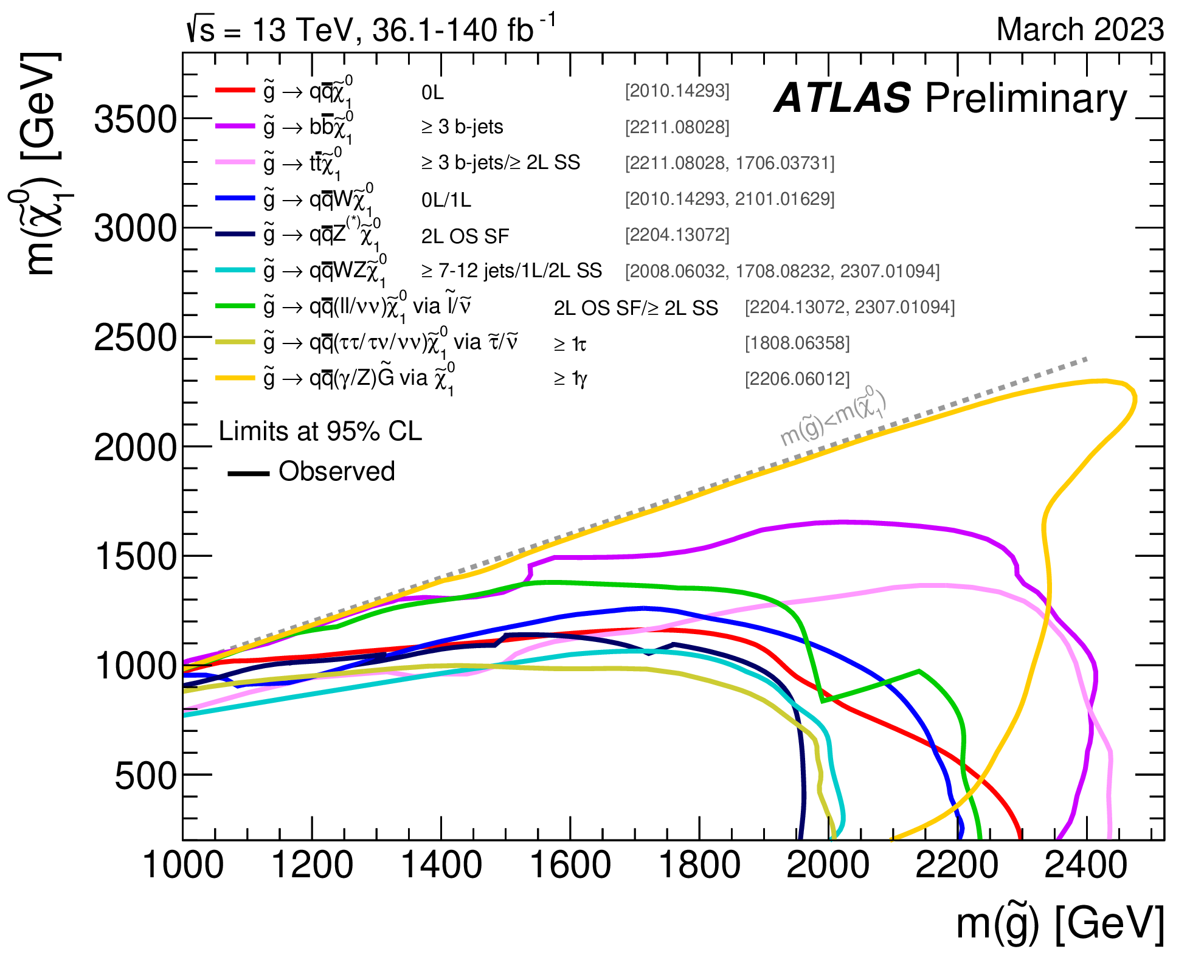}
  \includegraphics[height=0.3\textheight]{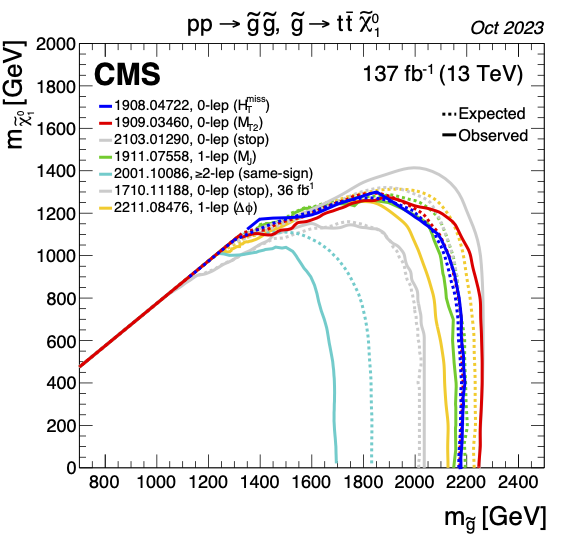}
\caption{Summary of ATLAS and CMS run 2 gluino pair production search limits
  in the context of various simplified models.
    \label{fig:lhc_gl}}
\end{center}
\end{figure}

We will not review {\it all} LHC SUSY analyses, but rather only a subset that
has the most significant impact on contemporary understanding of WSS.
We begin with Fig. \ref{fig:lhc_gl} which shows {\it a}) the ATLAS
and {\it b}) the CMS results from various searches for gluino pair production
followed by assumed decay modes (listed in the colored legends).
The results are plotted in the $m_{\tg}$ vs. $m_{\tchi_1^0}$ plane, and the arXiv number for each search report is also listed.
We will see in Sec. \ref{sec:nat} that third generation, especially top-squark
contributions to the weak scale can be substantial whilst first/second
generation contributions are Yukawa-suppressed.
The expected lighter top-squarks, whose masses are even more suppressed/split by substantial squark mixing, and the large top Yukawa coupling, all work to enhance gluino decays to $t\tst_1$ 
or $t\bar{t}\tchi$ final states\cite{Baer:1990sc,Baer:1991xs,Baer:1998bj}
(where $\tchi$ denotes any of the neutralinos (or charginos) in the case of $\tg\to t\bar{b}\tchi_{1,2}^-$ decays).
Meanwhile, gluinos contribute to the weak scale at 2-loop level so their naturalness bounds are rather weak, and they can easily range up to the 6-9 TeV level\cite{Baer:2018hpb}.
The upshot is that $\tg$ decay to third generation particles is typically highly favored in generic situations, and these decay modes are of particular interest. From the ATLAS results, the purple curves show a reach to
$m_{\tg}\sim 2.4$ TeV for light $m_{\tchi_1^0}$, and so these bounds we expect to
obtain, more or less. They both require $\ge 3$ $b$-jets in the final state.
The CMS results are only shown for $\tg\to t\bar{t}\tchi_1^0$,
but using different final state configurations based on the number of isolated leptons. Their bounds typically range up to $m_{\tg}\sim 2.2$ TeV unless
one moves into the $m_{\tg}\sim m_{\tchi_1^0}$ compressed region
(which is not to be expected in natural models where light higgsinos with mass $\sim 100-350$ GeV are expected).

\subsubsection{Squark and top-squark pair production}

\begin{figure}[tbp]
\begin{center}
  \includegraphics[height=0.3\textheight]{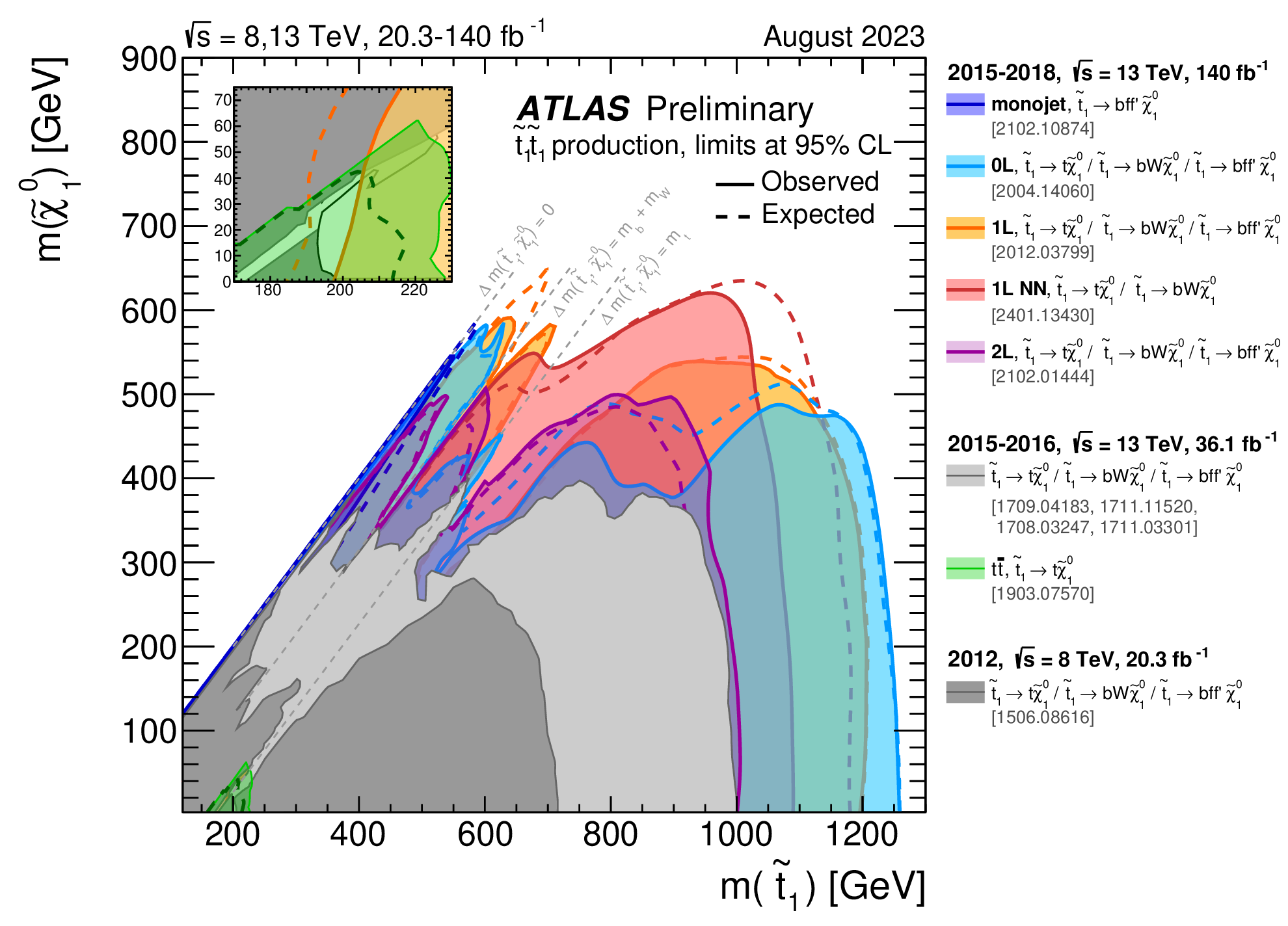}
  \includegraphics[height=0.3\textheight]{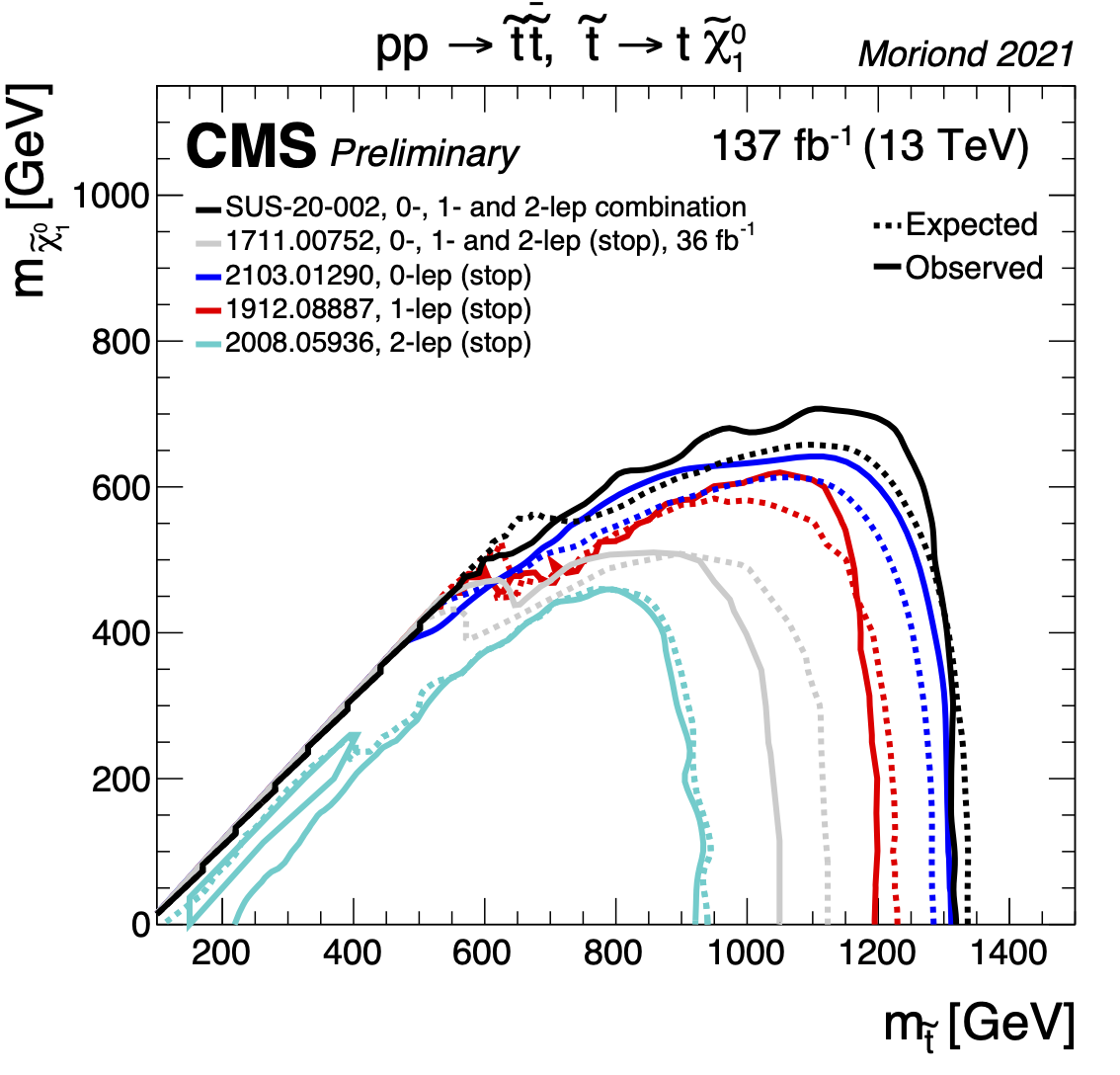}
\caption{Summary of ATLAS and CMS run 2 top-squark pair production search limits
  in the context of various simplified models.
    \label{fig:lhc_t1}}
\end{center}
\end{figure}

First/second generation squark/slepton contributions to the weak scale are tiny for TeV-scale masses and so they can range up to 30-40 TeV 
without affecting naturalness.  
Furthermore, lack of flavor and CP violation may imply that these
actually live in the tens-of-TeV range\cite{Dine:1993np,Cohen:1996vb,Arkani-Hamed:1997opn} 
(decoupling solution to the SUSY flavor and CP problems), 
so it may not be surprising that they have not been detected.
More meaningful bounds come from LHC top-squark searches, where $m_{\tst_1}$
is typically bounded by 2-3 TeV from naturalness\cite{Baer:2015rja}.
The ATLAS/CMS limits for these sparticles are shown in Fig. \ref{fig:lhc_t1}
for several assumed decay modes, usually $\tst_1\to t\tchi_1^0$,
and final state configurations. For $m_{\tchi_1^0}\alt 500$ GeV, these limits
require $m_{\tst_1}\agt 1.2-1.3$ TeV. These limits should also roughly hold
if other expected decays such as $\tst_1\to b\tchi_1^+$ and $t\tchi_2^0$ are included. 

A good deal of initial effort was expended on searching the
compressed region where $m_{\tst_1}\simeq m_t+m_{\tchi_1^0}$, motivated by
earlier naturalness measures that required much lighter top-squarks. 
These compressed regions have been largely excluded by dedicated
experimental searches. Such light top-squarks in the compressed region 
would also be hard to reconcile with the rather large measured value of
$m_h\simeq 125$ GeV which prefers heavier, TeV-scale top squarks.

\subsubsection{Gaugino pair production}

\begin{figure}[tbp]
\begin{center}
  \includegraphics[height=0.4\textheight]{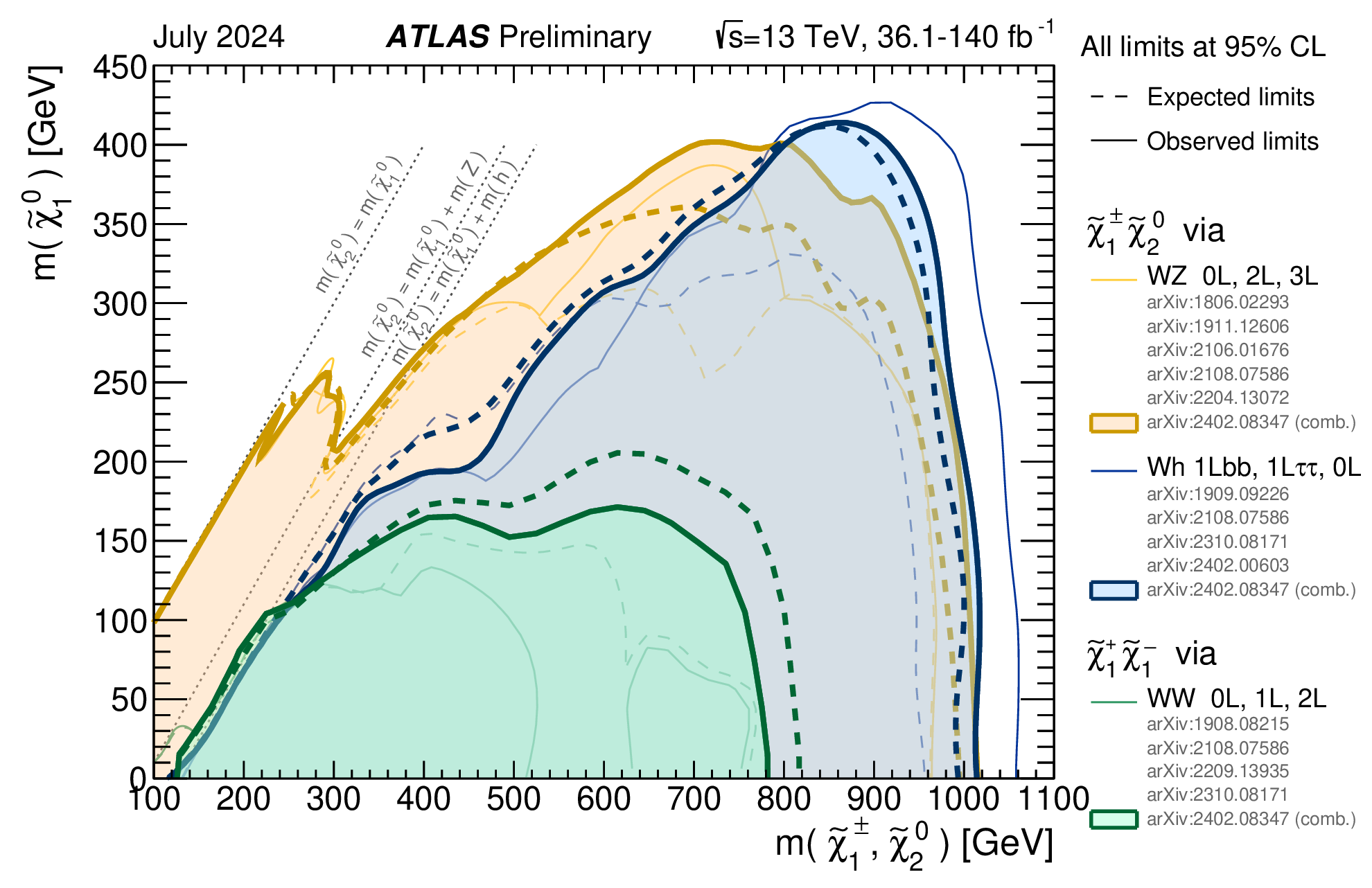}
  \caption{Summary of ATLAS search for gaugino pair production
    in LHC run 2.
        \label{fig:atlas_w1z2}}
\end{center}
\end{figure}

Gaugino pair production is also a lucrative target for LHC searches.
Rather strong limits have been placed on $\tchi_1^\pm\tchi_2^0$ production
where higgsinos are assumed heavy and the produced electroweakinos (EWinos)
are wino-like with $\tchi_1^0$ being bino-like.
The dominant decays are then likely
to be $\tchi_2^0\to h\tchi_1^0$ and $\tchi_1^\pm \to W\tchi_1^0$ leading to a
$Wh+\eslt$ final state\cite{Baer:2012ts}.
Search limits require $m_{\tchi_1^\pm}\agt 1$ TeV for $m_{\tchi_1^0}\alt 250$ GeV
as shown in Fig. \ref{fig:atlas_w1z2}.
The ``golden'' trilepton channel\cite{Baer:2012wg} which was once favored
for Tevatron SUSY searches $\tchi_1^\pm\tchi_2^0\to 3\ell+ \eslt$ gives similar limits, but would require the 
$\tchi_2^0\to h\tchi_1^0$ (spoiler) decay mode
to be somehow suppressed.

\subsubsection{Higgsino pair production}

Since the previously mentioned gaugino pair production search scenarios require lighter gauginos with heavier higgsinos, they now seem rather
implausible since $\mu\sim m_{weak}\sim 100$ GeV is required for
naturalness\cite{Chan:1997bi}.
Simply put, the $\mu \hat{H}_u \hat{H}_d$ term in the MSSM
Lagrangian is SUSY conserving rather than (soft) SUSY breaking.
This leads to the famous SUSY $\mu$ problem: why isn't $\mu\sim m_P$
whereas phenomenology requires $\mu\sim m_{weak}$ (twenty solutions
to the SUSY $\mu$ problem are reviewed in Ref. \cite{Bae:2019dgg}).
It also implies that $\mu$ feeds mass to the $W$, $Z$ and $h$ particles
as well as the partner higgsinos:
all should have mass $\sim 100-350$ GeV range otherwise
a LHP ensues. Given present limits, this often means that the four
higgsinos $\tchi_{1,2}^0$ and $\tchi_1^\pm$ should be the lightest
supersymmetric particles all with mass $\sim \mu$ and with mass splittings
of order $\Delta m\sim 5-20$ GeV\cite{Giudice:1995np}.
This implies that higgsino pair production should be a lucrative target for
LHC SUSY searches. With the higgsino-like
$\tchi_1^0$ as lightest SUSY particle (LSP), then the visible energy from
heavier higgsino decay should be very soft owing to the small
inter-higgsino mass splittings. Thus, light higgsinos could be produced
at considerable rates at LHC but the softness of their visible decay products
makes direct searches more difficult.
It was proposed in Ref. \cite{Baer:2011ec} to use a soft dimuon trigger
to try to capture higgsino pair production $pp\to\tchi_1^0\tchi_2^0$ followed by
$\tchi_2^0\to\tchi_1^0\ell^+\ell^-$ at the LHC. A characteristic signature
would be that the signal dilepton excess would be kinematically bounded
in invariant mass by $m(\ell^+\ell^- )<m_{\tchi_2^0}-m_{\tchi_1^0}$.

\begin{figure}[tbp]
\begin{center}
  \includegraphics[height=0.25\textheight]{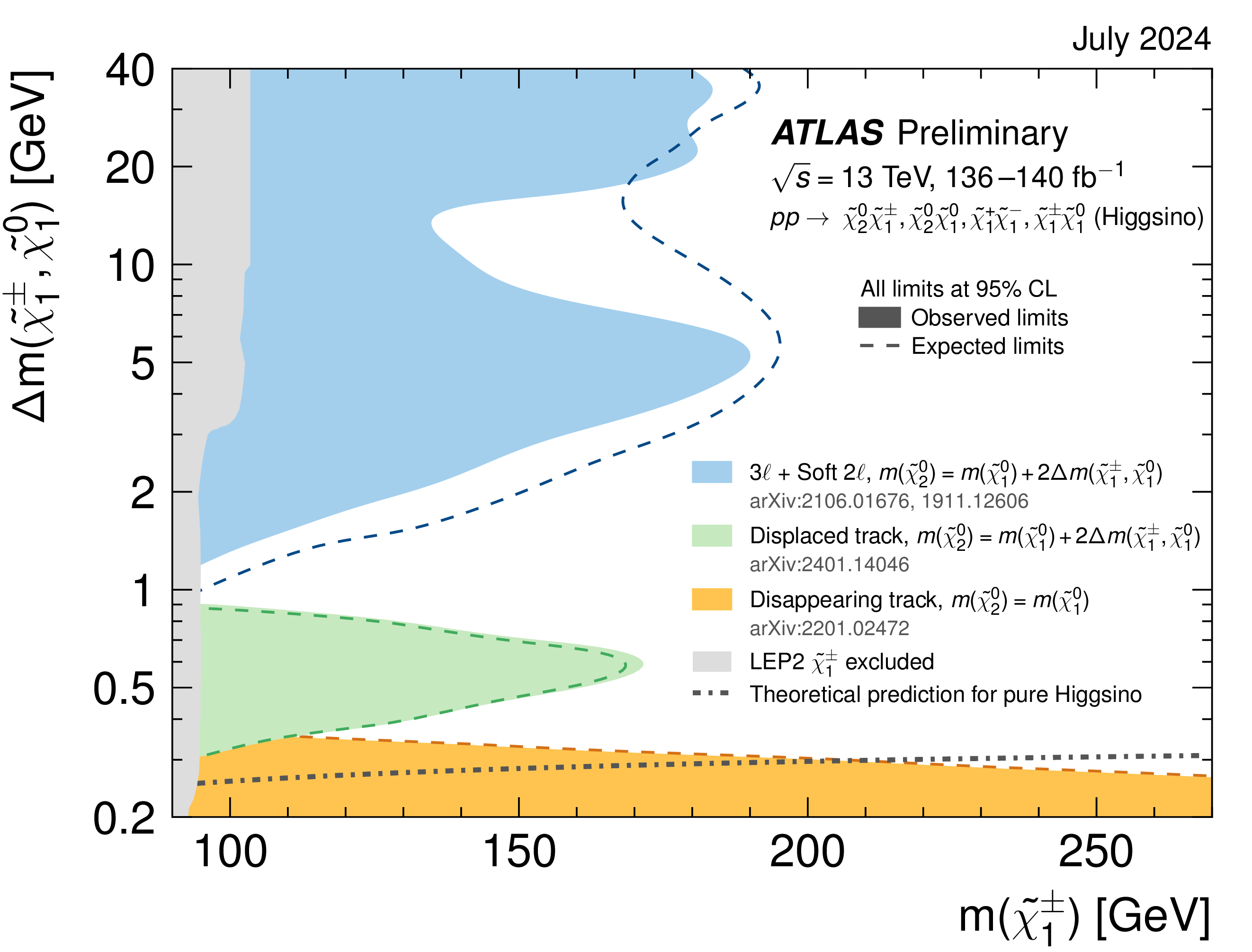}
  \includegraphics[height=0.25\textheight]{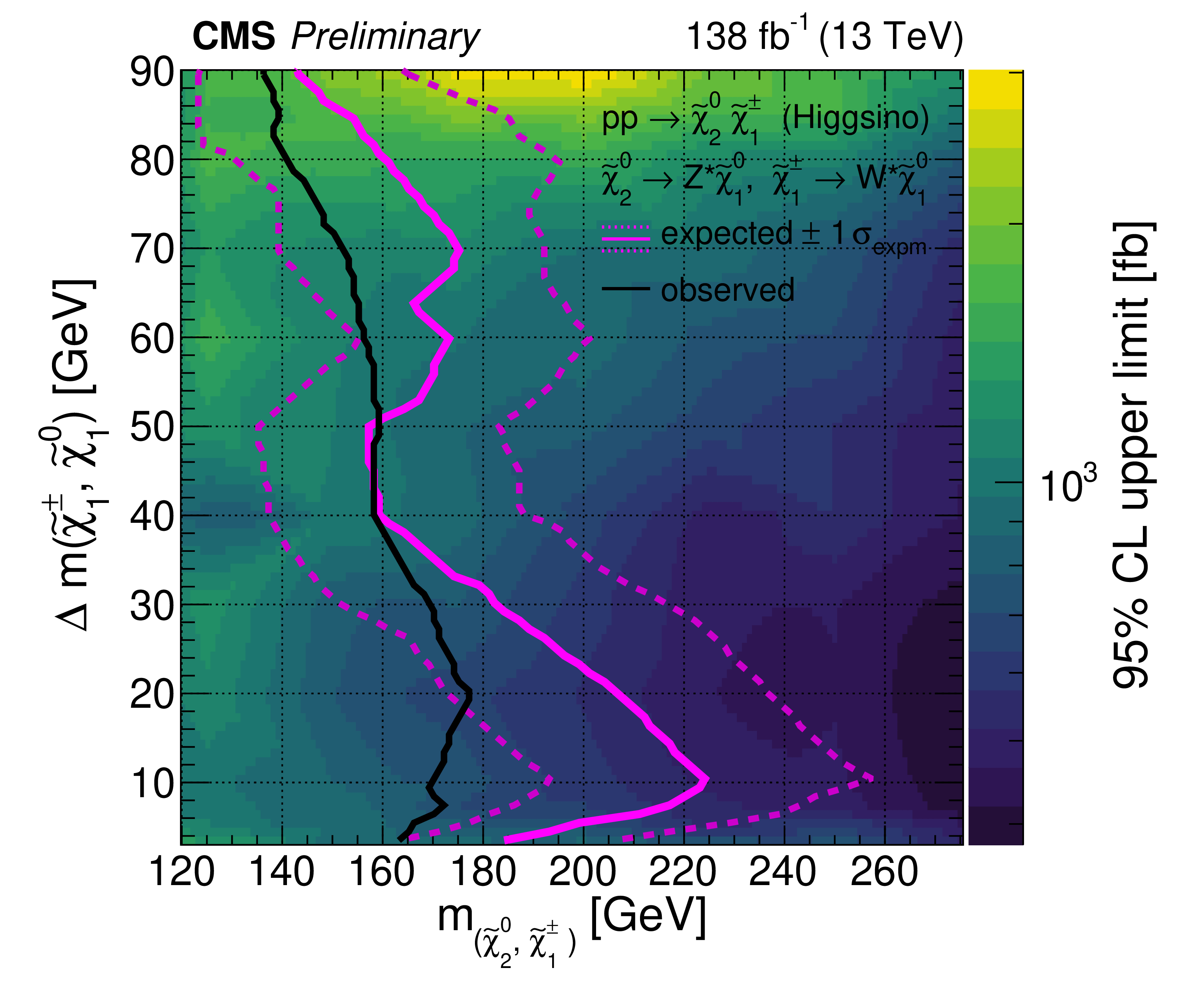}
\caption{Summary of ATLAS and CMS run 2 higgsino pair production 
$pp\to\tchi_2^0\tchi_1^0$ and $\chi_1^\pm\tchi_2^0$ search limits
  assuming $\tchi_2^0\to\tchi_1^0\ell\bar{\ell}$ decay.
    \label{fig:lhc_ho}}
\end{center}
\end{figure}

An alternative approach\cite{Han:2014kaa,Baer:2014kya,Han:2013usa} is to
search for $pp\to \tchi_1^0\tchi_2^0 g$ where a hard initial state gluon
or quark radiation would provide a trigger and also boost the
final state leptons to higher $p_T$ values.
Dominant backgrounds include $\tau\bar{\tau}j$,
$t\bar{t}$ and $WWj$ production.
Reconstructing the $\tau$s and requiring $m_{\tau\tau}^2<0$, 
or alternatively stronger angular cuts\cite{Baer:2021srt}, helps reduce the
ditau background. Both ATLAS\cite{ATLAS:2019lng} and CMS\cite{CMS:2021edw} have performed searches in this channel
and results are shown in Fig. \ref{fig:lhc_ho}.
The discrepancy between the expected and observed limit contours helps
show that each experiment actually has an approximate $2\sigma$ excess
in this channel. It will be exciting to see if this excess persists
or is diluted with upcoming data from LHC Run 3 and HL-LHC!

\subsubsection{Gaugino pair production with decay to light higgsinos}

\begin{figure}[tbp]
\begin{center}
  \includegraphics[height=0.4\textheight]{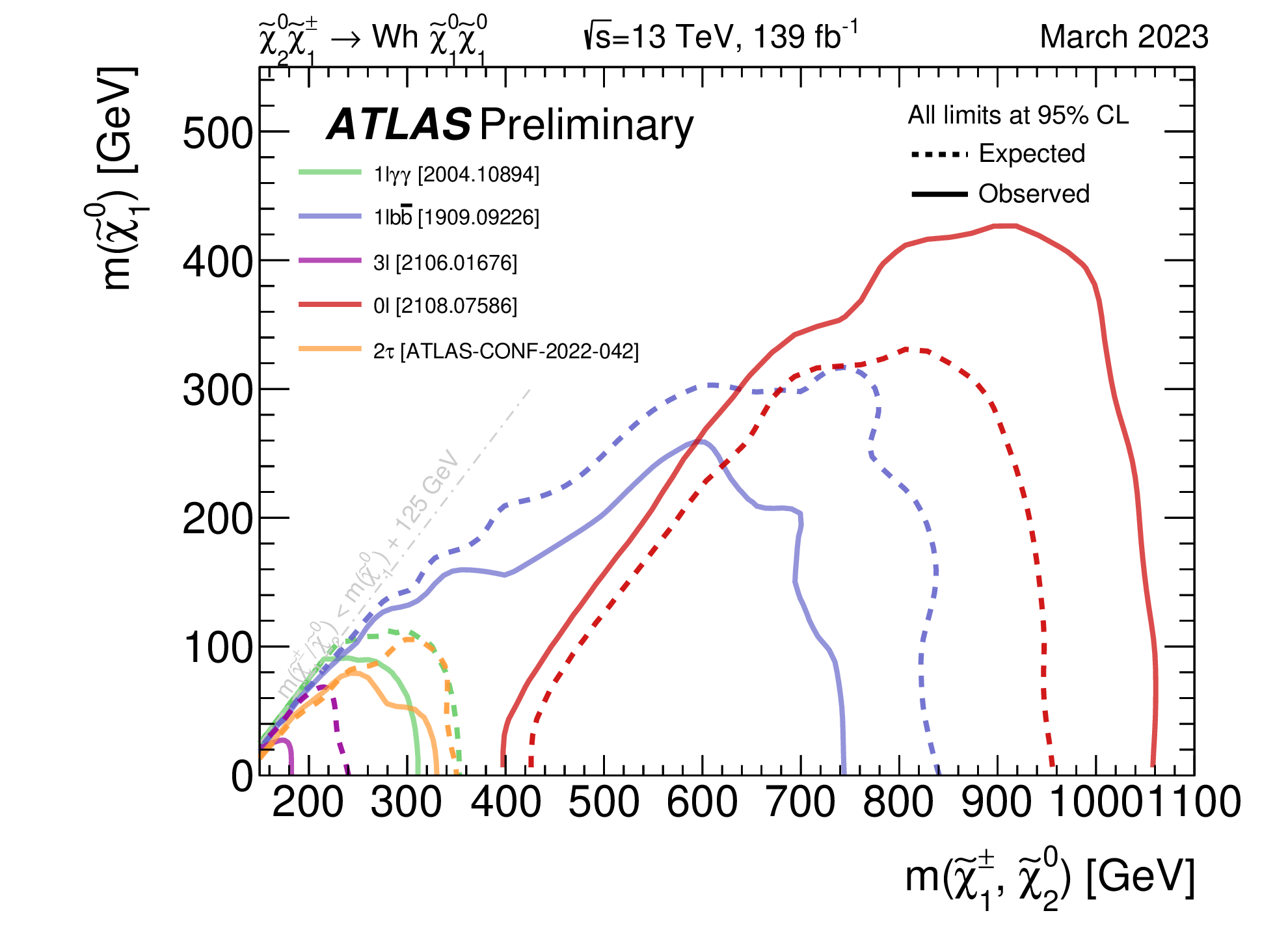}
  \caption{Summary of the ATLAS search for gaugino pair production
    in LHC run 2.
        \label{fig:atlas_gaugino}}
\end{center}
\end{figure}

Gaugino pair production (especially wino pairs) followed by decay to light
higgsinos also offers a lucrative SUSY discovery channel.
In this case, one may search for $pp\to \tchi_2^\pm\tchi_{3,4}^0$ where
$\tchi_2^+\to W^+\tchi_{1,2}^0$ and $\tchi_4^0\to W^\pm\tchi_1^\mp$
or $h\tchi_{1,2}^0$.
A search by ATLAS was made, assuming boosted hadronic $W$ (or $Z$ or $h$)
boson decays and the results are shown in Fig. \ref{fig:atlas_gaugino}
(red curve). For $m_{\tchi_1^0}\sim 200$ GeV, the range
$m_{\tchi_4^0}\sim 550-1050$ GeV is excluded. The lower range below 550 GeV is
allowed because the hadronic vector boson decays are not sufficiently boosted (although other search channels such as the listed $\ell b\bar{b}$ 
channel may exclude below 550 GeV).

Alternatively, one can search for leptonic $W$ decays in this channel,
which leads to a very low background same-sign diboson final state
(SSdB)\cite{Baer:2013yha}: $pp\to\tchi_2^\pm\tchi_4^0\to W^\pm W^\pm+\eslt$.
These events are quite different from same-sign dileptons arising from
gluino pair production in that they would be relatively jet-free aside from
the usual initial state radiation (ISR).
Due to the two leptonic branching fractions, these signal rates
would be expected to lie at the few event level for LHC Run 2, but they
should become increasingly important at LHC Run 3 and HL-LHC\cite{Baer:2017gzf}.

\subsubsection{RPV SUSY searches}

In contrast to the SM, the general MSSM admits both trilinear and bilinear
superpotential terms which lead to baryon (B) and lepton (L) number
violating interactions (for a review of $R$-parity violating processes, see,
{\it e.g.}, Ref's. \cite{Barger:1989rk,Dreiner:1997uz}).
There are very strong limits on some of these terms.
In particular, from the lifetime of the proton one can deduce
that the product of RPV couplings
\be
\lambda^\prime_{11k}\lambda^{\prime\prime}_{11k}\alt 10^{-26}\left(
\frac{m_{\td_k}}{100\ {\rm GeV}}\right)^2
  \label{eq:rpv}
  \ee
  where $k$ is a generation label.
  These sorts of limits motivate the assumption of $R$-parity conservation (RPC),
  where $R=(-1)^{3(B-L)-2s}$. Some consequences of RPC are that sparticles
  would be produced in pairs at collider experiments, that sparticles decay to
  other sparticles and that the lightest $R$-odd particle is absolutely stable
  (and can thus serve as a possible dark matter candidate).

  With 45 possible trilinear RPV couplings available, many LHC searches for
  RPV SUSY have   been performed, all with null results.
  (Some RPV SUSY searches from the ATLAS group are listed in
  Fig. \ref{fig:atlas}.)
  While it is logically possible that $R$-parity is violated, one would
  have to grapple with why some couplings could be large while others respect
  bounds like Eq. \ref{eq:rpv}.
  In many studies, the RPC seems ad-hoc. However, recent examination of discrete
  $R$-symmetries, which can arise from superstring compactifications, show that
  these can provide an underlying mechanism forbid direct RPV terms and 
  generate the $U(1)_{PQ}$ symmetry needed for the strong CP problem\cite{Baer:2018avn}.
  In this case, both axion and WIMP dark matter can emerge from the same underlying symmetry condition. However, recent work\cite{Baer:2025oid} implementing a $\mathbb{Z}_{4,8}^R$ symmetry needed to solve the SUSY $\mu$ problem shows that RPV couplings at the 
  $\lambda,\ \lambda^\prime,\ \lambda^{\prime\prime}\sim (f_a/m_P)\sim 10^{-7}$ level from non-renormalizable operators may appear which preserve the collider $\eslt$ signals but cause the relic WIMPs to decay shortly before BBN occurs in the early universe. 
  
\subsection{Status of SUSY after ton-scale noble liquid WIMP searches}
\label{ssec:wimp}

It is often touted that one of the great successes of SUSY under RPC is
that it predicts a dark matter candidate, usually the lightest neutralino,
which would be a weakly-interacting massive particle,
or WIMP\cite{Ellis:1983ew}.
Other possible SUSY DM candidates include axinos from the
PQ-extended MSSM\cite{Covi:1999ty}, the gravitino\cite{Steffen:2007sp},
or even hidden sector dark matter\cite{Acharya:2016fge}

\begin{figure}[tbp]
\begin{center}
  \includegraphics[height=0.5\textheight ,angle =-90]{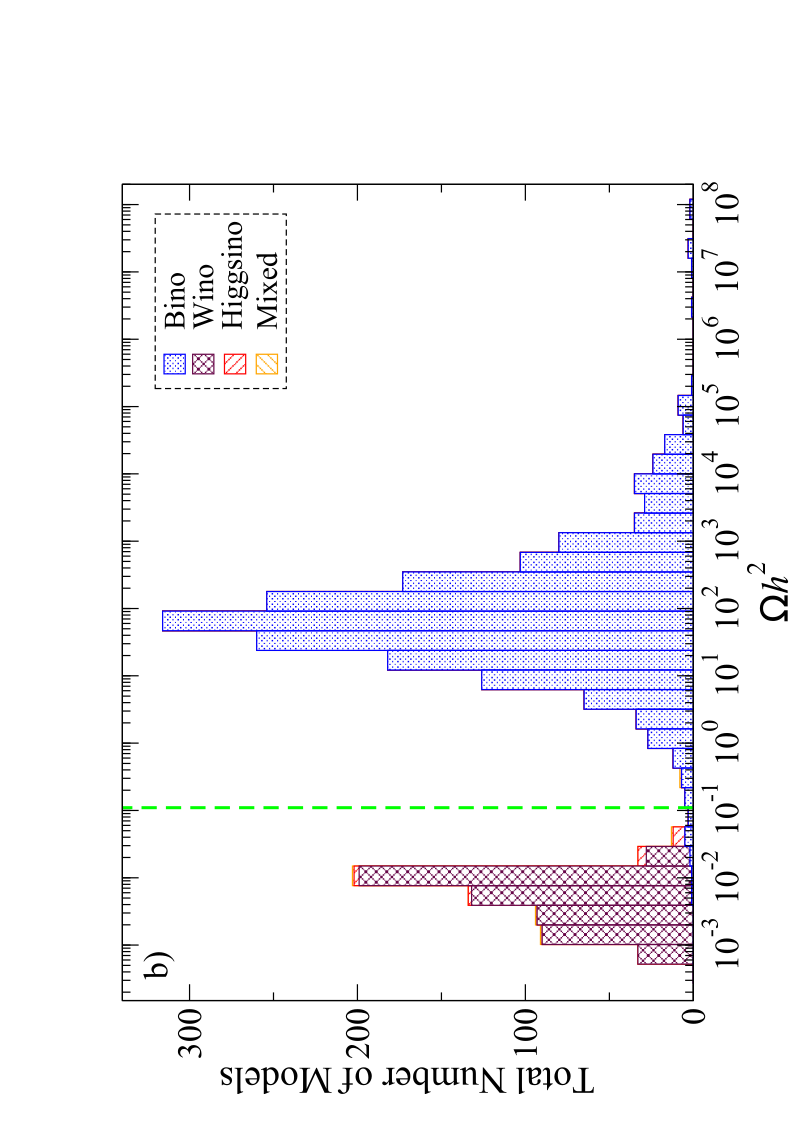}
  \caption{Neutralino relic abundance from a scan over the 19 parameter SUGRA model with $\mu <500$ GeV\cite{Baer:2010wm}.
        \label{fig:Oh2}}
\end{center}
\end{figure}

The SUSY WIMP is sometimes motivated by the so-called ``WIMP miracle''\cite{Baltz:2006fm}
in that the calculated thermally-produced (TP) dark matter abundance turns  out to be near the measured value in our universe ($\Omega_{\tchi_1^0}^{TP}h^2\simeq 0.12$).
This is not exactly true as shown in Fig. \ref{fig:Oh2}, where we see
that generically bino-like WIMPs are thermally overproduced and wino- or
higgsino-like WIMPs are thermally underproduced: one must have just the right
higgsino-gaugino admixture (well-tempered neutralino\cite{Arkani-Hamed:2006wnf,Baer:2006te}, WTN) or rely on additional
WIMP annihilation conditions (co-annihilation, resonance annihilation)
to match the measured dark matter abundance.\footnote{The WIMP miracle is obtained in the MSSM for sleptons in the 50 GeV range\cite{Baer:1995nc} (which is now excluded).}
In addition, an axion may be required to solve the strong CP problem of QCD
and this would bring in an additional dark matter candidate, one that is
produced quite differently via coherent field oscillations\cite{Baer:2014eja}.

\begin{figure}[tbp]
\begin{center}
  \includegraphics[height=0.4\textheight]{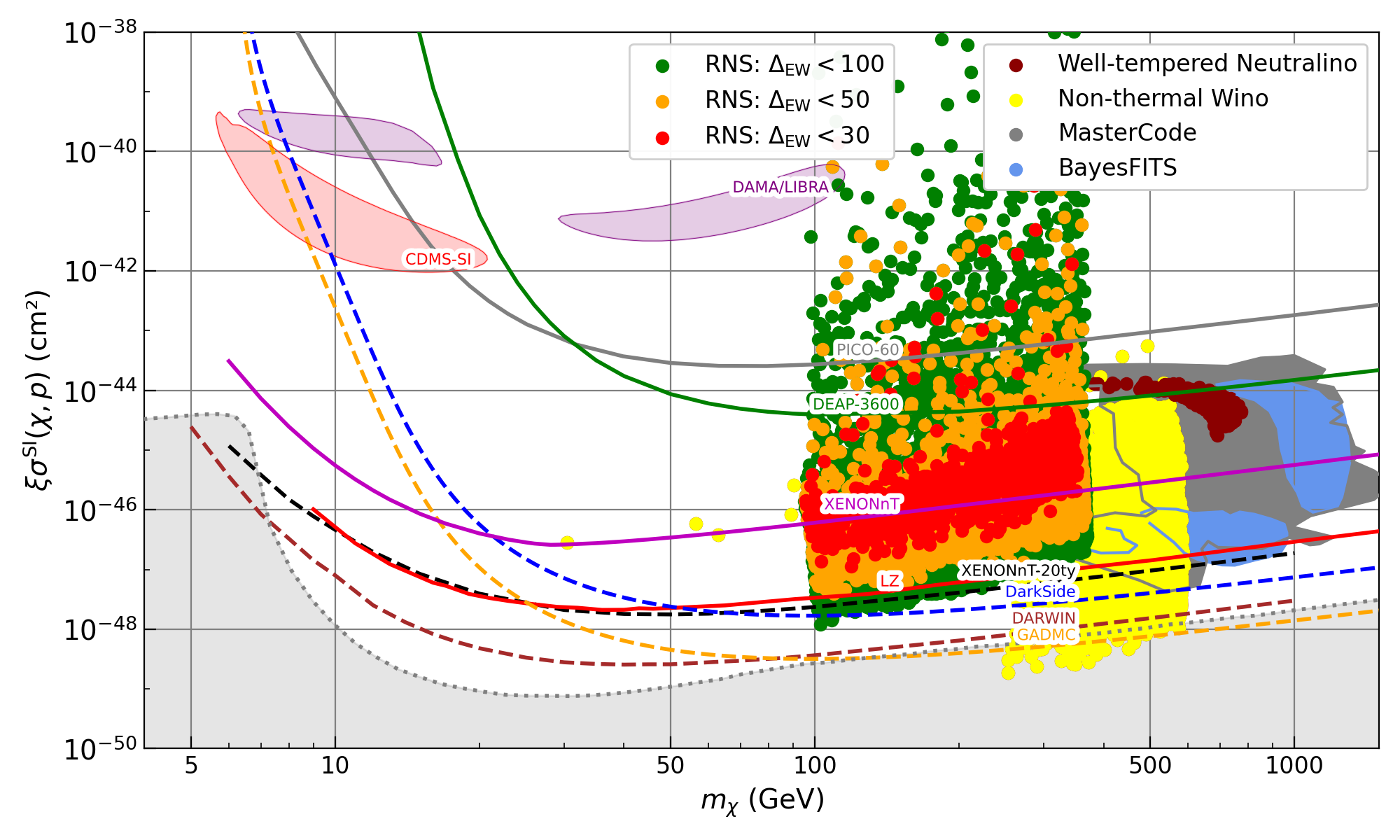}\\
    \includegraphics[height=0.4\textheight]{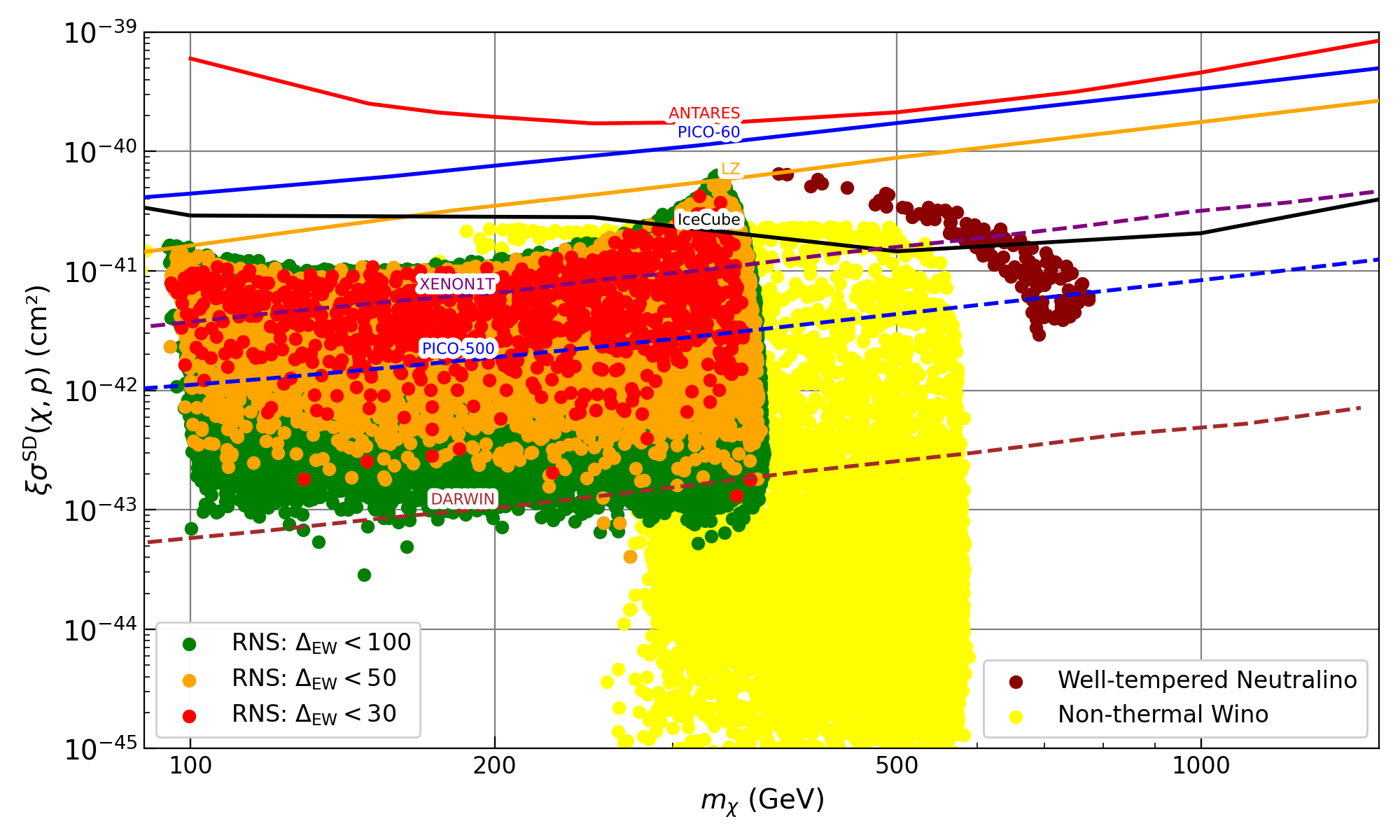}
    \caption{Summary of {\it a}) spin-independent (SI)
      and {\it b}) spin-dependent (SD) search limits on various
      supersymmetric models along with future projections.
      Dashed curves denote expected reaches of future experiments.
    Updated results from Ref. \cite{Baer:2016ucr}.
      \label{fig:sig_sig}}
\end{center}
\end{figure}

In Fig. \ref{fig:sig_sig}{\it a}), we show exclusion limits on {\it spin-independent}
(SI) WIMP-nucleon scattering cross sections vs. $m(wimp)$ from a variety of
recent WIMP search experiments (solid lines) including
LZ (2024)\cite{LZCollaboration:2024lux}.
The expected rate from radiatively-driven natural SUSY models (RNS) is tempered by a factor
$\xi\equiv \Omega_{\tchi_1^0}^{TP}h^2/0.12$
to account for cases where the WIMP relic abundance forms only a fraction
of the total dark matter (the remainder may be composed for instance of axions).
This only occurs for the RNS curves in the plot where higgsino-like WIMPs
are thermally underproduced and the remaining abundance is assumed to be SUSY DFSZ\cite{Dine:1981rt} axions.
We also show projections of future WIMP search experiments (dashed lines).
And we show predictions from a variety of SUSY models.
Of particular note are the maroon dots labeled WTN, which shows that
well-tempered neutralinos\cite{Arkani-Hamed:2006wnf} (as occur in the focus point region of the
CMSSM model\cite{Feng:2000gh}) are now ruled out by several orders of magnitude.
Also, the TP natural SUSY higgsino-like WIMP models now appear
ruled out, but only under the assumption of thermally-produced relic abundance.
In these models with mixed axion/WIMP dark matter, there exist a variety
of non-thermal processes which can enhance or diminish the relic abundance
from its thermally produced value\cite{Baer:2014eja}. And the aforementioned
implementation of discrete $R$-symmetries\cite{Baer:2025oid} allows for the $\tchi_1^0$ to all
decay via tiny RPV couplings before the onset of BBN leaving a universe with 
all-axion CDM.

In Fig. \ref{fig:sig_sig}{\it b}), we show exclusion limits on {\it spin-dependent}
(SD) WIMP-nucleon scattering cross sections vs. $m(wimp)$ from a variety of
recent WIMP search experiments (solid lines).
The previous strongest limits came from a combination of
PICO-60\cite{PICO:2019vsc} for $m_{\tchi_1^0}\alt 190$ GeV and IceCube high energy neutrino
searches from $\tchi\tchi\to WW$ in the Sun\cite{IceCube:2016yoy}.
For IceCube, the rates mainly depend on the Sun's ability to sweep up
WIMPs via $\tchi p$ scattering as the Sun traverses its orbit in the Milky Way.
In this case, the natural models (orange- and red-shaded regions)
have not yet been accessed by SD WIMP searches.
The recent LZ2024 results\cite{LZCollaboration:2024lux} now provide the
strongest limits for SD WIMP direct detection, but have just started to
exclude the upper portions of orange and red RNS parameter space.

\begin{figure}[tbp]
\begin{center}
  \includegraphics[height=0.4\textheight]{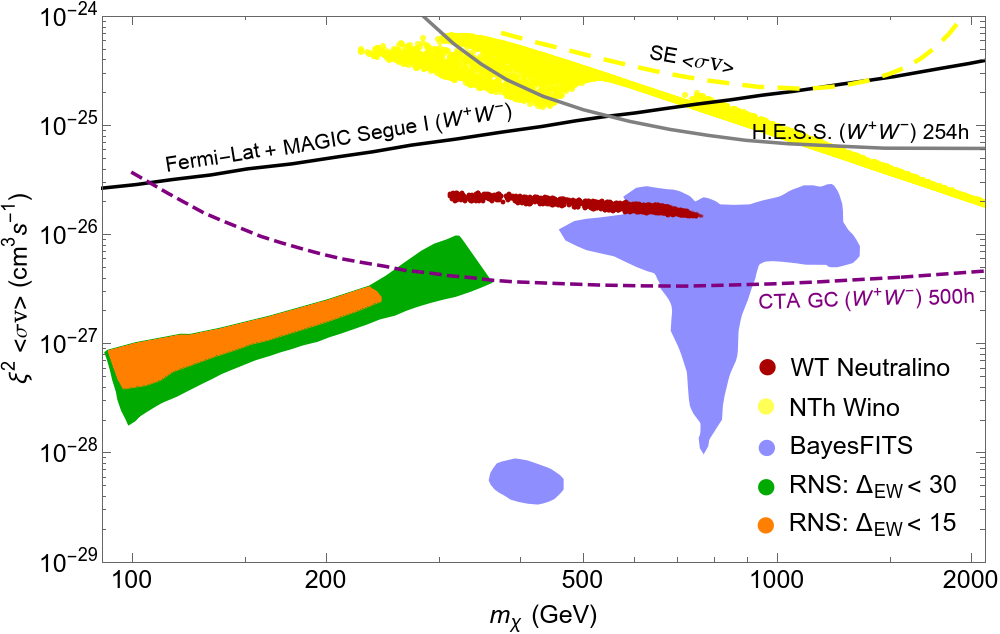}
  \caption{Summary of WIMP annihilation search limits on various
    supersymmetric models along with future projections.
    From Ref. \cite{Baer:2016ucr}.
        \label{fig:sigv}}
\end{center}
\end{figure}

In Fig. \ref{fig:sigv}, we show limits from searches for 
$\tchi\tchi\to {\rm SM\ particles}$
annihilation in various cosmic locations.
The overall rate is now tempered by a factor $\xi^2$ since one needs
$\tchi\tchi$ annihilation. This reduces the expected rates of natural
light higgsino-pair annihilation to well below experimental limits.
The non-thermally produced wino-like WIMPs expected from models like
mAMSB\cite{Moroi:1999zb} seem to be largely if not totally excluded
by a combination of Fermi-LAT/MAGIC\cite{MAGIC:2016xys}searches in dwarf spheroidal galaxies at
the lower mass range and from HESS\cite{HESS:2018cbt} searches at the high end.
The dashed yellow line denotes the expected rate if one includes Sommerfeld
enhancement (SE) in the annihilation rate.
In this latter case, 100\% wino-like WIMP dark matter would be completely
excluded owing to its large wino-wino annihilation rate\cite{Cohen:2013ama,Fan:2013faa,Baer:2016ucr}. 

\section{On the naturalness issue}
\label{sec:nat}

The lack of appearance of supersymmetric matter at the LHC has engendered
a {\it naturalness crisis} within the HEP community\cite{Craig:2013cxa,Lykken:2014bca,Dine:2015xga}. During pre-LHC times,
it was widely expected that superparticles would exist with mass values
comparable to $m_{weak}\sim 100$ GeV\cite{Ellis:1986yg,Barbieri:1987fn,Dimopoulos:1995mi,Anderson:1994tr,Chankowski:1997zh,Feng:2013pwa}.
There was already concern for a LHP after
the non-appearance of charginos at LEP2\cite{Barbieri:2000gf}.

\subsection{Log-derivative measure $\Delta_{p_i}$}
\label{ssec:DBG}

The above-mentioned works were all based on the log-derivative naturalness measure
\be
\Delta_{p_i}\equiv max_i \Bigg|\frac{\partial\log m_Z^2}{\partial\log p_i}\Bigg| =
max_i\Bigg|\frac{p_i}{m_Z^2}\frac{\partial m_Z^2}{\partial p_i}\Bigg|
\label{eq:Dpi}
\ee
introduced in \cite{Ellis:1986yg} and where the $p_i$ are 
the $i=1-N$ fundamental parameters of the theory in question.
The starting point of these papers was usually the tree-level
minimization condition of the MSSM scalar potential:
\be
m_Z^2/2=\frac{m_{H_d}^2-m_{H_u}^2\tan^2\beta}{\tan^2\beta -1}-\mu^2 
\label{eq:mzs1}
\ee
where the $m_{H_{u,d}}^2$ are the weak scale Higgs soft terms, $\mu$ is the
(SUSY conserving) superpotential Higgs mass term and $\tan\beta \equiv v_u/v_d$
is the ratio of Higgs field vevs.
Considering the pMSSM\cite{Arbey:2012bp} (19 free SUSY parameters 
all defined at the weak scale),
and taking the squared mass parameters as fundamental, then
\be
\Delta_{p_i}(pMSSM) \simeq 2\times max\left\{ m_{H_d}/\tan\beta ,m_{H_u}, \mu \right\}/m_Z^2
\ee
and we would expect $\mu\sim m_Z$ and $m_{H_u}^2$ --
which usually gets driven negative at the EW scale in order to trigger radiative electroweak symmetry breaking (REWSB) -- to be $|m_{H_u}^2|\sim m_Z^2$. Since the squared mass parameters are fundamental, we define $m_{H_{d,u}}=\sqrt{|m_{H_{d,u}}^2|}$.
Alternatively, one may expand $m_{H_u}^2$ (and $\mu$) in terms of
high-scale parameters and take these as the $p_i$ (see {\it e.g.}
the expansion Eq. 28 in Ref. \cite{Feng:2013pwa}).
In the works cited, the $p_i$ were usually taken as GUT-scale parameters
of gravity-mediated SUSY breaking models such as $m_0$, $m_{1/2}$, $A_0$
etc. with some argument as to whether the top- or other Yukawa couplings
should be included. The derived limits for $\Delta_{p_i}\alt 30$
from Ref's \cite{Barbieri:1987fn,Dimopoulos:1995mi} are given in
Table \ref{tab:nat} where we see that indeed sparticles should lie
typically below the 500 GeV value.
%
\begin{table}\centering
\begin{tabular}{lcc}
\hline
sparticle & $\Delta_{p_i}$ & $\Delta_{EW}$ \\
\hline
$m_{\tg}$   & 600 & 6000  \\
$m_{\tq,\tell}$ & 500 & 40,000 \\
$m_{\tst_1}$ & 450 & 3000 \\
$\mu$ & 350 & 350 \\
\hline
\end{tabular}
\caption{Upper limits on sparticle masses in GeV from naturalness at the
  $\sim 3\%$ level under $\Delta_{p_i}$ and $\Delta_{EW}$.
}
\label{tab:nat}
\end{table}

A criticism of $\Delta_{p_i}$ is that it is unclear what exactly the $p_i$ should be\cite{Baer:2013gva}. 
In many SUSY theories, the soft parameters merely
{\it parameterize our ignorance} of the origin of the soft terms and
consequently the (unknown) correlations that surely exist between them are ignored. 
For instance, in the CMSSM model, if one applies the universality of
scalar masses condition, then large cancellations occur in the expansion of
$m_{H_u}^2(m_{weak})$ such that TeV-scale sfermions are actually allowed in the so-called focus point region\cite{Feng:1999mn,Feng:2013pwa}.

In specific SUSY breaking models, {\it all} the soft terms are correlated,
and so even more cancellations can occur.
For example, in the stringy dilaton-dominated SUSY breaking model, 
one finds $m_0=m_{3/2}$  with $m_{1/2}=\sqrt{3}m_{3/2}$ 
and $A_0=-m_{1/2}$. In this case, with $m_{3/2}$ as the free
soft breaking parameter, then the expansion of $m_{H_u}^2 (weak)$
collapses back to some multiple of $m_{3/2}$ and 
$\Delta_{p_i}\to \Delta_{EW}$. Similarly, the soft terms of AMSB and 
GMSB are all computable in terms of a single parameter.
By ignoring correlations amongst the soft terms, then
$\Delta_{p_i}$ can overestimate finetuning by factors of up to 
$10^3$\cite{Baer:2023cvi}
as compared with the EW finetuning measure $\Delta_{EW}$ (defined below).

\subsection{High scale Higgs mass measure $\Delta_{HS}$}
\label{ssec:DHS}

An alternative measure of finetuning (actually inconsistent with $\Delta_{p_i}$) comes from expanding
\be m_h^2\simeq \mu^2+ m_{H_u}^2(m_{weak})\sim m_{H_u}^2(m_{GUT})+\delta m_{H_u}^2
\ee
and then defining
\be
\Delta_{HS}\equiv \delta m_{H_u}^2/m_h^2 .
\ee
In the literature, it is common to estimate $\delta m_{H_u}^2$
in terms of high scale parameters, but using a single step
integration of the RGE:
\be
\frac{dm_{H_u}^2}{dt}=\frac{2}{16\pi^2}\left(-\frac{3}{5}g_1^2M_1^2-3g_2^2M_2^2
+\frac{3}{10}g_1^2S+3f_t^2X_t\right) 
\ee
where $t=\log(Q^2)$, $Q$ is the renormalization scale, $X_t=m_{Q_3}^2+m_{U_3}^2+m_{H_u}^2+A_t^2$ and
$S=m_{H_u}^2-m_{H_d}^2+Tr [{\bf m}_Q^2-{\bf m}_L^2-2{\bf m}_U^2+{\bf m}_D^2+{\bf m}_E^2]$. 
Setting the gauge couplings, $m_{H_u}^2$ and $S$ to zero, one can
integrate to obtain an expression for $\delta m_{H_u}^2$ 
depending on third generation soft terms that seems to imply that naturalness requires three sub-TeV
third generation squarks\cite{Kitano:2006gv,Papucci:2011wy}
(now excluded by LHC). 

One issue here is that $\delta m_{H_u}^2$ is actually {\it not} independent of $m_{H_u}^2(m_{GUT})$ thus violating a 
finetuning rule that requires the tuning occur between 
independent parameters\cite{Baer:2013gva}. 
Another issue is that the requirement of small $\delta m_{H_u}^2$
typically leads to $m_{Hu}^2$ {\it not} running to negative (or small) values 
so that EW symmetry is typically not broken. In the context of the string landscape, this would lead to anthropically forbidden pocket universes.
The $\Delta_{HS}$ measure also leads to overestimates of finetuning by factors of up to $10^3$ as compared to $\Delta_{EW}$. 

\subsection{Electroweak finetuning measure $\Delta_{EW}$}
\label{ssec:DEW}

The notion of {\it practical naturalness} -- that an observable ${\cal O}$
is practically natural if all {\it independent} contributions to ${\cal O}$
are comparable to or less than ${\cal O}$ -- has been used in particle physics
to draw meaningful naturalness bounds on various quantities. 
For instance, by requiring the charm quark induced contribution to the $K_L-K_S$
  mass difference to be less than $\Delta m_K$, Gaillard and
  Lee\cite{Gaillard:1974hs} were
  able to predict $m_c\sim 1-2$ GeV in the months before the charm quark was discovered.

The measure $\Delta_{EW}$, based on practical naturalness,  was introduced  
in Ref. \cite{Baer:2012up} and is based on the MSSM scalar potential
minimization condition
\be
m_Z^2/2=\frac{m_{H_d}^2+\Sigma_d^d-(m_{H_u}^2+\Sigma_u^u)\tan^2\beta}{\tan^2\beta -1}-\mu^2
\label{eq:mzs}
\ee
where the $\Sigma_u^u$ and $\Sigma_d^d$ terms contain over 40
one\cite{Baer:2012cf}  and two\cite{Dedes:2002dy} loop corrections
based on the Coleman-Weinberg method\cite{Coleman:1973jx}.
For instance,
\be
\Sigma_u^u (\tst_{1,2})= \frac{3}{16\pi^2} F(m_{\tst_{1,2}}) \times\left[f_t^2-g_Z^2
  \mp\frac{f_t^2A_t^2-8g_Z^2(\frac{1}{4}-\frac{2}{3}x_W)\Delta_t}{m_{\tst_2}^2-m_{\tst_1}^2}\right]
\ee
where $\Delta_t= (m_{\tst_L}^2-m_{\tst_R}^2)/2+m_Z^2\cos 2\beta(\frac{1}{4}-\frac{2}{3}x_W)$, $g_Z^2=(g^2+g^{\prime 2})/8$, $x_W=\sin^2\theta_W$,
$F(m^2)=m^2\log((m^2/Q^2)-1)$ and $Q^2\simeq m_{tst_1}m_{\tst_2}$.
The measure $\Delta_{EW}$ is defined as
\be
\Delta_{EW}\equiv max|\rm terms\ on\ RHS\ of\ Eq.\ \ref{eq:mzs}|/(m_Z^2/2)
\ee
(The computer code
  DW4SLHA, available at {\it dew4slha.com}  by
  D. Martinez\cite{Baer:2021tta},
  computes $\Delta_{EW}$ (and $\Delta_{p_i}$ and $\Delta_{HS}$)
  for any input SUSY Les Houches Accord file\cite{Skands:2003cj}.)
Eq. \ref{eq:mzs} is used in SUSY spectra generators to
finetune contributions such that $m_Z=91.2$ GeV so this is the
(hidden) place in spectra computations where finetuning actually occurs. 
An advantage of $\Delta_{EW}$
is that it is model independent: whatever model one uses to compute a
particular spectra, one gains the same value of $\Delta_{EW}$, whether it be
the pMSSM or some high scale model (see {\it e.g.} Ref's \cite{vanBeekveld:2016hug,vanBeekveld:2019tqp,VanBeekveld:2021tgn} for
$\Delta_{EW}$ values in thorough scans of the pMSSM model). 
This is of course not true of the other naturalness measures.
Also, $\Delta_{EW}$ is the most conservative of the three naturalness
measures in that $\Delta_{EW}$ is necessarily smaller than the other two measures\cite{Mustafayev:2014lqa}.

\subsection{Comparing naturalness measures in SUSY model parameter space}

\begin{figure}[tbp]
\begin{center}
\includegraphics[height=0.24\textheight]{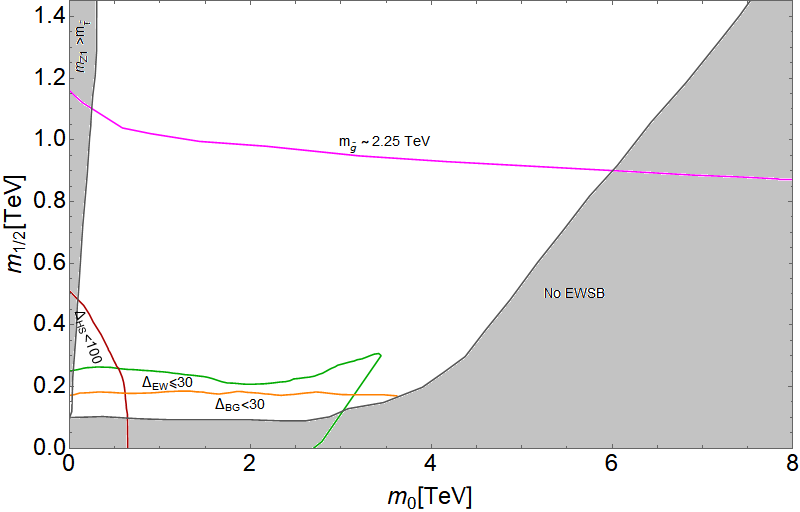}
\includegraphics[height=0.24\textheight]{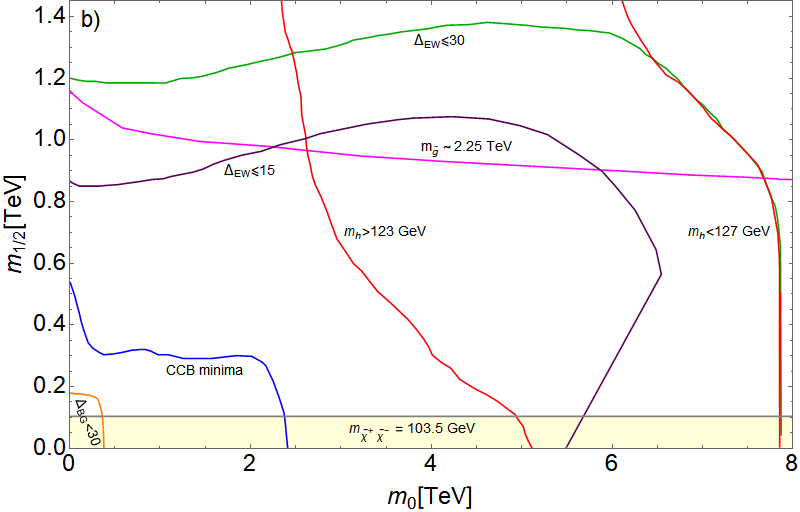}
\caption{The $m_0$ vs. $m_{1/2}$ plane of {\it a}) the mSUGRA/CMSSM model 
with $A_0=0$ and {\it b}) the NUHM2 model with $A_0=-1.6 m_0$, $\mu =200$ GeV 
and $m_A=2$ TeV. For both cases, we take $\tan\beta =10$.
We show contours of various finetuning measures along with Higgs mass 
contours and LEP2 and LHC Run 2 search limits.
\label{fig:m0mhf}}
\end{center}
\end{figure}

In Fig. \ref{fig:m0mhf}{\it a}), we show contours of the various measures in
the $m_0$ vs. $m_{1/2}$ plane of the CMSSM model with $\tan\beta =10$
and $A_0=0$, with $\mu >0$\cite{Baer:2019cae}. We also show the contour
of $m_{\tg}=2.25$ TeV, below which is excluded by LHC gluino pair searches.
In this case, one would conclude that natural SUSY within the context of the CMSSM is excluded.
From the plot, we see that $\Delta_{HS}$ prefers the lower-left corner
while $\Delta_{p_i}$ and $\Delta_{EW}$ prefer $m_{1/2}\alt0.25$ TeV
but do allow for $1-3$ TeV sfermion masses. However, if we proceed
to the two-extra-parameter non-universal Higgs model (NUHM2)\cite{Ellis:2002iu}
as shown in Fig. \ref{fig:m0mhf}{\it b}) with $\mu = 200$ GeV, $A_0=-1.6 m_0$
and $m_A=2$ TeV, then $\Delta_{HS}$ shrinks into the lower-left where
charge-or-color breaking minima (CCB) are found while $\Delta_{EW}$
allows far more parameter space, even well-beyond LHC gluino limits,
and furthermore, the bulk of the $\Delta_{EW}$ natural p-space has
$m_h\sim 123-127$ GeV, in accord with data. The enlarged natural
parameter space is due to the freedom to take $m_{H_u}^2(m_{GUT})$
as large as possible such that it is driven via RGEs to small
$m_{H_u}^2(weak)$, a feature called radiatively-driven naturalness,
or RNS\cite{Baer:2012up}. The scalar mass non-universality is to be expected
in gravity-mediated SUSY breaking models\cite{Soni:1983rm,Kaplunovsky:1993rd,Brignole:1993dj},
whereas the assumed universality of the CMSSM is typically ad-hoc.

Upper bounds on sparticle masses can also be calculated using
$\Delta_{EW}$\cite{Baer:2015rja,Baer:2017pba}. We list some of these
in Table \ref{tab:nat}. It is noteworthy that under $\Delta_{EW}\alt 30$,
the $m_{\tg}$ upper bound has risen to $6$ TeV since gluinos only contribute to
the weak scale at two-loop level. The top-squark upper bound has moved to 3 TeV, well-beyond current LHC limits.
The mass limits on first/second generation squarks/sleptons has risen to
$\sim 40$ TeV since their contributions to the weak scale are suppressed by
tiny Yukawa couplings.
The $\mu$ parameter remains $\alt 350$ GeV, so light higgsinos are still
required (this was emphasized by Chan {\it et al.} already in the 1990's\cite{Chan:1997bi}). The upshot is that under the more conservative, model-independent
$\Delta_{EW}$ measure, much of the allowed SUSY parameter space is now
well-beyond LHC lower limits, and there is much of SUSY parameter space
still left to explore.

\subsection{An exception to low $\mu$ naturalness: strong scalar sequestering}

There is one set of models where large $\mu$ can be reconciled with
naturalness.
These models, known as hidden sector scalar sequestering (HSS)\cite{Murayama:2007ge,Perez:2008ng,Kim:2009sy,Martin:2017vlf,Baer:2024zvr},
posit a strongly interacting nearly superconformal hidden sector
which is active between some high scale $m_*\sim m_P$ and an intermediate scale
$M_{int}$ where the superconformal symmetry and supersymmetry are broken, and where
the hidden sector is integrated out leaving only the visible sector at
$Q<M_{int}$.
Under HSS, at scales $M_{int}<Q<m_*$, the non-holomorphic soft terms can be
driven to tiny values
\be
m_{soft}(M_{int})\sim (M_{int}/m_*)^\Gamma m_{soft}(m_*)
\ee
by hidden sector effects, where the exponent $\Gamma$ includes a combination
of classical and anomalous dimensions of hidden sector fields and is not
calculable since the hidden sector becomes strongly interacting. 
Thus, scalar soft masses and the $B\mu$ bilinear term can get crunched to
$\sim 0$ at  $Q\sim M_{int}$ for $\Gamma\sim 1$ or more.
The holomorphic soft terms are protected from such running via
non-renormalization theorems.
If the $\mu$ term is generated from non-holomorphic Giudice-Masiero
SUGRA terms, then the $\mu$ term and Higgs soft terms combine so that
instead the combinations $m_{H_{u,d}}^2+\mu^2$ gets crunched to zero.
Thus, as $Q\sim M_{int}$, $\mu^2$ and
$m_{H_{u,d}}^2$ are no longer independent: at $Q=M_{int}$, then
$m_{H_{u,d}}^2\simeq -\mu^2$.

Phenomenologically, two cases of interest were identified.
The first is  strong scalar sequestering (the PRS scheme\cite{Perez:2008ng})
where superconformal running dominates MSSM running for $Q>M_{int}$.
In this case $m_\phi^2$, $m_{H_{u,d}}^2+\mu^2$ and $B\mu$ are all
$\sim 0 $ at $Q\sim M_{int}$.
The second SPM scheme\cite{Martin:2017vlf} includes medium strength scalar sequestering
where MSSM and HSS running are comparable and where instead the
non-holomorphic soft terms are driven to (calculable) fixed points.
These two cases were scrutinized in Ref's \cite{Martin:2017vlf}
and \cite{Baer:2024zvr}.

For the PRS scheme, the intermediate scale boundary conditions
$m_\phi^2\sim 0$ with $m_{H_{u,d}}^2\sim -\mu^2$ almost always leads to
unacceptable EW symmetry breaking (unbounded from below (UFB) or charge
and color breaking (CCB) minima, or at best a charged LSP.
For the SPM scheme with universal gaugino masses, then viable spectra are
only generated for very large gaugino masses which end up lifting top-squark
EW contributions $\Sigma_u^u (\tst_{1,2})$ to large values so that the
models are still finetuned. By invoking non-universal gaugino masses
with larger $M_1,\ M_2>M_3$, then viable spectra can be found which are
natural, but still have large $\mu$.

\section{Expectations for SUSY from the string landscape: stringy naturalness}
\label{sec:string}

String theory offers perhaps the only plausible route forward for
unifying the SM with a quantum theory of
gravity\cite{Green:1987sp,Green:1987mn,Polchinski:1998rr}.
Consistent string theories
require either 10 or 11 spacetime dimensions $M_4\times K$,
so it is expected that the additional 6 or 7 space dimensions are curled up
on some compact manifold $K$ with size of order $10^{-17}$ cm.
For Ricci-flat manifolds $K$ with special holonomy, then the presence of a
conserved Killing spinor indicates a remnant spacetime SUSY on $M_4$ below the
compactification scale $m_c$\cite{Candelas:1985en}.
The $4-d$ laws of physics on $M_4$
then depend on the geometrical and topological properties of the space $K$.
Acharya argues that the only stable compactifications are compact
manifolds with a Ricci flat metric of special holonomy, which then
maintains a remnant spacetime SUSY in $4-d$\cite{Acharya:2019mcu}.

The number of allowed manifolds vastly increased  with the
realization of flux compactifications\cite{Bousso:2000xa,Douglas:2006es},
and the possibilities may range from $10^{500}-10^{272,000}$\cite{Ashok:2003gk,Taylor:2015xtz}.
The huge number of possible flux vacua provides a setting for Weinberg's anthropic
solution to the cosmological constant problem\cite{Weinberg:1987dv}, where our universe is but one
{\it pocket universe} within the eternally inflating multiverse\cite{Linde:2015edk}.
However, with so many possible compact spaces, one may despair of ever finding the
exact one which describes physics in our universe.

\subsection{Sparticle and Higgs mass distributions from the landscape}

A way forward has been proposed by Douglas and others by applying statistical
methods to the string landscape\cite{Douglas:2004zg,Douglas:2004qg}. 
Along with understanding the origin of the
hierarchy $\rho_{vac}\sim \Lambda_{CC}m_P^2\ll m_P^4$, one may apply statistical methods to
understand the origin of the gauge hierarchy. We assume a {\it friendly}
portion of the landscape\cite{Arkani-Hamed:2005zuc}:
those vacua which lead to the MSSM (plus possible
hidden sector fields) in the $4-d$ theory, but with undetermined SUSY
breaking scale (likely broken dynamically so that all orders of
scales might develop\cite{Affleck:1984xz}).
The differential distribution of string vacua is given by
\be
dN_{vac}=f_{SUSY}\cdot f_{EWSB}\cdot dm_{hidden}^2
\label{eq:dNvac}
\ee
where $f_{SUSY}$ denotes the distribution of the SUSY breaking scale on the
landscape and $f_{EWSB}$ denotes an anthropic selection for the magnitude
of the derived value of the weak scale.

Since nothing in string theory favors any one SUSY breaking scale over
another, the SUSY breaking $F_i$ fields are assumed distributed
randomly as complex numbers whilst the SUSY breaking $D_j$ fields are
distributed as real numbers. The overall SUSY breaking scale is
$m_{hidden}^4=\sum_i F_i F_i^\dagger +\sum_j D_jD_j$, leading to a
power-law distribution\cite{Douglas:2004qg,Susskind:2004uv,Arkani-Hamed:2005zuc}
\be
f_{SUSY}\sim m_{soft}^n
\label{eq:msoft}
\ee
where $n\equiv 2n_F+n_D-1$ and where $n_F$ are the number of $F_i$ breaking fields, $n_D$ are the number
of $D$-breaking fields contributing to the overall SUSY breaking scale and $m_{soft}\sim m_{hidden}^2/m_P$.
(The factor 2 comes from the $F_i$ being distributed as complex numbers.)
Note for the textbook case of SUSY breaking via a single $F$ term field,
there is already a linear draw to large soft terms on the landscape.

With the landscape statistically favoring the largest soft breaking terms,
one may wonder why one expects $m_{soft}\sim m_{weak}$.
Agrawal {\it et al.}\cite{Agrawal:1997gf}
argue that if the derived value for the weak scale  is much smaller or
much larger than its measured value in our universe, then the up-down
quark mass differences would become skewed such that complex nuclei would not
form, and we would be left with a universe with only protons, or only neutrons.
Thus, the anthropic condition is that the anthropically-allowed universes
have weak scales which lie within the so-called ABDS window:
\be
0.5 m_{weak}^{OU}< m_{weak}^{PU}<(2-5)m_{weak}^{OU}
\label{eq:ABDS}
\ee
where $m_{weak}^{PU}$ is the weak scale within each allowed pocket universe
and $m_{weak}^{OU}$ is the measured value of $m_{weak}$ in our universe.
This becomes predictive if one assumes that gauge and Yukawa couplings arise
via dynamics, but only quantities with mass values -- like $\Lambda_{CC}$ and $m_{hidden}$ -- 
scan in the multiverse\cite{Arkani-Hamed:2005zuc}.

To implement a multiverse simulation, we adopt the NUHM3 model
as expected in gravity-mediated SUSY breaking, and scan over parameter values
\bea
m_0(1,2)&:& 0-60\ {\rm TeV}\\ 
m_0(3)&:& 0.1-10\ {\rm TeV}\\  
m_{H_u}&:& m_0(3)-2m_0(3)\\
m_{H_d}(\sim m_A) &:& 0.3-10\ {\rm TeV}\\
m_{1/2}&:& 0.5-3\ {\rm TeV}\\
-A_0&:& 0-50\ {\rm TeV}\\
\mu_{GUT}&:& \ fixed\\
\tan\beta &:& 3-60
\eea
The soft terms are all scanned according to  $f_{SUSY}\sim m_{soft}^{\pm 1}$
while $\mu$ is fixed at a natural value $\mu =150$ GeV.
For $\tan\beta$, we scan uniformly.
The goal is to take scan upper limits beyond those imposed by $f_{EWSB}$
so the plot upper bounds do not depend on scan limits.
We also scan using $n=-1$ which might instead show a case inspired by
dynamical SUSY breaking wherein
the weak scale is uniformly distributed over all decades of possible values.
The lower limits for the $n=-1$ case are selected in accord with
previous scans for $n=1$ with a draw to large soft terms just for
consistency. If we lower the lower bound scan limits,
then the $n=-1$ histograms will migrate to what becomes even worse
discord with experimental limits.
At the end of the day, we must rescale (not finetune) the weak scale
back so that  $m_Z=91.2$ GeV to gain predictions for {\it our universe}.

\begin{figure}[tbh]
\begin{center}
\includegraphics[height=0.22\textheight]{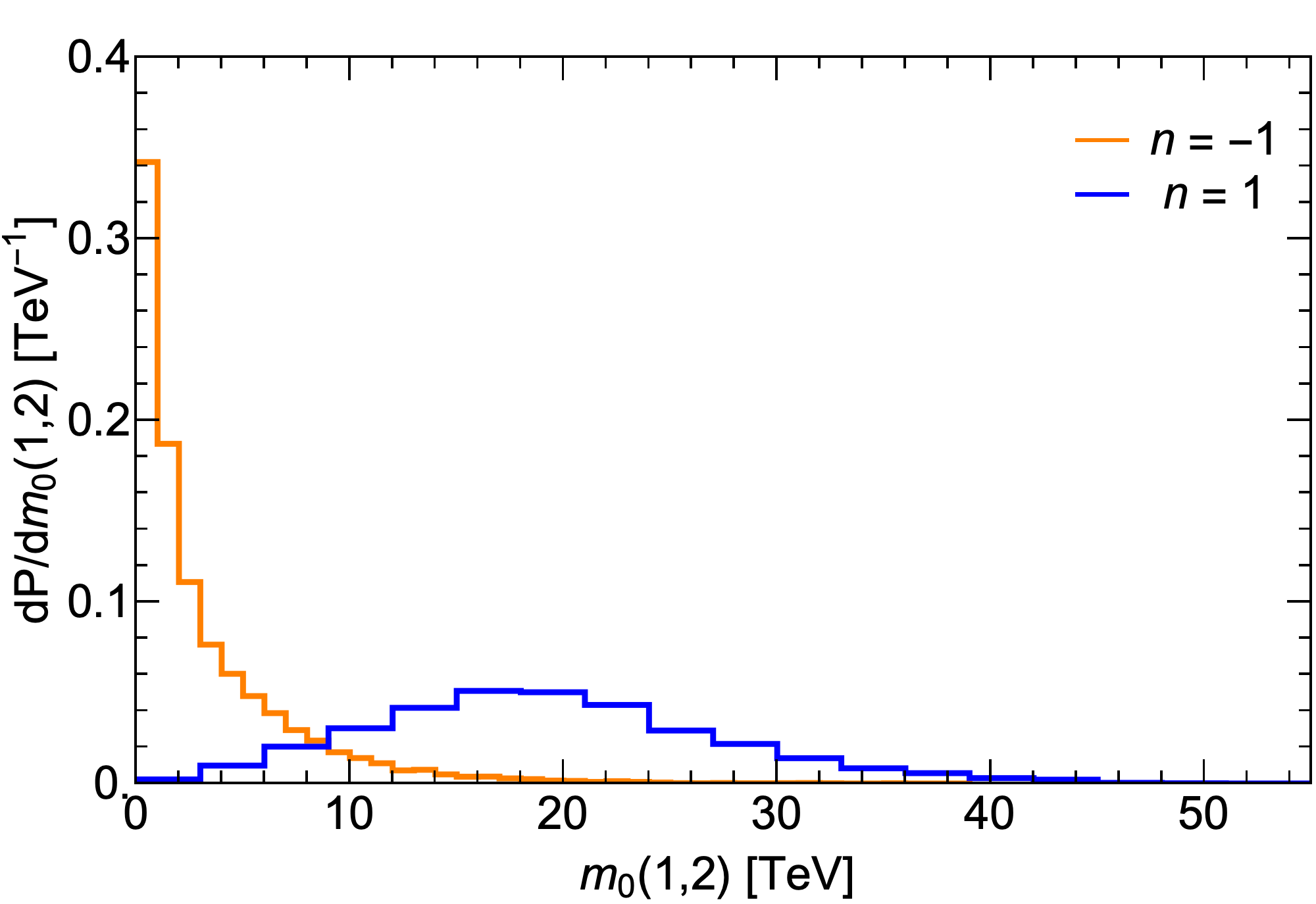}
\includegraphics[height=0.22\textheight]{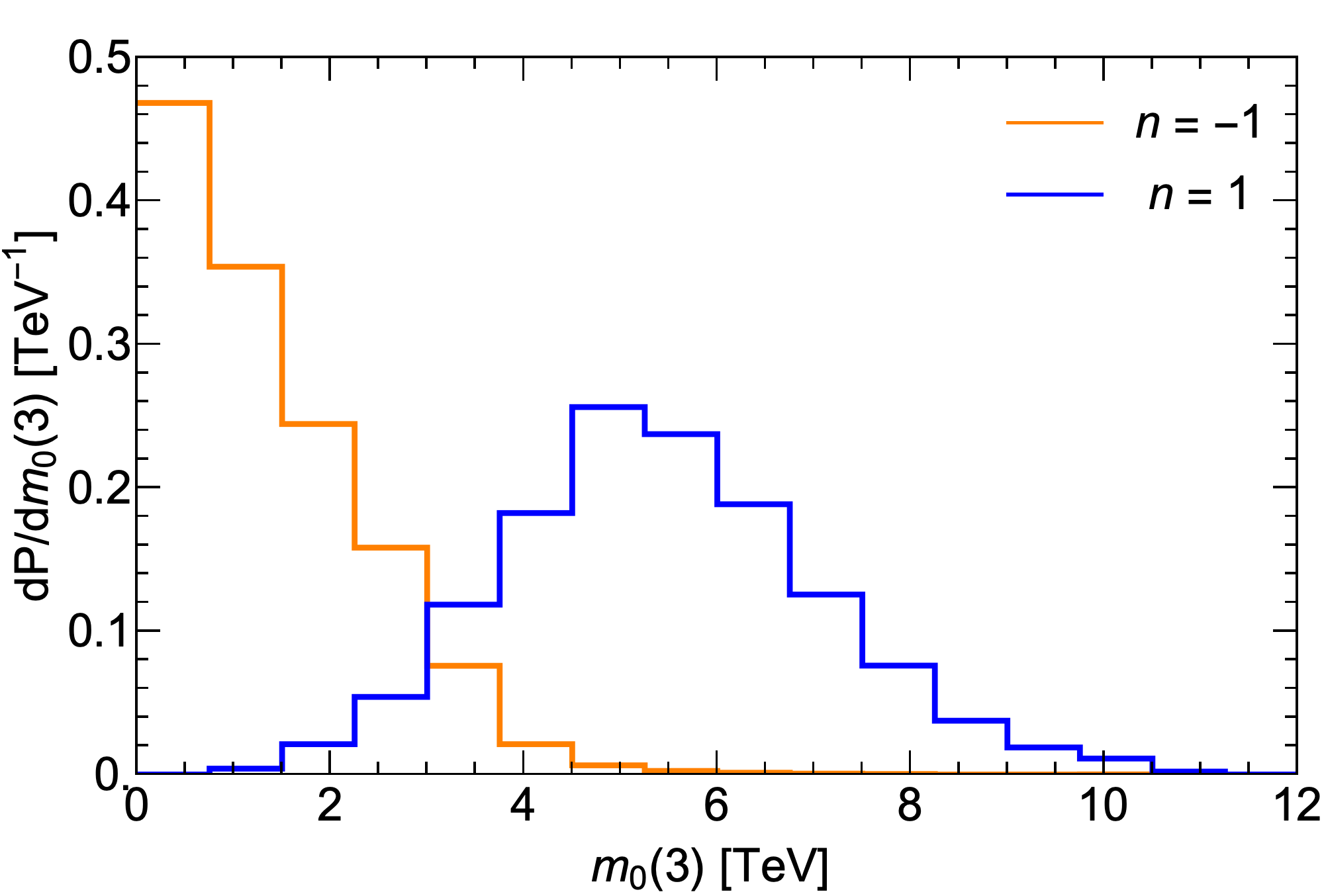}\\
\includegraphics[height=0.22\textheight]{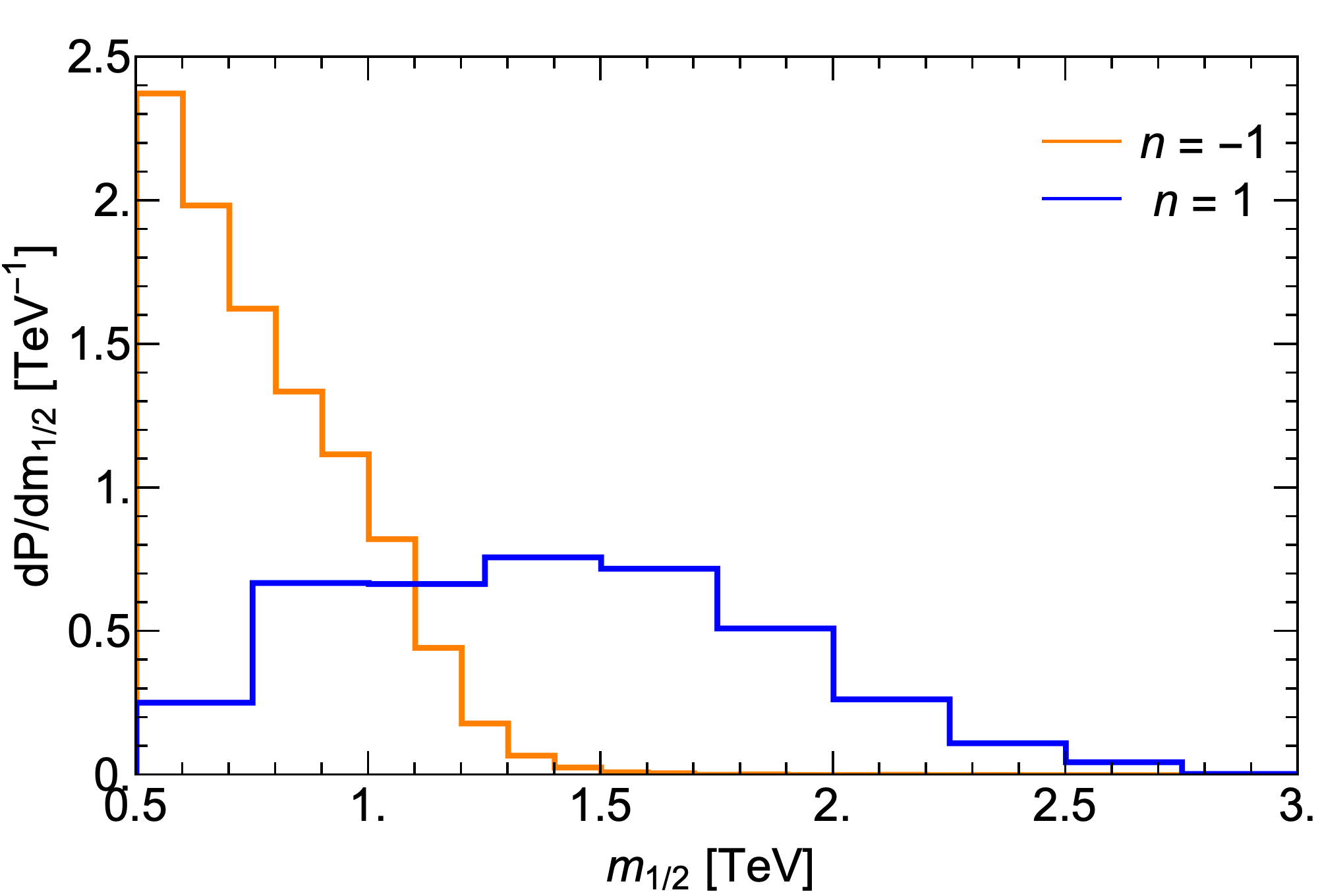}
\includegraphics[height=0.22\textheight]{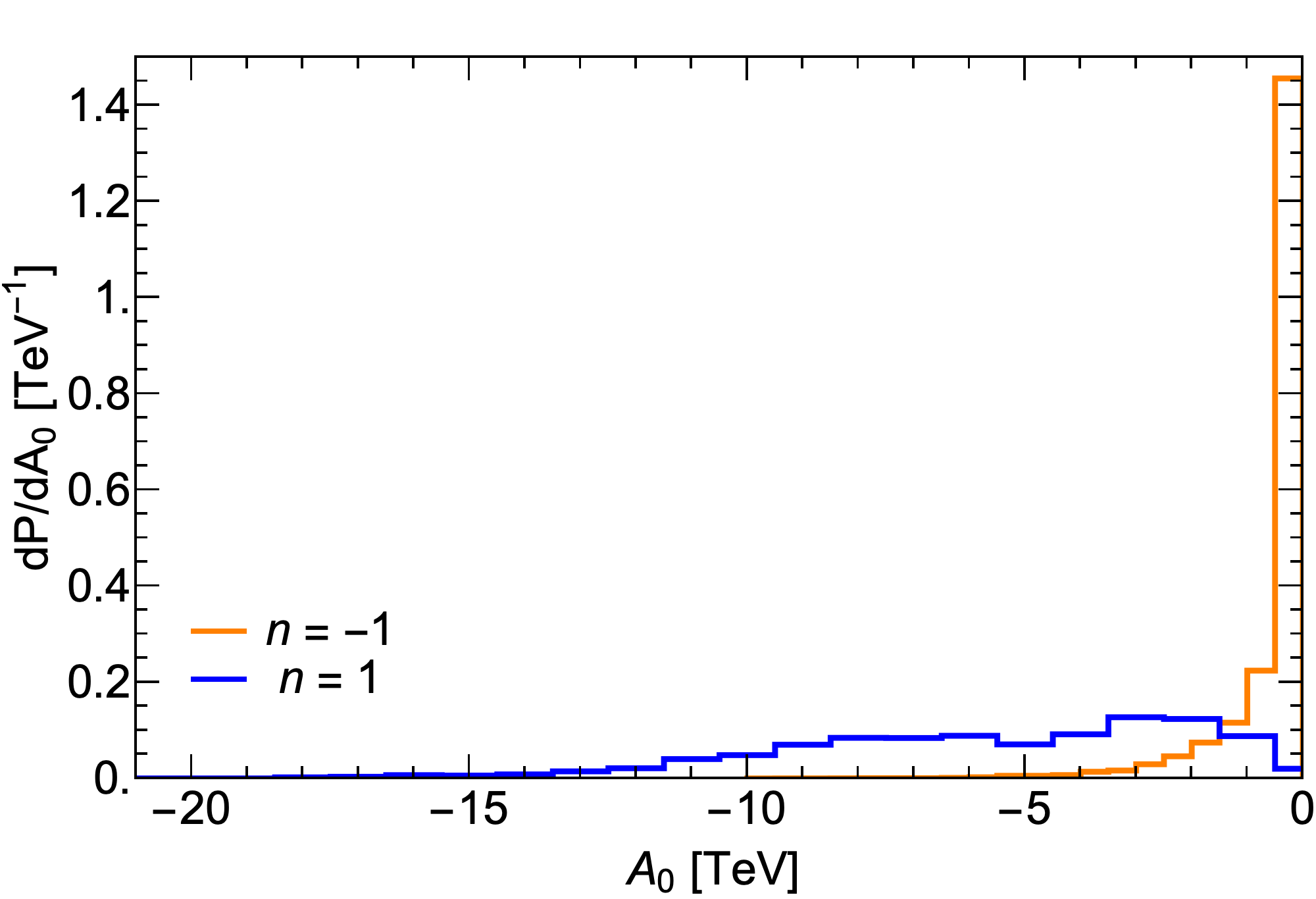}
\caption{Probability distributions for NUHM3 soft terms
{\it a}) $m_0(1,2)$, {\it b}) $m_0(3)$, {\it c}) $m_{1/2}$ and
{\it d}) $A_0$ from the $f_{SUSY}=m_{soft}^{\pm 1}$ distributions of soft terms
in the string landscape with $\mu =150$ GeV.
\label{fig:land4}}
\end{center}
\end{figure}

In Fig. \ref{fig:land4}, we show putative landscape distributions for
various NUHM3 parameters. In frame {\it a}), we show the distribution
for first/second generation scalar masses $m_0(1,2)$. For $n=1$, then we see
the probability distributions peaks around $m_0(1,2)\sim 20$ TeV but extends
as high as $\sim 45$ TeV. Such large first/second generation scalar masses
provide the decoupling/quasi-degeneracy soluton to the SUSY flavor and CP problems\cite{Baer:2019zfl}. 
In contrast, for $n=-1$ then the distribution is sharply peaked
near 0 as expected. In frame {\it b}), we show the distribution in
third generation scalar soft mass $m_0(3)$.
Here, the $n=1$ distribution peaks at 5 TeV but runs as high as 10 TeV.
The $n=-1$ distribution again peaks at zero, which will lead to very light
third generation squarks. The distribution in $m_{1/2}$ shown in frame {\it c})
for $n=1$ peaks around $m_{1/2}\sim 1.5$ TeV leading to gaugino masses
typically beyond the present LHC limits. For $n=-1$, the distribution peaks at
low $m_{1/2}$ leading to gauginos that are typically excluded.
In frame {\it d}) we see the distribution in trilinear soft term $-A_0$.
For $n=1$, the distribution has a double peak structure with most values
in the multi-TeV range leading to large stop mixing and consequently
cancellations in the $\Sigma_u^u(\tst_{1,2})$ and uplift of $m_h$ to
$\sim 125$ GeV. For $n=-1$, then $A_0$ peaks around zero, and we expect
little stop mixing and lighter values of $m_h$.

\begin{figure}[tbh]
\begin{center}
\includegraphics[height=0.22\textheight]{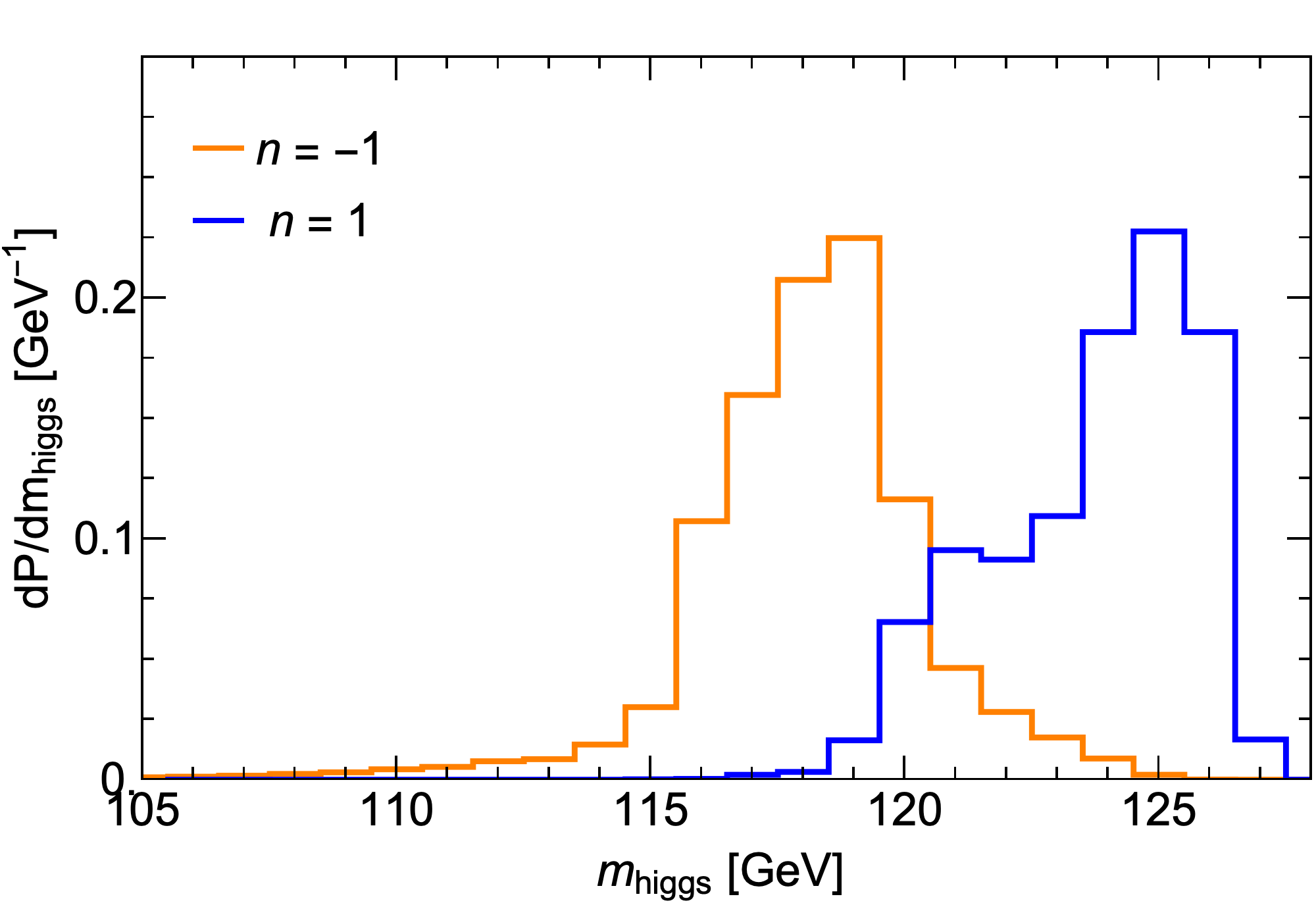}
\includegraphics[height=0.22\textheight]{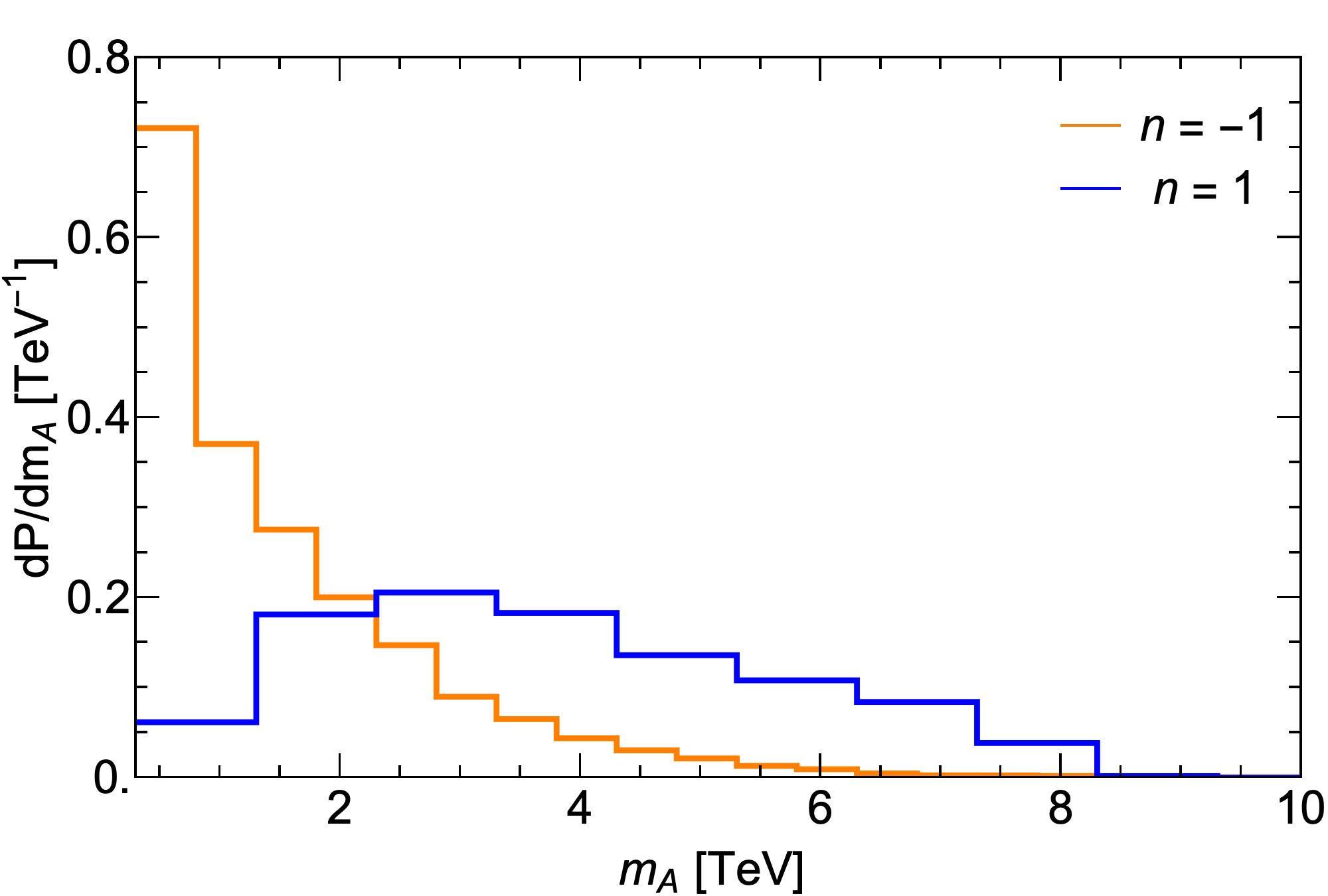}\\
\caption{Probability distributions for light Higgs scalar mass
{\it a}) $m_h$ and pseudoscalar Higgs mass {\it b}) $m_A$
from the $f_{SUSY}=m_{soft}^{\pm 1}$ distributions of soft terms
in the string landscape with $\mu =150$ GeV.
\label{fig:higgs}}
\end{center}
\end{figure}

In Fig. \ref{fig:higgs}, we plot the landscape distributions for light and
heavy SUSY Higgs boson masses. In frame {\it a}), for $n=1$ we see a
distribution with a strong peak around $m_h\sim 124-126$ GeV in accord
with data. The distribution cuts off for $m_h\agt 127$ GeV because
otherwise the $\Sigma_u^u(\tst_{1,2})$ contributions become too large leading
to too large a value of $m_{weak}$ beyond the ABDS window. For $n=-1$, the
distribution peaks at $m_h\sim 118$ GeV with really no significant
probability beyond $m_h\sim 124$ GeV. This essentially rules out the $n=-1$
case. In frame {\it b}), the distribution in heavy pseudoscalar mass $m_A$
is shown. For $n=+1$, the distribution peaks at $m_A\sim 2.5$ TeV with
a distribution extending as high as $m_A\sim 8$ TeV. These values are well
beyond recent ATLAS search limits\cite{ATLAS:2020zms} from $H,\ A\to \tau\bar{\tau}$, which are plotted
in the $m_A$ vs. $\tan\beta$ plane. For $n=-1$, then we expect rather light
$m_A$, possibly at a few hundred GeV, leading to large light-heavy Higgs
mixing. This also seems in contradiction with LHC results which favor
a very SM-like light Higgs as expected in the decoupling limit.

\begin{figure}[tbh]
\begin{center}
\includegraphics[height=0.22\textheight]{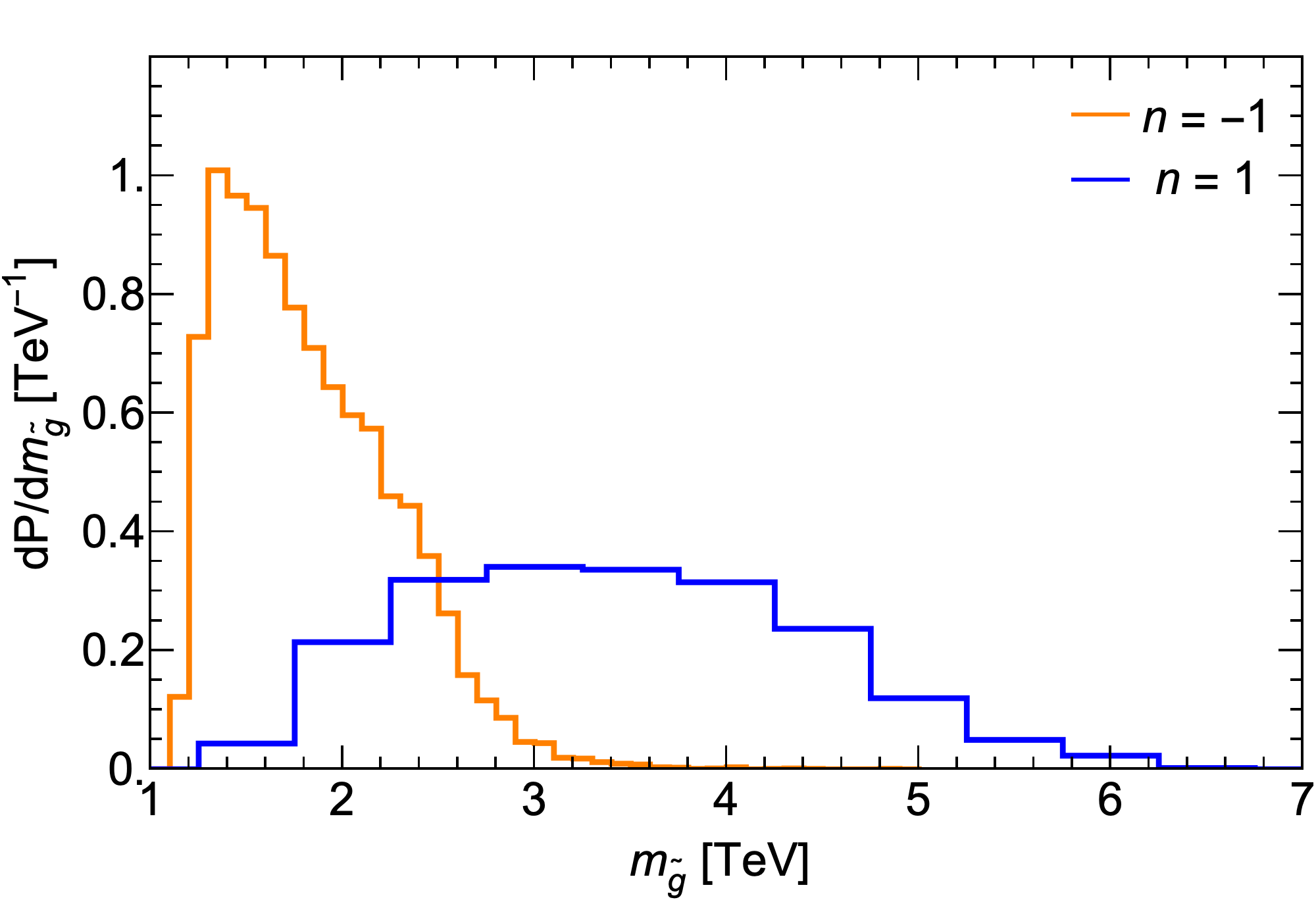}
\includegraphics[height=0.22\textheight]{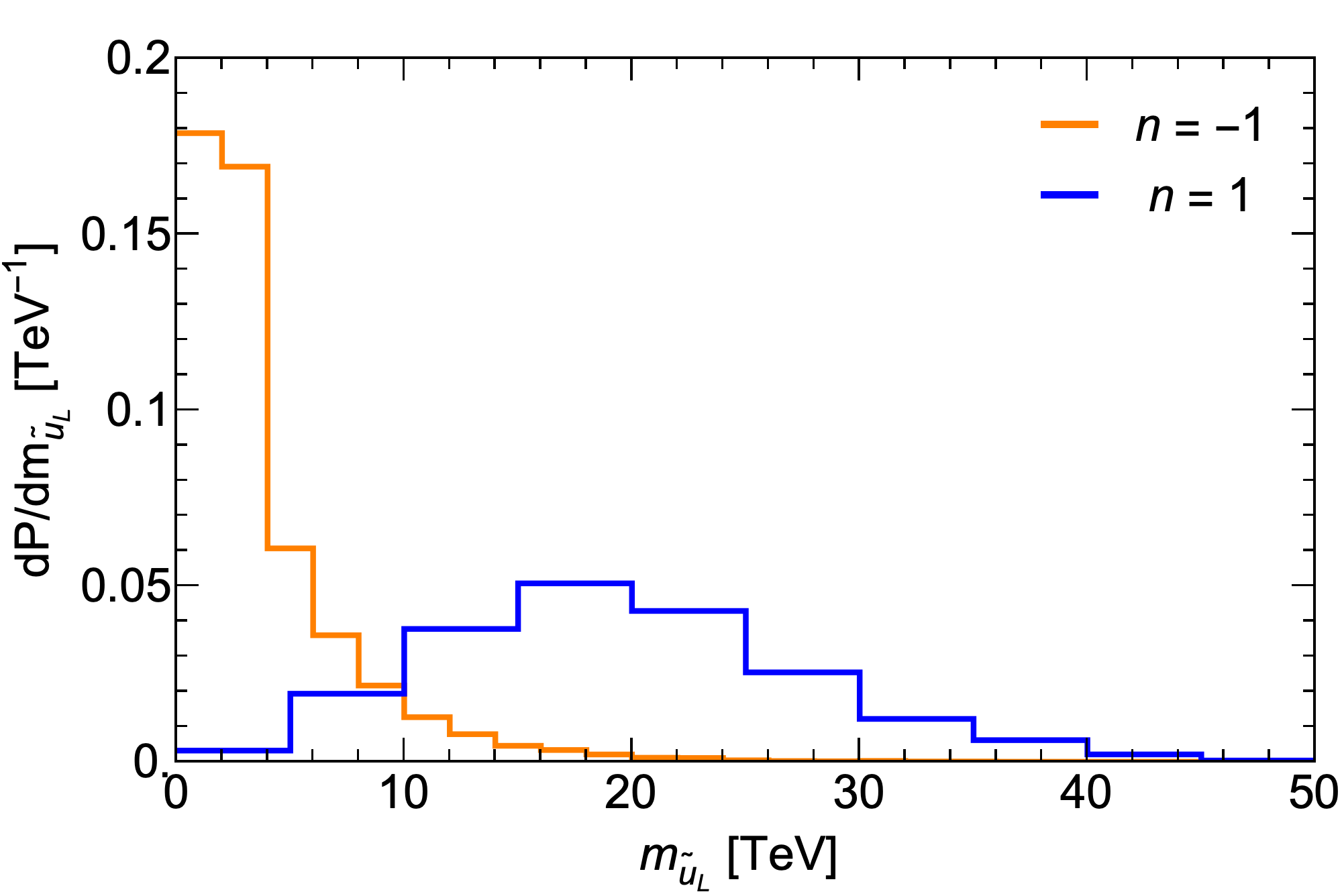}\\
\includegraphics[height=0.22\textheight]{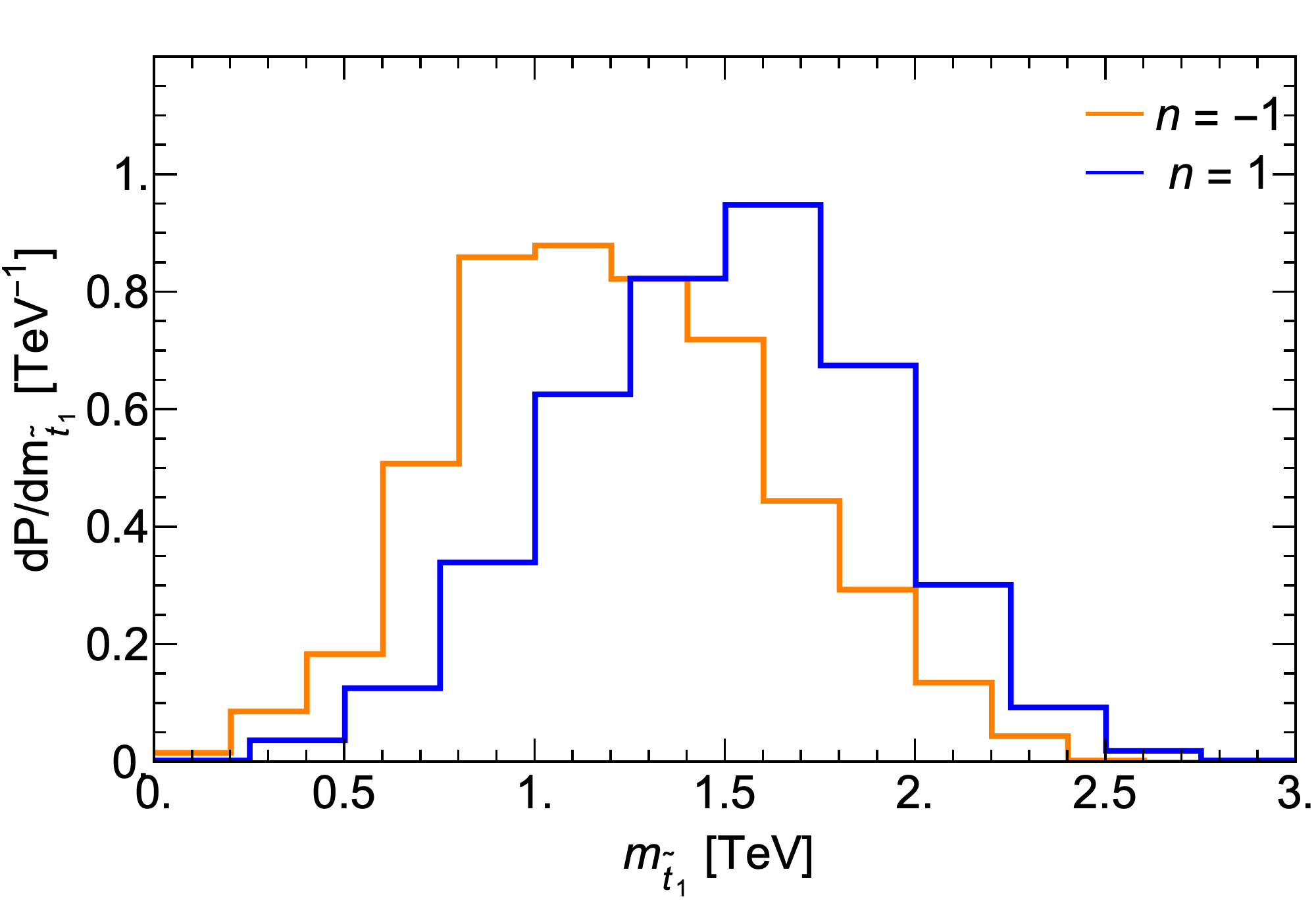}
\includegraphics[height=0.22\textheight]{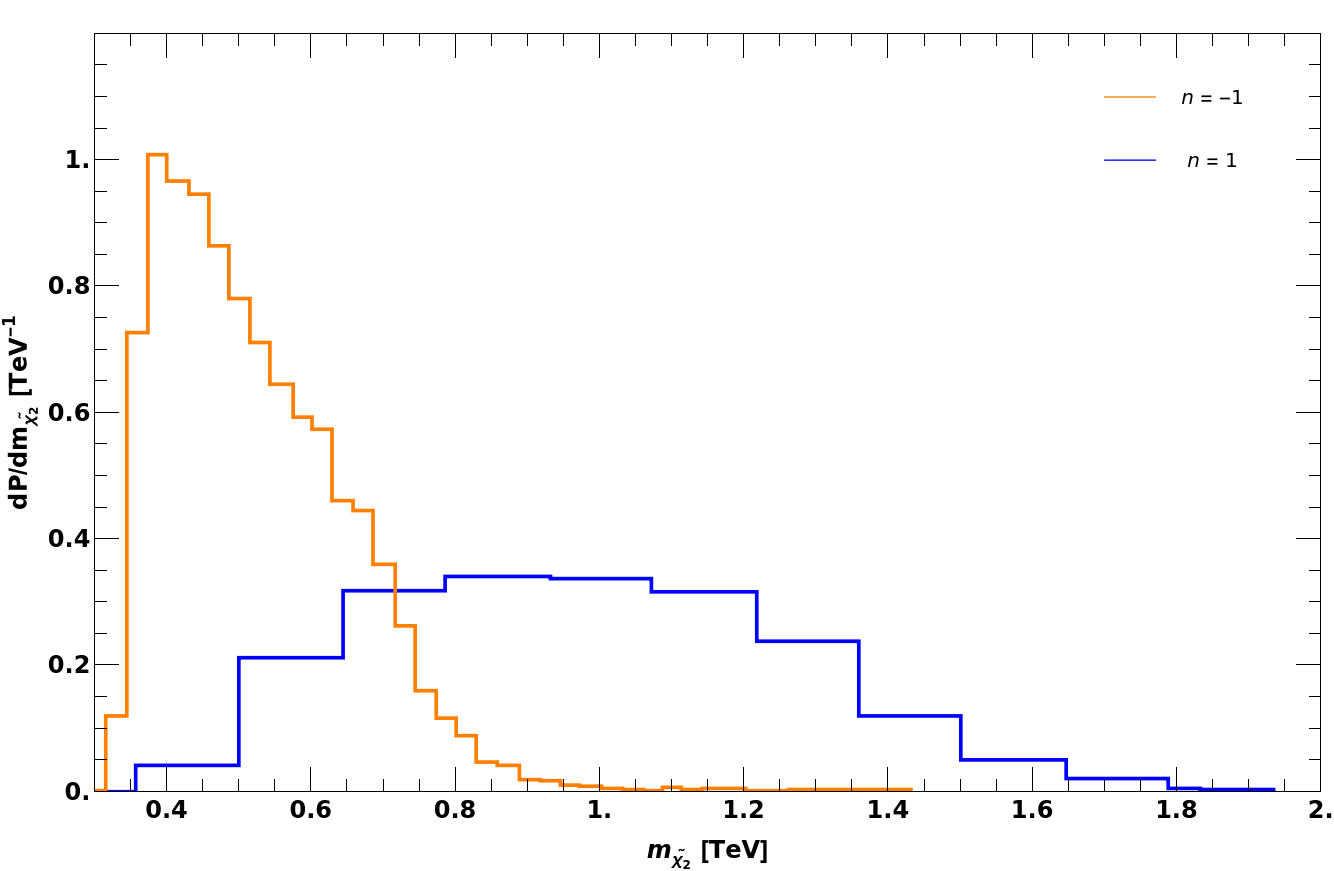}
\caption{Probability distributions for 
{\it a}) $m_{\tg}$, {\it b}) $m_{\tu_L}$, {\it c}) $m_{\tst_1}$ and
{\it d}) $m_{\tchi_2^\pm}$ from the $f_{SUSY}=m_{soft}^{\pm 1}$ distributions of soft terms
in the string landscape with $\mu =150$ GeV.
\label{fig:mass}}
\end{center}
\end{figure}

In Fig. \ref{fig:mass}, we plot the expected strongly interacting
sparticle mass distributions from the landscape.
In frame {\it a}), we see for $n=1$ that $m_{\tg}$ peaks around
$m_{\tg}\sim 2.5-4$ TeV which is well beyond current LHC limits which
require $m_{\tg}\agt 2.2$ TeV. The upper distribution edge extends as far
as $m_{\tg}\sim 6$ TeV. In contrast, for the $n=-1$ distribution,
then the bulk of probability is below 2.2 TeV, although a tail does
extend somewhat above present LHC bounds. In frame {\it b}), we show the distribution in first generation squark mass $m_{\tu_L}$.
For $n=1$, the distribution peaks around $m_{\tq}\sim 20$ TeV but extends to
beyond 40 TeV. For $n=-1$, then squarks are typically expected at
$m_{\tq}\alt 1-2$ TeV and one would have expected squark discovery at LHC
(although a tail extends into the multi-TeV range).
In frame {\it c}), we show the light top squark mass distribution $m_{\tst_1}$.
Here, the $n=1$ distribution lies mainly between $1<m_{\tst_1}<\sim 2.5$ TeV
whereas LHC searches require $m_{\tst_1}\agt 1.1$ TeV. For $n=-1$, then
somewhat lighter stops are expected although there still is
about a 50\% probability to lie beyond LHC bounds on $m_{\tst_1}$.
In frame {\it d}), we show the distribution in $m_{\tchi_2^\pm}$. For $n=1$,
we expect the wino mass $m_{\tchi_2^\pm}\sim 0.5-1.6$ TeV whilst for $n=-1$ then we expect instead that $m_{\tchi_2^\pm}\sim 0.1-1$ TeV.

\subsection{Stringy naturalness}
\label{ssec:stringy}

For the case of the string theory landscape, in Ref. \cite{Douglas:2004zg} 
Douglas has introduced the concept of {\it stringy naturalness}:
\begin{quotation}
{\bf Stringy naturalness:} the value of an observable ${\cal O}_2$ 
is more natural than a value ${\cal O}_1$ if more 
{\it phenomenologically viable} vacua lead to  ${\cal O}_2$ than to ${\cal O}_1$.
\end{quotation}

We can compare the usual naturalness measure $\Delta_{p_i}$
to what is expected from stringy naturalness in the $m_0$ vs. $m_{1/2}$
plane\cite{Baer:2019cae}.
We generate SUSY soft parameters
in accord with Eq.~\ref{eq:dNvac} for values of $n=2n_F+n_D-1=1$ and 4.
The more stringy natural regions of parameter space are denoted by the higher
density of sampled points.
\begin{figure}[!htbp]
\begin{center}
\includegraphics[height=0.4\textheight]{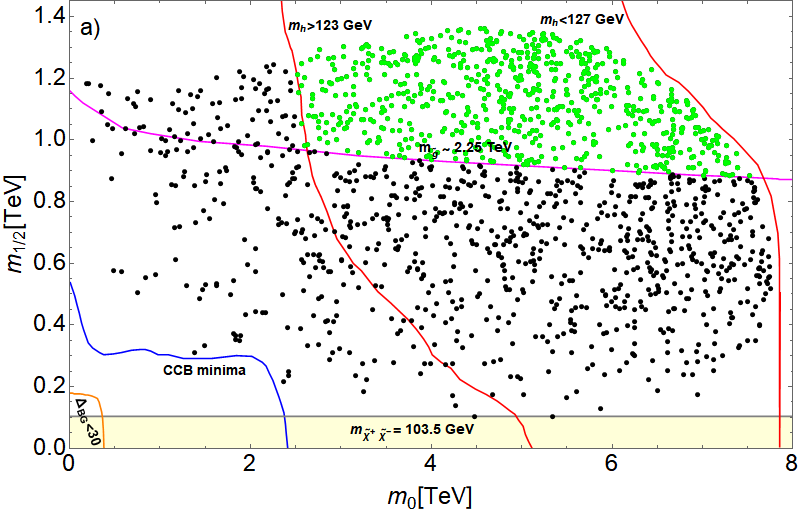}
\caption{The $m_0$ vs. $m_{1/2}$ plane of the NUHM2 model 
with $A_0=-1.6 m_0$, $\mu =200$ GeV and $m_A=2$ TeV and
an $n=1$ draw on soft terms,
The higher density of points denotes greater stringy naturalness.
The LHC Run 2 limit on $m_{\tg}>2.25$ TeV is shown by the magenta curve.
The lower yellow band is excluded by LEP2 chargino pair search limits.
The green points are LHC-allowed while black are LHC-excluded.
\label{fig:m0mhfn1}}
\end{center}
\end{figure}

In Fig. \ref{fig:m0mhfn1}, we show the stringy natural regions for the case
of $n=1$. 
Of course, no dots lie below the CCB boundary since such minima must be vetoed
as they likely lead to an unlivable pocket universe. 
Beyond the CCB contour, the solutions are in accord with livable vacua. 
However, now the density of points {\it increases} with increasing 
$m_0$ and $m_{1/2}$ (linearly, for $n=1$), showing that the more stringy 
natural regions lie at the 
{\it highest} $m_0$ and $m_{1/2}$ values which are consistent with 
generating a weak scale within the ABDS bounds. 
Beyond these bounds, the density of points of course drops to zero 
since contributions to the weak scale exceed its measured value by
at least a factor of 4. 
There is some fluidity of this latter bound
so that values of $\Delta_{EW}\sim 20-40$ might also be entertained. 
The result that stringy naturalness for
$n\ge 1$ favors the largest soft terms (subject to $m_Z^{PU}$ not ranging too far from
our measured value) stands in stark contrast to conventional naturalness
which favors instead the lower values of soft terms. 
Needless to say, the stringy natural
favored region of parameter space is in close accord with LHC results in that
LHC find $m_h=125$ GeV with no sign yet of sparticles.

\subsection{Natural SUSY emergent from the landscape}
\label{ssec:emerge}

Our goal here is to build a toy simulation of our friendly neighborhood of
the string landscape. We generate the soft terms for the NUHM4 model
according to an $n=1$ linear power-law selection. 
While this favors the largest possible soft terms,
the anthropic veto $f_{EWSB}$ places an upper bound on such terms because
usually large soft terms lead to too large a value of $m_{weak}^{PU}$
beyond the ABDS window. 
The trick is to take the upper bound on scan limits
beyond the upper bound posed by $f_{EWSB}$. 
However, in some cases larger soft
terms are {\it more} apt to generate vacua within the ABDS window.
A case in point is
$m_{H_u}^2$: the smaller its value, the deeper negative it runs to unnatural
values at the weak scale, while as it gets larger, it barely runs
negative: EW symmetry is barely broken. As its high scale value becomes
even larger, it doesn't run negative by $m_{weak}$, and EW symmetry is typically not properly
broken -- such vacua failing to break the EW symmetry are vetoed. Also, for small $A_0$,
the $\Sigma_u^u(\tst_{1,2})$ terms can be large. When $A_0$ becomes large
negative, then cancellations occur in $\Sigma_u^u(\tst_{1,2})$ such that
these loop corrections then lie within the ABDS window: large negative
weak scale $A$ terms make $\Sigma_u^u(\tst_{1,2})$ more natural while
raising the light Higgs mass $m_h\sim 125$ GeV.

\begin{figure}[!htbp]
\centering
   \includegraphics[width=0.8\textwidth]{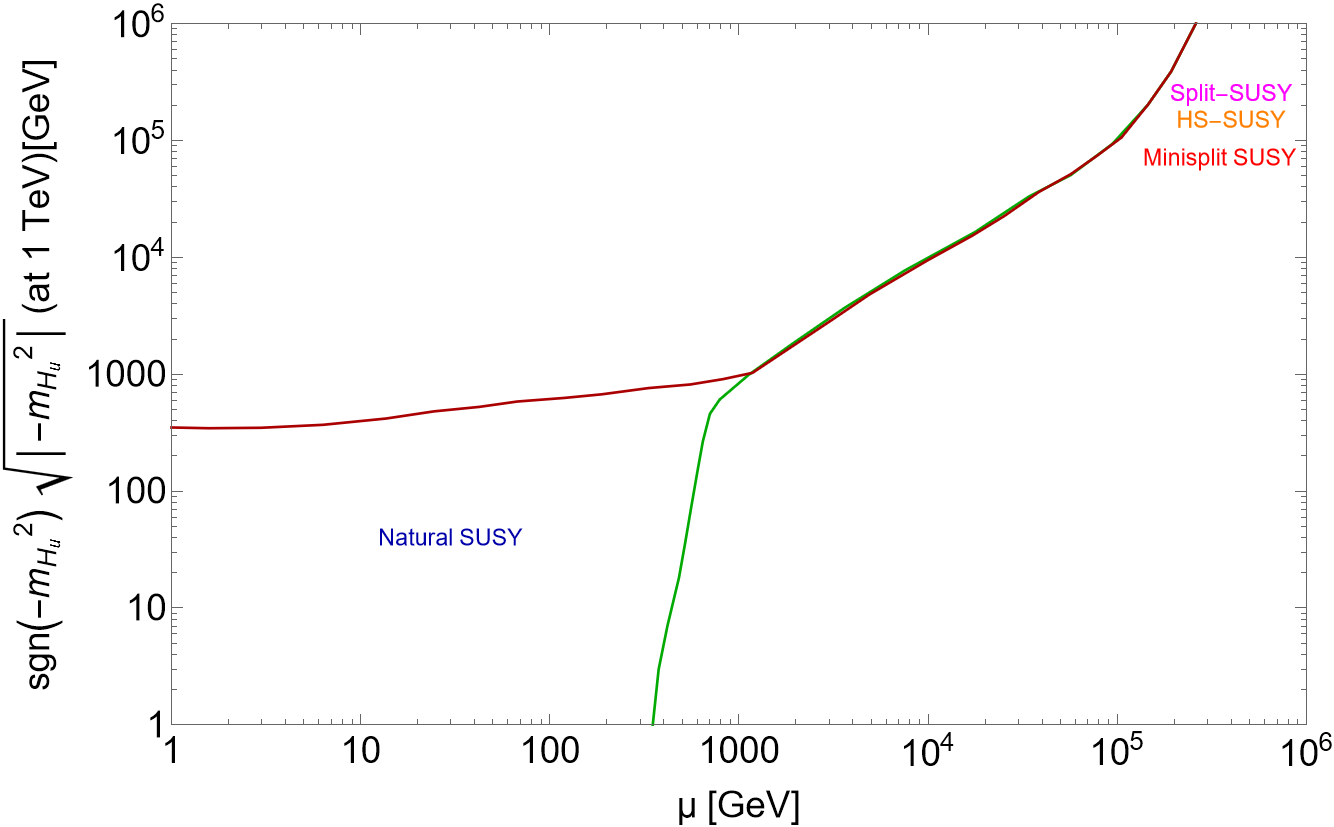}\quad
   \caption{The $\mu^{PU}$ vs. $\sqrt{-m_{H_u}^2(weak)}$ parameter
space in a toy model ignoring radiative corrections to the
Higgs potential. The region between red and green curves
leads to $m_{weak}^{PU}<4 m_{weak}^{OU}$ so that the atomic principle 
is satisfied.
}
    \label{fig:mu_mhu}   
\end{figure}

A plot of the weak scale values of $m_{H_u}$ and $\mu$ is shown in
Fig. \ref{fig:mu_mhu} (taken from Ref. \cite{Baer:2022wxe})
for the case where all radiative corrections --
some negative and some positive\cite{Baer:2021tta} -- lie within the ABDS window.
The ABDS window lies between the red and green curves.
Imagine playing darts with this target, trying to land your dart within
the ABDS window. There is a large region to the lower-left where both
$m_{H_u}$ and $\mu$ are $\alt 350$ GeV which leads to PUs with complexity.
Alternatively, if you want to land your dart at a point with $\mu\sim 1000$ GeV,
then the target space has pinched off to a tiny volume: the target space
is finetuned and your dart will almost never land there. The EW natural
SUSY models live in the lower-left ABDS window while finetuned SUSY models
with large $\Delta_{EW}$ lie within the extremely small volume between the red and green curves in the upper-right plane.
This tightly-constrained region is labeled by split
SUSY\cite{Arkani-Hamed:2004ymt}, high scale SUSY\cite{Barger:2005qy}
and minisplit\cite{Arvanitaki:2012ps}.

It is often said that landscape selection offers an alternative to naturalness
and allows for finetuned SUSY models. After all, isn't the CC finetuned?
However, from Fig. \ref{fig:mu_mhu} we see that models with EW naturalness
(low $\Delta_{EW}$) have a far greater relative probability to emerge from
landscape selection than do finetuned SUSY models.

\begin{figure}[!htbp]
    \centering
    \includegraphics[width=0.8\textwidth]{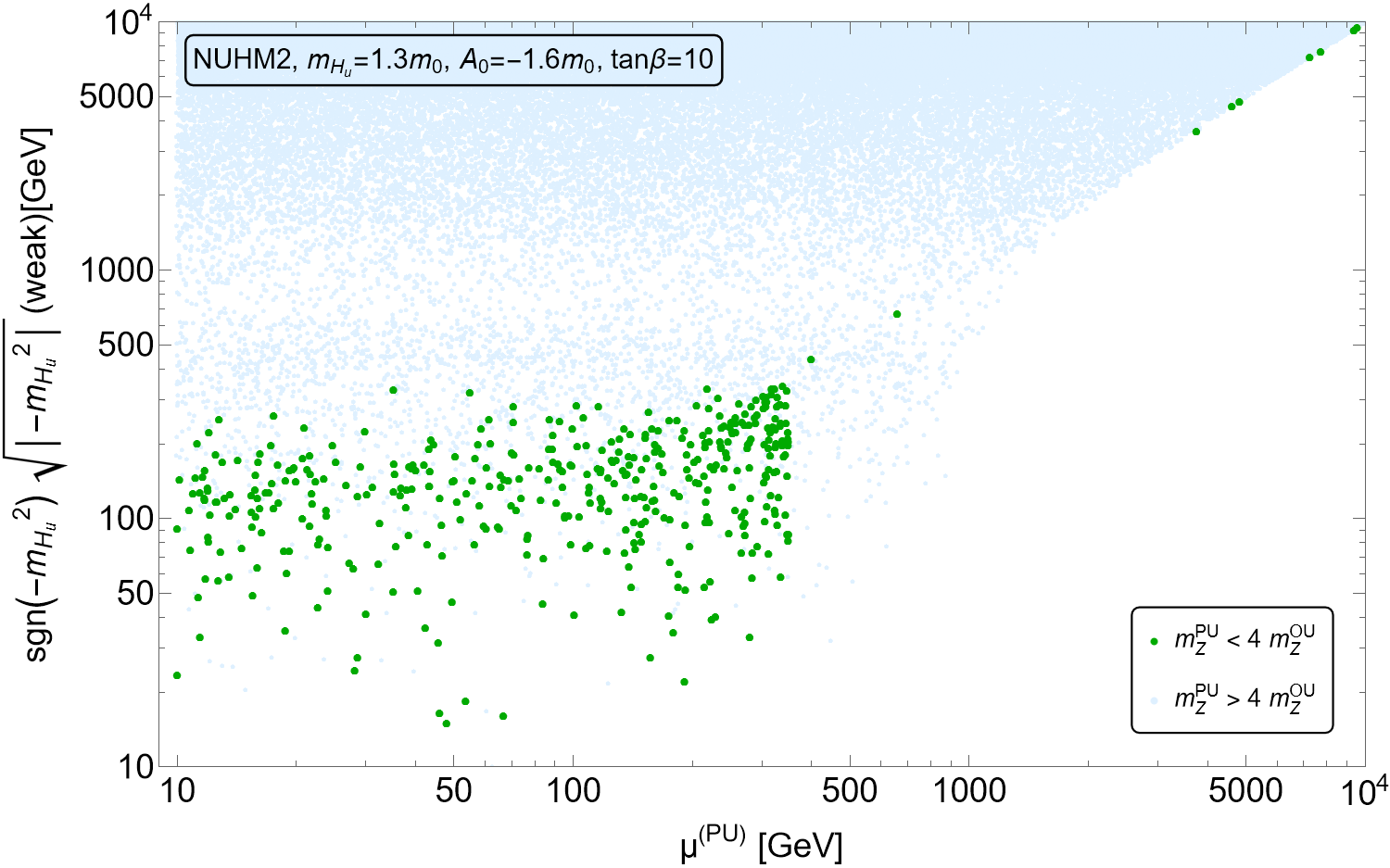}\quad
    \captionof{figure}{The value of $m_{H_u}(weak)$ vs. $\mu^{PU}$
The green points denote vacua with appropriate
EWSB and with $m_{weak}^{PU}<4 m_{weak}^{OU}$ so that the atomic principle 
is satisfied. Blue points have $m_{weak}^{PU}>4 m_{weak}^{OU}$.
}
\label{fig:dots1}
\end{figure}
   
In Fig. \ref{fig:dots1} (from Ref. \cite{Baer:2022wxe}),
we perform a numerical exercise to generate high scale
SUSY soft terms in accord with an $n=1$ draw in $f_{SUSY}$. 
The green dots are viable vacua states with appropriate EWSB and
$m_{weak}^{PU}$ within the ABDS window.
While some dots do land in the finetuned region, the bulk of points
lie within the EW natural SUSY parameter space.

\begin{figure}[!htbp]
\begin{center}
\includegraphics[height=0.4\textheight]{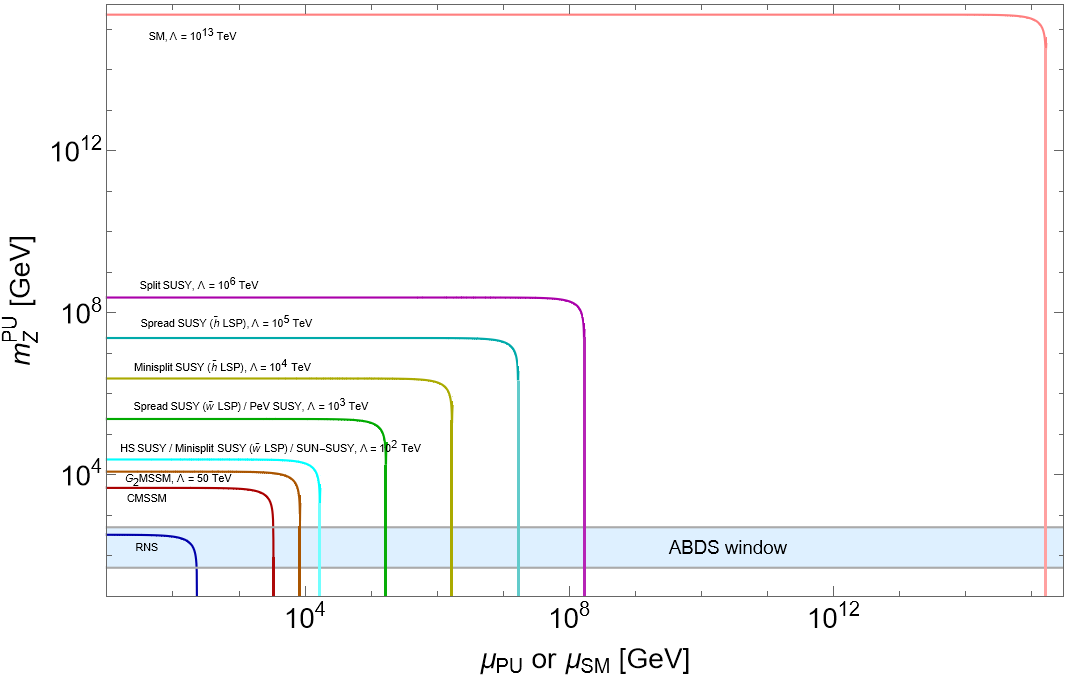}
\caption{Values of $m_Z^{PU}$ vs. $\mu_{PU}$ or $\mu_{SM}$ for
  various natural (RNS) and unnatural SUSY models and the SM.
  The ABDS window extends here from $m_Z^{PU}\sim 50-500$ GeV.
\label{fig:mzPU}}
\end{center}
\end{figure}

An alternative view is gained from Fig. \ref{fig:mzPU} from
Ref. \cite{Baer:2022dfc}. Here, we compute contributions to the scalar potential within a variety of SUSY models
including RNS (radiatively-driven natural SUSY\cite{Baer:2012up}),
CMSSM\cite{Kane:1993td}, G$_2$MSSM\cite{Acharya:2008zi},
high scale SUSY\cite{Barger:2007qb}, spread SUSY\cite{Hall:2011jd},
minisplit\cite{Arvanitaki:2012ps}, split SUSY\cite{Arkani-Hamed:2004ymt}
and the SM with cutoff $\Lambda =10^{13}$ TeV,
indicative of the neutrino see-saw scale\cite{Vissani:1997ys}.
The $x$-axis is either the SM $\mu$ parameter or the SUSY $\mu$ parameter
while the $y$-axis is the calculated value of $m_Z$ within the PU.
The ABDS window is the horizontal blue-shaded region.
For $\mu$ distributed as equally likely at all scales (the distribution's probability density integrates to a log),
then the length of the $x$-axis interval leading to $m_Z^{PU}$ within
the ABDS window can be regarded as a relative probability measure $P_\mu$ for
the model to emerge from the landscape. There is a substantial interval for
the RNS model, but for finetuned SUSY models, the interval is typically
much more narrow than the width of the printed curves.
We can see that finetuned models have only a tiny range of $\mu$ values
which allow habitation within the ABDS window.
Using the magic of algebra, the width of the $\mu$ intervals can be computed,
and the results are in Table \ref{tab:models}. Here, $P_\mu$ is to be considered as a {\it relative} probability. From the Table, we see that the SM is
about $10^{-27}$ times less likely to emerge as compared to RNS.
Minisplit is $10^{-4}- 10^{-8}$ times less likely to emerge (depending on the
version of minisplit).
Even the once-popular CMSSM model is $\sim 10^{-3}$ times less
likely than RNS to emerge from the landscape.
\begin{table}\centering
\begin{tabular}{lcccccc}
\hline
model & $\tilde{m}(1,2)$ & $\tilde{m}(3)$ & gauginos & higgsinos & $m_h$ & $P_\mu$ \\
\hline
SM & - & - & - & -& - & $7\cdot10^{-27}$ \\
CMSSM ($\Delta_{EW}=2641$) & $\sim 1$ & $\sim 1$ & $\sim 1$ & $\sim 1$ & $0.1-0.13$
& $5\cdot 10^{-3}$ \\
PeV SUSY & $\sim 10^3$ & $\sim 10^3$ & $\sim 1$ & $1-10^3$ &
$0.125-0.155$ & $5\cdot 10^{-6}$ \\
Split SUSY & $\sim 10^6$ & $\sim 10^6$ & $\sim 1$ & $\sim 1$ & $0.13-0.155$
& $7\cdot 10^{-12}$ \\
HS-SUSY & $\agt 10^2$ & $\agt 10^2$ & $\agt 10^2$ & $\agt 10^2$ & $0.125-0.16$
& $6\cdot 10^{-4}$ \\
Spread ($\tilde{h}$LSP) & $10^{5}$  & $10^5$ & $10^2$ & $\sim 1$ & $0.125-0.15$ & $9\cdot 10^{-10}$ \\
Spread ($\tilde{w}$LSP) & $10^{3}$ & $10^{3}$ & $\sim 1$ & $\sim 10^2$ & $0.125-0.14$  & $5\cdot 10^{-6}$ \\
Mini-Split ($\tilde{h}$LSP)& $\sim 10^4$ & $\sim 10^4$ & $\sim 10^2$ & $\sim 1$  & $0.125-0.14$ & $8\cdot10^{-8}$ \\
Mini-Split ($\tilde{w}$LSP)& $\sim 10^2$ & $\sim 10^2$ & $\sim 1$ & $\sim 10^2$ & $0.11-0.13$ & $4\cdot 10^{-4}$ \\
SUN-SUSY  & $\sim 10^2$ & $\sim 10^2$ & $\sim 1$ & $\sim 10^2$  & $0.125$
& $4\cdot 10^{-4}$ \\
G$_2$MSSM  & $30-100$ & $30-100$ & $\sim 1$  & $\sim 1$  & $0.11-0.13$
& $2\cdot 10^{-3}$ \\
RNS/landscape & $5-40$  & $0.5-3$ & $\sim 1$ & $0.1-0.35$ & $0.123-0.126$
& $1.4$ \\
\hline
\end{tabular}
\caption{A survey of some unnatural and natural SUSY models
  along with general expectations for sparticle and Higgs
  mass spectra in TeV units.
  We also show relative probability measure $P_\mu$ for the model to emerge
  from the landscape.
  For RNS, we take $\mu_{min}=10$ GeV.
}
\label{tab:models}
\end{table}

\subsection{SUSY parameter space with decoupled sfermion masses}

The string landscape is expected to statistically prefer large soft breaking terms over small soft terms, provided the weak scale lies within the ABDS window. One of the consequences of this is that first and second generation
sfermion masses are expected to lie in the tens-of-TeV range, where the upper
bound on these comes from 2-loop RGE terms which depend on gauge couplings and so are generation independent\cite{Arkani-Hamed:1997opn}. 
As remarked earlier, this can provide 
a mixed decoupling/quasi-degeneracy solution\cite{Baer:2019zfl} to the SUSY flavor and CP
problems which respects naturalness as envisioned early-on by
Dine {\it et al.}\cite{Dine:1993np} and Cohen {\it et al.}\cite{Cohen:1996vb}. It also offers a reason why the lower-left portion of $m_0(3)$ vs. $m_{1/2}$ parameter space is disfavored theoretically. 

\begin{figure}[htb!]
    \centering
    \begin{subfigure}{\textwidth}
        \centering
        \includegraphics[height=0.35\textheight]{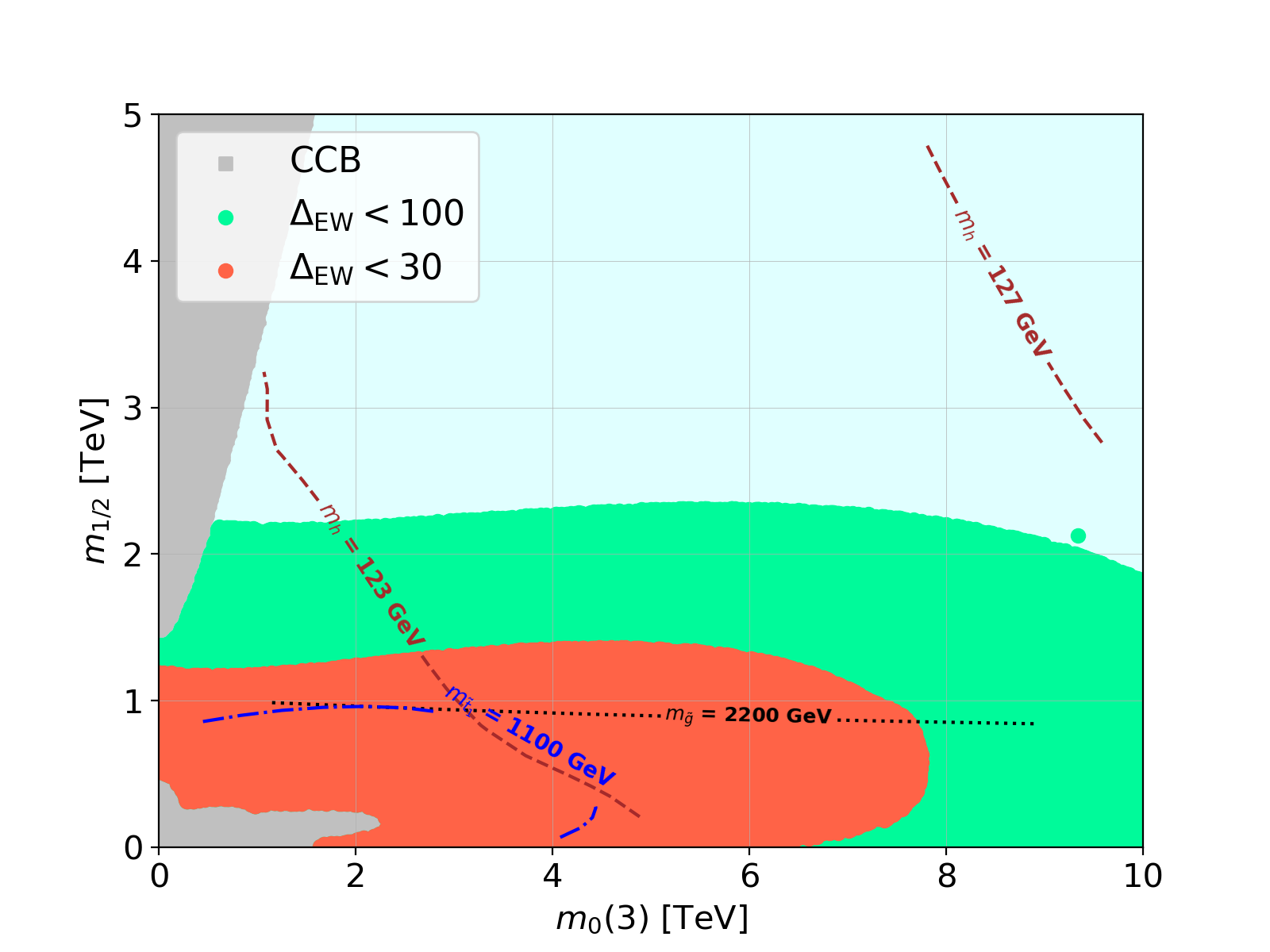}
    \end{subfigure}
    \begin{subfigure}{\textwidth}
        \centering
        \includegraphics[height=0.35\textheight]{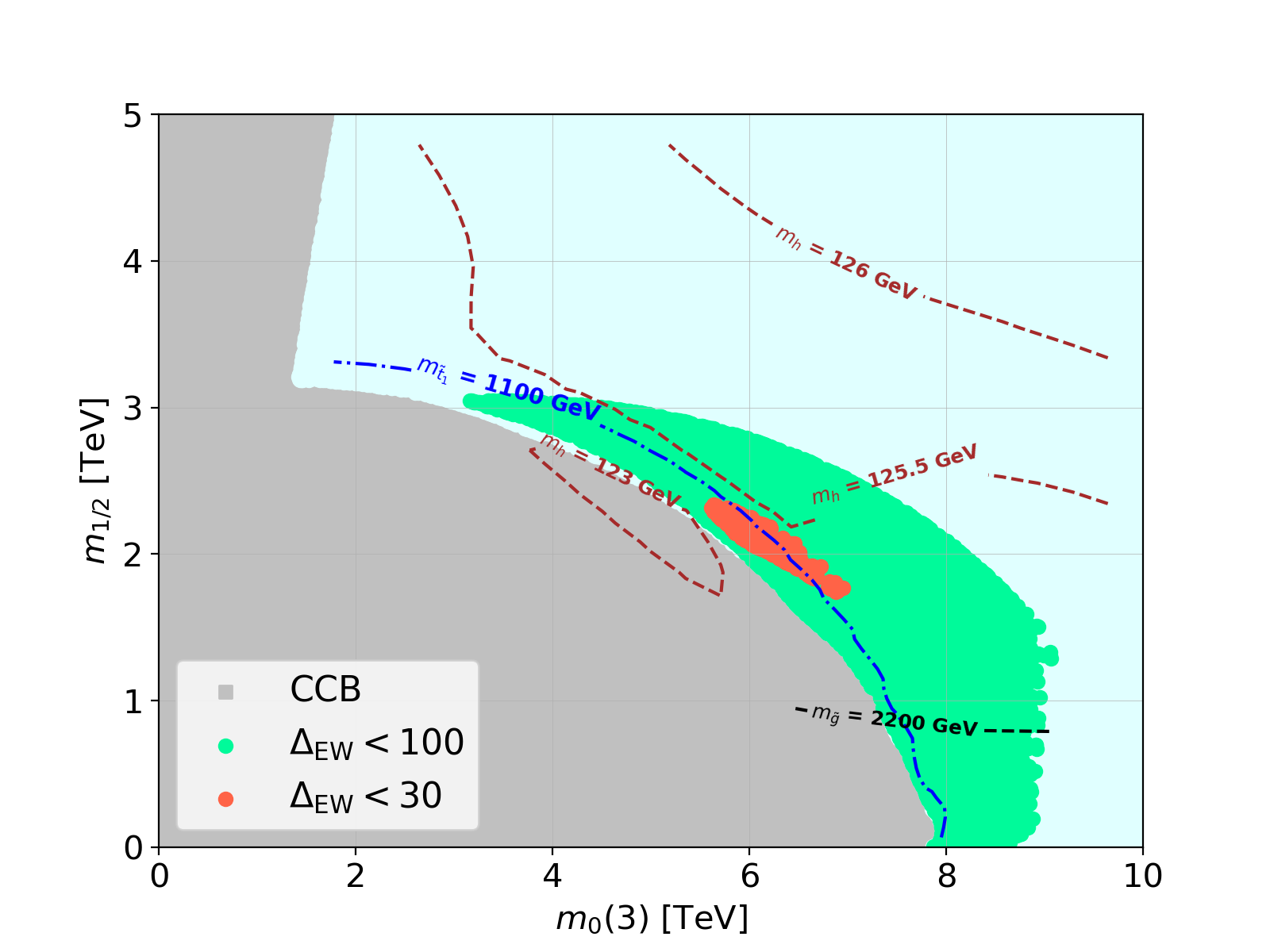}
    \end{subfigure}
    \caption{The $m_0(3)$ vs. $m_{1/2}$ parameter space of 
          {\it a}) the NUHM2 model with $m_0(1,2)=m_0(3)$ and
          $A_0=-1.6m_0(3)$ and
          {\it b}) the NUHM3 model with $m_0(1,2)=30$ TeV 
          $A_0=-m_0(3)$. For both frames we take $\mu =200$ GeV, 
          $m_A=2$ Tev and $\tan\beta =10$. 
                \label{fig:pspace1}}
\end{figure}

In Fig. \ref{fig:pspace1}{\it a}), we plot the $m_0(3)$ vs. $m_{1/2}$ parameter space
of the NUHM2 model for $A_0=-1.6 m_0(3)$ with $m_A=2$ TeV and $\mu =200$ GeV and
$\tan\beta =10$. From the plot\cite{Baer:2024hpl}, we see the natural region with 
$\Delta_{EW}\alt 30$ is mostly excluded by the ATLAS/CMS bounds on $m_{\tg}\alt 2.2$ TeV. However, if we plot the similar parameter space with
$m_0(1,2)=30$ TeV in NUHM3, as in frame {\it b}) with $A_0=-m_0(3)$, 
then we see that the lower-left portion of p-space develops CCB minima
because the 2-loop RG contributions from large first/second generation sfermions drives top-squark soft terms to negative values. There is still
a remaining region of natural parameter space (shaded orange) but it lies at
$m_{1/2}\sim 2$ TeV with $m_0(3)\sim 4-6$ TeV. This points to another reason
why it would have been unlikely for LHC to have already discovered SUSY.

\section{Update on dark matter in natural SUSY models}
\label{sec:DM}

As noted earlier, natural SUSY models with a diminished
abundance of thermally-produced light higgsinos now seem to be ruled out
by recent LZ2024 search limits\cite{LZCollaboration:2024lux}
for SI WIMP-nucleon scattering. However, natural SUSY (and stringy)
models contain a variety of non-thermal processes which may augment or
diminish the neutralino relic abundance or even change the nature of
dark matter. (See recent work in Ref. \cite{Baer:2025oid} on {\it all axion dark matter with no WIMPs} from SUSY.)
Other unnatural SUSY models still have allowed regions of parameter space
with a bino-like LSP due in part to the presence of {\it blind spots},
or cancellations in the SI DD rate, in some regions of parameter
space\cite{Ellis:2000ds,Baer:2003jb}, as seen in Fig. \ref{fig:sig_sig}{\it a}).

While EW natural SUSY models are characterized by light higgsinos with
mass $m_{\tilde{\chi}_1^0}\sim 100-350$ GeV, one must also introduce
the axionic solution to the strong CP problem in order to address naturalness in the QCD $\theta$ parameter.
The DFSZ axion model includes two Higgs doublets, and so fits in well with
SUSY, which also requires two Higgs doublets. The supersymmetrized DFSZ
model includes a superpotential term
\be
W\ni \lambda_\mu S^2H_uH_d/m_P
\ee
which solves the SUSY $\mu$ problem under the Kim-Nilles 
mechanism\cite{Kim:1983dt}
when the PQ charged field $S$ develops a vev of order the PQ scale
$f_a\sim 10^{11}$ GeV: then a weak scale $\mu$ term develops
with 
\be
\mu =\lambda_\mu f_a^2/m_P
\ee
which is of order $m_{weak}$ for $f_a\sim 10^{11}$ GeV.

Actually, in this sort of setup, SUSY can help solve the so-called
{\it axion quality problem} where higher order non-renormalizable operators
must be suppressed to very high order lest they produce
$\bar{\theta}_{QCD}\agt 10^{-10}$. The assumption of a  global $U(1)_{PQ}$ could help,
but global symmetries are not respected by quantum gravity.
A way forward is to hypothesize a discrete, intrinsically supersymmetric,
$R$ symmetry wherein the superspace $\theta_a$ coordinates carry non-trivial
$R$-charge. Such discrete $R$ symmetries are expected to arise in various
forms from string compactifications from 10 to 4 spacetime
dimensions\cite{Nilles:2017heg}.
Anomaly-free discrete $R$ symmetries which are consistent
with GUT representations have been categorized by
Lee {\it et al.}\cite{Lee:2011dya}, and consist
of ${\mathbb Z}_4^R$, ${\mathbb Z}_6^R$, ${\mathbb Z}_8^R$, ${\mathbb Z}_{12}^R$
and ${\mathbb Z}_{24}^R$.
A model which solves the axion quality problem and provides an origin for
both $U(1)_{PQ}$ and $R$-parity is ${\mathbb Z}_{24}^R$\cite{Baer:2018avn,Bhattiprolu:2021rrj}.
Under this discrete $R$-symmetry, the superpotential is given by
\bea
W&\ni &f_uQH_uU^c+f_dQH_d D^c+f_{\ell}LH_dE^c+f_{\nu}LH_uN^c\nonumber \\
& +& fX^3Y/m_P+\lambda_\mu X^2 H_uH_d/m_P+M_NN^cN^c/2 .
\eea
The $R$-charge assignments (listed in Ref. \cite{Baer:2018avn}) forbid $R$-parity violating operators and also
dangerous higher dimensional proton decay operators, but allow for a
SUSY $\mu$ term to be developed along with see-saw neutrino masses.
For ${\mathbb Z}_{24}^R$ and under $R$-charge
assignments $Q_R(X)=-1$ and $Q_R(Y)=5$, then the lowest order
PQ violating superpotential operators allowed are 
$X^8Y^2/m_P^7$, $Y^{10}/m_P^7$ and $X^4Y^6/m_P^7$. 
These operators lead to PQ breaking terms in the scalar potential suppressed
by powers of $(1/m_P)^8$. For instance, the term $X^8Y^2/m_P^7$  leads to
$V_{PQ}\ni 24f\lambda_3^*X^2YX^{*7}Y^{*2}/m_P^8+h.c.$ which is sufficiently 
suppressed by enough powers of $m_P$ so that a high quality axion emerges
under the criteria from Kamionkowski {\it et al.}\cite{Kamionkowski:1992mf}.
Thus, Peccei-Quinn symmetry emerges as an
{\it accidental, approximate global symmetry} arising from the more
fundamental discrete ${\mathbb Z}_{24}^R$ symmetry which is quantum gravity
compatible. (This is an excellent example
of how SUSY solves several problems arising in axion physics.)
In the SUSY DFSZ model, the $a\gamma\gamma$ coupling becomes suppressed due to higgsinos circulating in the $a\gamma\gamma$ triangle diagram\cite{Bae:2017hlp}.
This means SUSY DFSZ axions may be far more difficult to detect compared to axions from non-SUSY models.

If discrete $R$-symmetries are used to forbid the $\mu$ parameter (thus solving the first part of the SUSY $\mu$ problem) then they may also forbid RPV terms in the superpotential while generating the global $U(1)_{PQ}$ needed for the axionic solution to the strong CP problem. 
However, the RPV terms may re-arise from non-renormalizable
operators involving the above $X$ and $Y$ fields which develop vevs under SUSY 
breaking\cite{Baer:2025oid}.
This will generate suppressed RPV couplings of order $\lambda\sim (f_a/m_P)^N$ 
where $N$ may be as low as 1 for $\mathbb{Z}_4^R$ and $\mathbb{Z}_8^R$ 
(hence $\lambda,\ \lambda^\prime,\ \lambda^{\prime\prime}\sim 10^{-7}$ (additional suppression is needed to obey bounds from proton decay)). 
These small of RPV couplings
would imply an unstable LSP with lifetime $\tau_{\tchi}\sim 10^{-3}-10$ s so 
that collider $\eslt$ signatures remain intact but where all the $\tchi$ 
generated in the early universe decay away shortly before BBN, thus leaving a universe with axion-only cold dark matter and no WIMPs.

In SUSY axion models, computation of the DM relic abundance is more
complicated than in SUSY alone due to the presence of a spin-0 saxion
field $s$ and a spin-$1/2$ axino field $\tilde{a}$ which are needed to fill out the axion superfield:
\be
\hat{a}\equiv (s+i a)/\sqrt{2}+\sqrt{2}\ta \theta +F_a\theta\theta
\ee
where $F_a$ is the accompanying auxiliary field.
Under gravity-mediation, the saxion is expected to obtain a soft
SUSY breaking mass $m_s\sim m_{3/2}$ as is the axino
$m_{\ta}\sim m_{3/2}$\cite{Chun:1995hc,Kim:2012bb}.
The axino can be produced thermally\cite{Brandenburg:2004du,Strumia:2010aa}, even though it is unlikely to enter
thermal equilibrium, and then (cascade) decays to the LSP plus visible
states via a delayed decay suppressed by $f_a$.
The saxion can be produced thermally but also by coherent scalar field oscillations (CO) and decays to pairs of SM or 
SUSY particles\cite{Bae:2013hma}: for
$m_s>2m_{sparticle}$ and saxion decay after neutralino freeze-out, saxion production in the early universe augments LSP
production while for $m_s<2m_{sparticle}$, the saxion decays only to SM particles and can
result in {\it entropy dilution} of any stable relics that are around at the
saxion decay temperature $T_D^s$.
The full DM relic density calculation requires simultaneously solving eight
coupled Boltzmann equations, including thermal and CO-produced axions
and gravitino production and decay\cite{Bae:2014rfa}.
Roughly, in natural SUSY with thermally underproduced higgsino LSPs,
for smaller $f_a$ values there is mainly axion dark matter, but for large
$f_a$, then late saxion and axino decays can (over)-augment the LSP
abundance\cite{Bae:2013bva,Bae:2013hma}.
One must also be careful of overproduction of dark radiation from $s\to aa$
decays\cite{Bae:2013qr}, parametrized by the number of effective neutrino species in the early universe $\Delta N_{eff}$
(which is at present in accord with SM predictions).
For a recent review of non-thermal dark matter production, see {\it e.g.}
Ref. \cite{Baer:2014eja} and Ref. \cite{Kane:2015jia}.

Another plausible method for non-thermal dark matter production in the
early universe comes from the presence of moduli fields that arise from string compactifications:
gravitationally-coupled scalar fields with no classical potential.
These give rise to the so-called cosmological moduli problem (CMP)
wherein a CO-produced modulus field $\phi$ could dominate the
energy density of the universe at early times, but then late-decay to
SM and SUSY pairs: if the decay occurs after the onset of BBN, then it
would destroy the successful BBN predictions from standard
cosmology\cite{Coughlan:1983ci,deCarlos:1993wie,Banks:1993en}. Modulus field decays can also
overproduce LSP dark matter, gravitinos or dark radiation, or could
reduce the expected thermal dark matter production rate via entropy dilution.
For recent work in the context of natural SUSY, see \cite{Bae:2022okh,Baer:2022fou,Baer:2023bbn}.

Candidates for $R$-parity conserving SUSY dark matter,
other than the lightest neutralino, also exist.
They include:
\bi
\item A gravitino LSP $\tG$.
  The gravitino can be thermally produced\cite{Bolz:2000fu,Pradler:2006qh}
or produced via SUSY cascade decays in the early universe. Usually
one expects $m_{\tG}\equiv m_{3/2}>m_{sparticle}$ in gravity-mediation, but
theoretical prejudice could be wrong. In this case, sparticle production
at LHC would lead to lightest-visible-SUSY-particle (LVSP) production followed
by late decays to the gravitationally coupled gravitino.
One would then of course expect no DD or IDD WIMP signals.
\item An axino $\ta$ LSP. The axino mass is even more model-dependent
  than the gravitino, although it is expected to be $\sim m_{3/2}$,
  although this could be wrong. Axinos would be both thermally and
  non-thermally produced in the early universe. The LVSP decay to
  axino+SM states would be suppressed by powers of $f_a$.
  Again, one would expect no DD or IDD WIMP signals.
  See {\it e.g.} \cite{Covi:1999ty,Choi:2013lwa,Baer:2008yd}.
\item In SUSY axion models, then one necessarily has mixed axion/LSP
  dark matter, so one mustn't ignore this multi-component dark matter scenario.
\item In stringy models, where hidden sectors are plentiful, one
  may also have hidden sector dark matter which could include a portal coupling to the SM or MSSM. This scenario is recently considered in Ref's \cite{Acharya:2016fge,Acharya:2017kfi}.
\ei

A final possibility is to invoke $R$-parity violation\cite{Dreiner:1997uz}. In this case, the LSP
decays into SM states and there is no SUSY WIMP dark matter.
For this possibility, one might expect the DM to be comprised of perhaps axions.
Here one would have to reconcile $R$-parity violation with 
the previously mentioned strong constraints
on such couplings from proton decay.

\section{SUSY models on the plausibility meter}
\label{sec:models}

Next, we examine the status of a variety of previously popular SUSY models
in light of recent LHC and WIMP detection results. 

\subsection{Implausible: CMSSM, GMSB, mAMSB, split-SUSY, inoMSB, WTN}

\subsubsection{CMSSM}

The constrained minimal supersymmetric standard model\cite{Kane:1993td}
(CMSSM) is defined by its unified soft terms at the scale of gauge
coupling unification $m_{GUT}\simeq 2\times 10^{16}$ GeV.
It is defined by its minimal particle content (MSSM) and SM gauge symmetry, and
its choice of GUT scale boundary conditions which define the parameter space:
\be
m_0,\ m_{1/2},\ A_0,\ \tan\beta ,\ sign (\mu )\ \ \ (CMSSM)
\ee
where $m_0$ are the unified scalar mass soft terms, $m_{1/2}$ are the
unified gaugino  masses and $A_0$ are the unified trilinear couplings.
The bilinear soft term $B_\mu$ has been traded for the ratio of Higgs
field vevs $\tan\beta\equiv v_u/v_d$.
The CMSSM was certainly the most popular SUSY model in the pre-21st century
SUSY community.
In those times, it was frequently taken as a simple expression of what
supergravity theory predicted, and was also called the minimal
supergravity (mSUGRA) model\cite{Arnowitt:1993qp,Arnowitt:2012gc}.
The unified gaugino masses are well-motivated
if the gauge kinetic function in the SUGRA Lagrangian is a linear function of
the same hidden sector singlet SUSY breaking field for each gauge group.
The unified scalar mass term $m_0$ is {\it not} well-motivated in SUGRA in that instead {\it non-universality} is 
expected\cite{Soni:1983rm,Kaplunovsky:1993rd,Brignole:1993dj}. 
For this reason,
we refrain from the mSUGRA label and instead use CMSSM in that the
assumed scalar mass universality is ad-hoc: it could occur under simplifying
conditions, but is not expected in general string compactifications.
Phenomenologically, universal matter scalar masses across the generations
provide a super-GIM mechanism solution to the SUSY flavor problem,
but in light of LHC results, the decoupling solution seems favored,
and also favored by rather general considerations from the string landscape\cite{Baer:2019zfl}.

Within the CMSSM parameter space, one can always move to large enough
$m_0$ and $m_{1/2}$ values to evade LHC sparticle mass bounds,
and one can move to large enough $A_0$ to gain $m_h\simeq 125$ GeV,
so in this sense CMSSM is not excludeable by LHC alone. However,
the remaining CMSSM parameter space which fulfills these conditions is
highly finetuned under $\Delta_{EW}$ and high $\Delta_{EW}$ means too large
a prediction of the weak scale unless one engages in an implausible
finetuning: $m_{weak}\sim m_Z\sqrt{\Delta_{EW}/2}$.
Thus, the bulk of remaining CMSSM p-space gives a bad prediction of
$m_{weak}\gg \sim 100$ GeV unless one finetunes (this finetuning is hidden
in spectra-generator codes where the value of $\mu$ is dialed/tuned such that
$m_Z=91.2$ GeV). A possible exception was the CMSSM focus point region
where $\mu$ could be small. However, this only occurs for $A_0\sim 0$ which then
causes the Higgs mass $m_h$ to fall well below its measured value.
Nowadays, given the ad-hocness of the scalar mass universality assumption,
it is not a big blow that CMSSM is excluded, up to implausible finetunings.

\subsubsection{Gauge mediated SUSY breaking (GMSB)}

In gauge mediated SUSY breaking (GMSB) models\cite{Dine:1995ag}, a messenger sector is introduced
as a mediary between hidden sector SUSY breaking and the visible sector
(MSSM) fields so that soft SUSY breaking terms arise via gauge rather than
gravitational interactions.
A success of these models is that they yield scalar masses
dependent on (generation independent) gauge couplings, and so yield a
universality solution to the SUSY flavor problem.
The soft SUSY breaking terms depend on the induced messenger sector
vev $\langle F_S\rangle$ rather than the hidden sector scale
$\langle F\rangle$ so that $\langle F\rangle$ need not lead to the weak scale
and the gravitino mass can be much smaller than the soft terms:
thus a gravitino LSP is common in GMSB models.
In GMSB, the trilinear $a$ terms arise at two-loop level and so are tiny
compared to scalar and gaugino masses. Thus, to gain $m_h\sim 125$ GeV,
very large top-squark masses in the tens-of-TeV range are required.
Imposing $m_h\simeq 125$ GeV along with LHC constraints,
the $\Sigma_u^u(\tst_{1,2})$ terms become very large leading to
$\Delta_{EW}\agt 10^3$\cite{Baer:2014ica}.
Thus, GMSB models seem highly implausible if not excluded.

\subsubsection{Minimal anomaly mediated SUSY breaking (mAMSB)}

Anomaly-mediated SUSY breaking models come in several different guises.
\begin{enumerate}
\item AMSB1\cite{Giudice:1998xp} is motivated by $4-d$ models of dynamical supersymmetry breaking
  where singlet hidden sector SUSY breaking superfields are scarce\cite{Affleck:1984xz}.
  In such a case, with ``charged'' SUSY breaking fields, scalars gain large masses but gauginos masses are forbidden at lowest order. Gauginos gain
  loop-suppressed masses proportional to the corresponding
  gauge group beta functions so that the wino is lightest of the neutralinos.
  These models motivate (the highly EW-finetuned models)
  PeV-scale SUSY\cite{Wells:2004di}, split SUSY\cite{Arkani-Hamed:2004ymt,Arkani-Hamed:2004zhs}
  and mini-split\cite{Arvanitaki:2012ps,Arkani-Hamed:2012fhg}
  (a version of split SUSY with lighter scalar masses
  that allows for $m_h\simeq 125$ GeV).
  The $A$-terms also arise at loop level.
  The minimal phenomenological AMSB (mAMSB) model\cite{Gherghetta:1999sw,Feng:1999hg} has parameter space
  \be
  m_0,\ m_{3/2},\ \tan\beta,\ sign (\mu) \ \ \ ({\rm mAMSB})
  \ee
The small $A$-terms
  mean that top-quarks in the tens-of-TeV regime are needed to generate
  spectra with $m_h\simeq 125$ GeV. Thus, in light of LHC Higgs mass and
  sparticle search constraints, these models all require
  $\Delta_{EW}\agt 10^2$\cite{Baer:2014ica,Baer:2024fgd},
  making them rather implausible (but see below
  for {\it natural AMSB}).
\item The AMSB2\cite{Randall:1998uk} requires extra spacetime dimensions with separate
  SUSY breaking and visible sector branes so that SUSY breaking is {\it sequestered} from the visible sector, and hence scalars, gauginos and $A$-terms all
  receive dominant loop-level values. In AMSB2, sleptons become tachyonic
  so that so-called {\it bulk} contributions to scalar masses $m_0$ are required for consistent phenomenology. This makes the p-space of AMSB2 similar to
  AMSB1. AMSB2 is motivated in part by a solution to the SUSY flavor problem
  wherein the AMSB contributions to matter scalars are generation independent.
  Since the phenomenological p-space is the same for AMSB1 and AMSB2, they both suffer the same finetuning issues in the mAMSB rendition.
  The mAMSB model with 100\% non-thermally-produced wino-like WIMP dark matter\cite{Moroi:1999zb} also seems ruled out by
  IDD searches with Sommerfeld enhanced annihilation, as in Fig. \ref{fig:sigv}.
\end{enumerate}

\subsubsection{PeV, split, minsplit, spread and high scale SUSY}

This class of models-- PeV\cite{Wells:2004di}, split\cite{Arkani-Hamed:2004ymt,Arkani-Hamed:2004zhs}, minisplit\cite{Arvanitaki:2012ps,Arkani-Hamed:2012fhg}, spread\cite{Hall:2011jd} and high-scale SUSY\cite{Barger:2005qy,Barger:2007qb,Elor:2009jp,Ellis:2017erg}-- 
began in the early 2000s when the string landscape picture arose leading people to
question whether naturalness could be replaced by anthropic selection of the weak scale from the landscape. These models are all characterized by all three generations of
scalar (squark and slepton) masses in the $10-10^6$ TeV regime. Such heavy scalar masses help to provide a decoupling solution to the SUSY flavor and CP problems but give enormous contributions to the weak scale via the $\Sigma_u^u(\tst_{1,2})$ 
and other terms in Eq. \ref{eq:mzs}. 
Thus, the initial motivation was to preserve the successes of gauge coupling unification and the prediction of a dark matter particle while abandoning naturalness. 
Once the Higgs mass was discovered at $m_h\simeq 125$ GeV, then the scalar masses were dialed down from $\sim 10^6$ TeV (which gave $m_h\sim 140-160$ GeV) 
to the 10-100 TeV range which could accommodate the 125 GeV value. The idea was that if anthropic selection could select a finetuned value of the cosmological constant, then why not also select the magnitude of the weak scale via anthropics.

It is now understood, as described in Sec. \ref{ssec:emerge}, that anthropic selection of a
weak scale around $m_{weak}\sim 100$ GeV also probabilistically selects for natural SUSY models (as described by low $\Delta_{EW}\alt 20-40$) over finetuned SUSY models. The reason is that the volume of finetuned model parameter space shrinks to tiny values on the landscape since only very finetuned parameter values can compensate for the enormous contributions to the weak scale. In contrast, for natural SUSY models where all contributions to the weak scale are comparable to the weak scale, then a comparatively large volume of parameter values are available to gain a pocket universe value for the weak scale that lies within the ABDS window.

\subsubsection{Gaugino mediated SUSY breaking/no-scale SUGRA}

The gaugino mediated SUSY breaking (inoMSB) models rely on an
extra-dimensional separation between visible and observable sectors.
MSSM chiral superfields are assumed to live on an extra-dimensional brane separated
from the visible sector brane but both live in the extra-dimensional bulk
where both gravity and gauge superfields propagate. Thus, these models
predict scalar masses $m_0\sim 0$ at the high/GUT scale while
gaugino masses are of order the weak scale since they couple directly to
the SUSY breaking brane. The parameter space is similar to
no-scale SUGRA GUT models\cite{Lahanas:1986uc}:
\be
m_{1/2},\ \tan\beta,\ sign (\mu )
\ee
defined at the compactification scale $M_c$ which can lie
above the GUT scale. The trilinear soft $A$-terms are also suppressed. 
From these boundary conditions, the LSP is usually the lightest stau
in violation of cosmological limits against charged stable exotic relics
so that above-the-GUT-scale running is invoked to avoid this constraint\cite{Schmaltz:2000gy}.
The small $A$-terms require stop masses in the tens-of-TeV range, so the
models which lead to $m_h\simeq 125$ GeV while avoiding LHC limits
are highly finetuned (thus implausible)\cite{Baer:2024fgd}.

\subsubsection{Well-tempered neutralino (WTN)}

SUSY models based on gravity-mediation may have a
{\it well-tempered neutralino}\cite{Arkani-Hamed:2006wnf,Baer:2006te}
(WTN) wherein the gaugino and higgsino components of the lightest
neutralino are adjusted to ``just-so'' values so that the relic abundance lies
at $\Omega_{\tchi_1^0}h^2\simeq 0.12$, between the two peaks of
Fig. \ref{fig:Oh2}.
The spin-independent direct detection rates for neutralino-quark scattering
depend on a Higgs exchange diagram where the coupling involves a product of
gaugino times higgsino components, so that the WTN always has a large
SI DD rate with $\sigma^{SI}(\tchi_1^0 p)\sim 10^{-44}$ cm$^2$ (see maroon dots
on Fig. \ref{fig:sig_sig}{\it a}).).
Thus, these models are presently ruled out by dark matter DD experiments.

\subsection{Plausible: NUHM2-4, nAMSB, GMM}

\subsubsection{Gravity-mediation described by NUHM models}

In gravity-mediated SUSY breaking, hidden sector singlet SUSY breaking results
in MSSM soft terms $m_{soft}\sim m_{3/2}\sim m_{hidden}^2/m_P$.
As noted previously, for the vast assortment of generic hidden sectors
expected from the string landscape, here the various moduli fields number
in the hundreds and non-trivial compactification manifolds are the rule and not the exception, then non-universal scalar masses are expected\cite{Soni:1983rm,Kaplunovsky:1993rd,Brignole:1993dj}. This means that the generic expectation
for SUGRA models is that $m_{H_u}\ne m_{H_d}\ne m_0(i)$ for generation index
$i=1-3$ (as opposed to the ad-hoc universality assumption of the CMSSM model)
since the Higgs and matter multiplets live in different GUT representations.
However, the intra-generational scalar mass universality {\it is}
well motivated in that each element of a generation fills out that
16-dimensional spinor representation of $SO(10)$\footnote{As emphasized by
  Nilles as the first of his five golden rules of string
  phenomenology\cite{Nilles:2004ej}. }.
Also, a perhaps underappreciated facet of SUGRA models is that generically
large trilinear $A$-terms $\sim m_{3/2}$ are expected which can lift the light
Higgs mass $m_h\to 125$ GeV in the maximal mixing scenario\cite{Carena:2002es}
even for natural top-squark mass values. Putting all the pieces together,
then we expect that generic gravity-mediation to be expressed by the
four-extra-parameter non-universal Higgs model (NUHM4) with parameter space
given by\cite{Matalliotakis:1994ft,Ellis:2002wv,Baer:2005bu}
\be
m_0(i),\ m_{H_u},\ m_{H_d},\ m_{1/2},\ A_0,\ \tan\beta
\ee
where it is convenient to trade the two GUT scale Higgs soft masses
$m_{H_u}$ and $m_{H_d}$ for the weak scale parameters $m_A$ and $\mu$ via the
scalar potential minimization conditions so that the NUHM4 parameter
space is given by
\be
m_0(i),\ m_{1/2},\ A_0,\ \tan\beta, \mu,\ m_A\ \ \ (NUHM4)
\ee
The NUHM4 model might be naively expected to suffer the SUSY flavor
problem due to the non-universal generations. However, the string landscape
pull to large independent soft terms can result in $m_0(1)\sim m_0(2)\gg m_0(3)$
according to their contributions to the weak scale, resulting in a
string landscape motivated\cite{Baer:2020dri} mixed quasi-degeneracy/decoupling
solution to the SUSY flavor and CP
problems wherein $m_0(1)\simeq m_0(2)\sim 10-40$ TeV with
$m_0(3)\sim 5-10$ TeV\cite{Dine:1993np,Cohen:1996vb}.
(The upper bounds on $m_0(1,2)$ are generation-independent and result from
two-loop RG contributions to their running which feeds into third generation
running\cite{Arkani-Hamed:1997opn,Baer:2024hpl}:
too large $m_0(1,2)$ drives $m_0(3)$ to tachyonic values resulting
in CCB minima for the scalar potential\cite{Baer:2024hpl}.)
For convenience, the first two generations can be taken degenerate for
large $m_0(1,2)$ leading to the 3-extra-parameter NUHM3 model.
If the first two generations don't affect phenomenology too much
({\it e.g.} for low mass sparticle LHC searches) then generational
universality can be assumed without too much cost, leading to NUHM2
with a single value $m_0\equiv m_0(1)=m_0(2)=m_0(3)$.

It is perhaps gratifying that under the original, simplest model of
SUSY breaking, with large trilinear soft terms, that plenty of
EW natural parameter space with $m_h\simeq 125$ GeV and sparticles beyond
LHC search constraints can be found (as depicted in Fig. \ref{fig:m0mhf}{\it b})).
In fact, the portion of NUHM parameter space with these features
{\it and sparticles beyond LHC search limits} is actually favored by the
general selection criteria of the string landscape, as shown in
Fig. \ref{fig:m0mhfn1}. From the point of view of stringy naturalness,
LHC Run 2 is seeing pretty much exactly what is expected from the
landscape expression of SUGRA EFTs.

\subsubsection{Natural anomaly-mediated SUSY breaking (nAMSB)}

We have remarked above that anomaly-mediated SUSY breaking models as
expressed by the mAMSB model are now highly implausible since they
are highly EW-finetuned for $m_h\sim 125$ GeV and also because wino-like WIMP
dark matter seems ruled out by IDD results.
However, this point of view merely results from oversimplifications in the
proposed EFT of the mAMSB model which posits universal bulk scalar masses
$m_0$ and no bulk $A$-terms.
The bulk scalar masses should arise from additional SUGRA contributions
to soft terms, which as we've seen, are expected to be non-universal
(as actually posited in the original RS paper\cite{Randall:1998uk}).
In addition, there is no reason to exclude the presence of bulk trilinear
contributions to the AMSB-induced $A$-terms.

Thus, a more theoretically acceptable AMSB EFT would be AMSB with the
non-universal bulk terms $m_{H_u}$, $m_{H_d}$ and $m_0(i)$, along with
bulk trilinears $A_0$.
One can again trade $m_{H_u}^2$ and $m_{H_d}^2$ for
$\mu$ and $m_A$ via the scalar potential  minimization conditions to
arrive at a more plausible AMSB EFT which has been dubbed
{\it natural} AMSB\cite{Baer:2018hwa} (nAMSB) since it
allows for EW-naturalness while accommodating $m_h\simeq 125$ GeV:
\be
m_0(i),\ m_{3/2},\ A_0,\ \mu,\ m_A,\ \tan\beta\ \ \ \ (nAMSB) .
\ee
For nAMSB, one can match one of the conditions for EW naturalness
by selecting models with $\mu\sim 100-350$ GeV and one can lift
$m_h\to 125$ GeV while allowing for natural top-squark contributions
to $\Sigma_u^u(\tst_{1,2})$ by selecting for larger bulk $A$-terms.
For nAMSB, the gaugino masses are as expected in usual AMSB models
with the wino as the lightest gaugino.
However, the higgsinos can be the lightest of the EWinos with mass less than
the winos, so in nAMSB one expects a higgsino-like WIMP as LVSP.

\begin{figure}[tbp!]
\begin{center}
  \includegraphics[height=0.4\textheight]{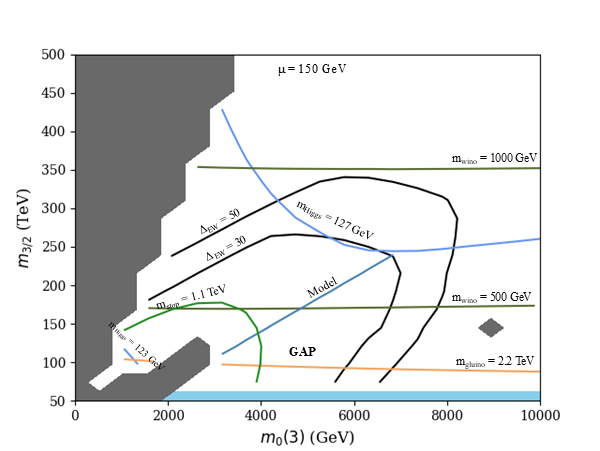}
    \caption{Plot of $m_0(3)$ vs. $m_{3/2}$ parameter space in the
      nAMSB model for $m_0(1,2)=2m_0(3)$, $A_0=1.2 m_0(3)$ and $\tan\beta =10$,
      with $m_A=2$ TeV and $\mu =150$ GeV. The gray regions contain a charged LSP or CCB minima.
\label{fig:m03m32}}
\end{center}
\end{figure}

For nAMSB arising from landscape selection of large $m_{3/2}$ and bulk
contributions\cite{Baer:2023ech}, then again one expects statistically
$m_h\simeq 125$ GeV but with gluinos and top-squarks
beyond the present LHC reach.
The nAMSB $m_{3/2}$ vs. $m_0(3)$ parameter space (from Isajet\cite{Paige:2003mg}) is shown in
Fig. \ref{fig:m03m32} for $A_0=1.2 m_0(3)$, $\tan\beta =10$, $m_A=2$ TeV
and $\mu =150$ GeV.
The LHC gluino mass limit is shown by the orange contour
and the $m_h=123$ and $127$ GeV contours are shown in blue, while black contours
show values of $\Delta_{EW}$. The region between the green contours
labelled by $m_{wino}=500$ and 1000 GeV is actually excluded by the ATLAS
search for boosted hadronically-decaying $W$ bosons from wino
production and decay\cite{ATLAS:2021yqv}.
The region labeled GAP below the $m_{wino}=500$ GeV contour
and above the $m_{\tg}=2.2$ TeV contour is EW-natural, has $m_h\sim 125$ GeV
and is LHC allowed.

\subsubsection{Natural generalized mirage mediation (GMM)}

A third well-motivated natural SUSY model comes from mirage mediation (MM)
models, where there are comparable AMSB and gravity-mediation contributions
to soft SUSY breaking terms\cite{Choi:2004sx,Choi:2005ge}. 
In usual mirage mediation models,
$m_{3/2}$ still sets the AMSB contributions to soft terms, but
model-dependent gravity mediation contributions also arise parametrized by
$\alpha$, where $\alpha =0$ gives pure AMSB and $\alpha \to\infty$ gives
the limit of pure gravity-mediation.
Additional soft term contributions to scalar masses involve
new parameters $c_m$ and $c_{m3}$ and trilinear contributions parametrized
by $a_3$ were posited in Choi {\it et al.} Ref. \cite{Choi:2005ge}.
These take on discrete values for the originally posited orbifold
compactification manifold of \cite{Choi:2005ge}.
The original MM models with discrete parameters were all shown to be highly
EW-finetuned for $m_h\simeq 125$ GeV in Ref. \cite{Baer:2014ica}.

\begin{figure}[t]
  \centering
  {\includegraphics[width=.48\textwidth]{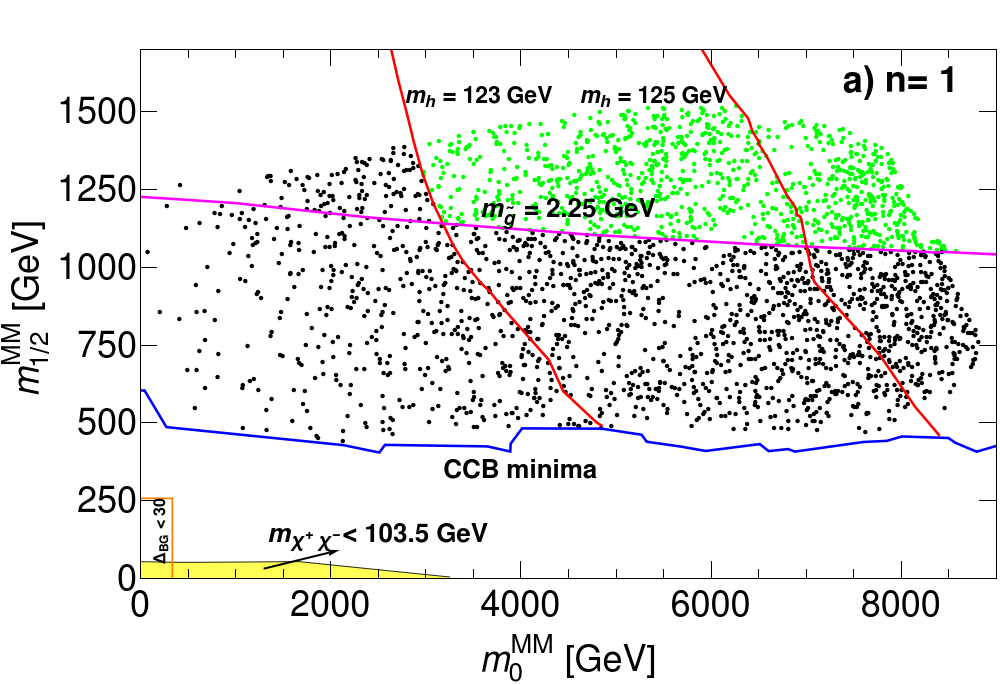}}
  \caption{For $m_{3/2}=20$ TeV, we plot 
the GMM parameter space in the $m_0^{MM}$ vs. $m_{1/2}^{MM}$ 
parameter space for $a_3=1.6\sqrt{c_{m}}$ with $c_{m3}=c_m$
and $\tan\beta =10$ with $m_A=2$ TeV. We plot for a 
landscape draw of $n=1$.
}  
\label{fig:GMMland}
\end{figure}

However, for the much more complicated Calabi-Yau flux compactifications
expected from the landscape, then these discrete parameters should be elevated
to continuous ones.
The result is the generalized mirage mediation model\cite{Baer:2016hfa}
(GMM) with p-space
\be
\alpha,\ m_{3/2},\ c_m,\ c_{m3},\ a_3,\ \tan\beta,\ \mu,\ m_A\ \ \ (GMM).
\ee
In the GMM model, the gaugino masses still unify at the mirage scale
\be
\mu_{mir}=m_{GUT}e^{-8\pi^2/\alpha}
\ee
but now the superpotential $\mu$ parameter can be selected to be in its natural
range and the gravity-mediated $A$-terms can lift $m_h\to 125$ GeV while
maintaining natural top-squark masses\cite{Baer:2016hfa}.
Landscape selection of large gravity-mediated contributions again statistically
prefer $m_h\sim 125$ GeV with sparticles largely beyond LHC Run 2 reach\cite{Baer:2019tee} as is illustrated in Fig. \ref{fig:GMMland} where again the
greater density of dots indicates greater stringy naturalness.
Here, the parameters are $m_0^{MM}\equiv\sqrt{c_m}\alpha(m_{3/2}/16\pi^2)$ and
$m_{1/2}^{MM}\equiv \alpha (m_{3/2}/16\pi^2)$ to correspond to the gravity-mediated
contributions to these parameters.

\subsection{Lessons from plausible SUSY models}

The focus above has been on supersymmetric models which allow for
the LHC results that $m_h\simeq 125$ GeV with sparticles beyond
LHC search limits, and that are also {\it natural} in that they predict
a value of the weak scale that is not too far from $m_{weak}\sim 100$ GeV
{\it without} invoking any sort of (implausible)
EW finetuning in Eq. \ref{eq:mzs}.
The latter requirement corresponds to EW finetuning measure
$\Delta_{EW}\alt 30$. Invoking these requirements divides SUSY models
in plausibility classes: implausible (CMSSM, GMSB, mAMSB, inoMSB, split, MM and WTN) and plausible
(NUHM2-4, nAMSB, nGMM).

\begin{figure}[htb!]
\centering
    {\includegraphics[height=0.4\textheight]{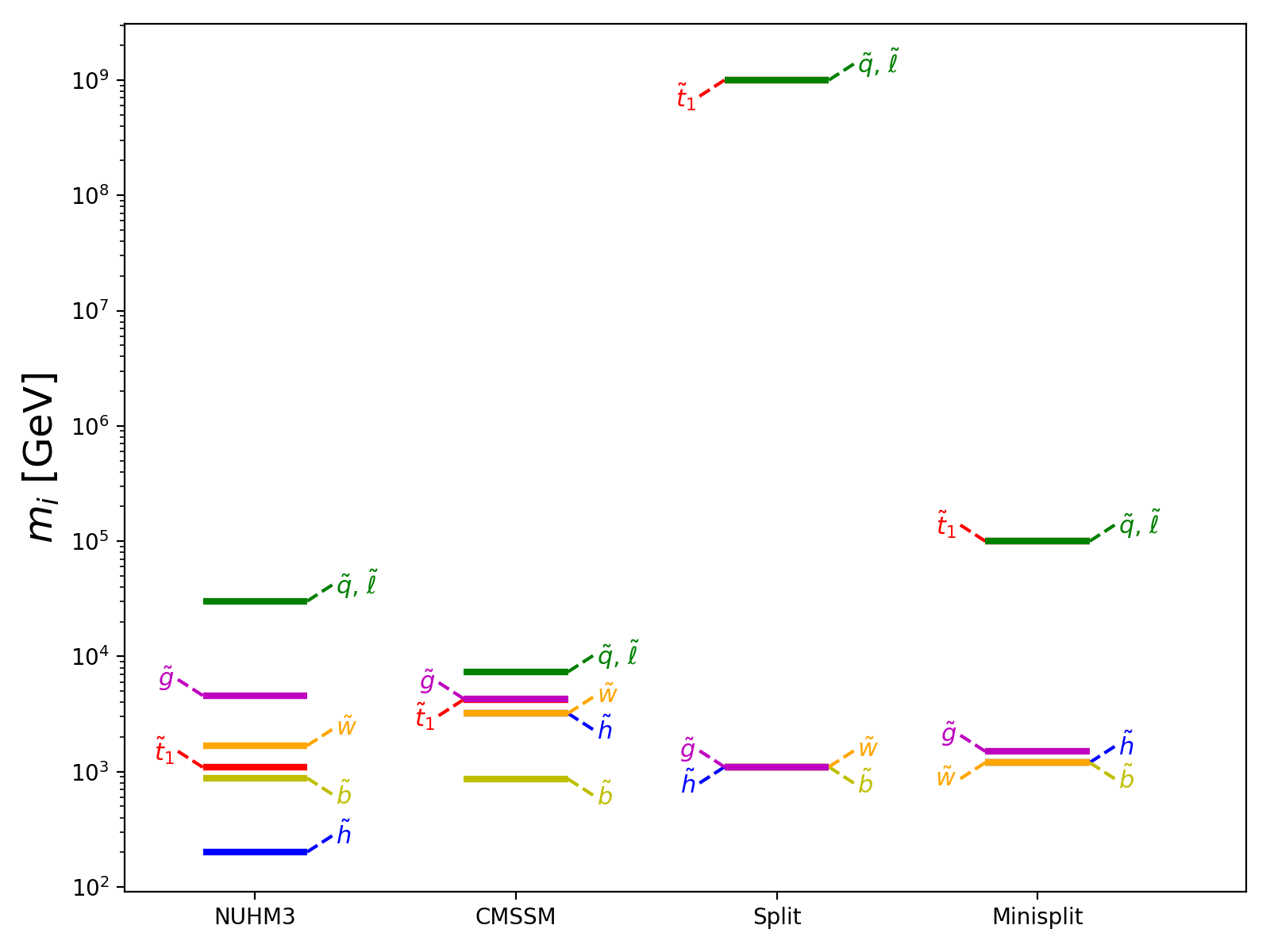}}
    \caption{Figurative plot of sparticle masses in the
      paradigm natural SUSY model along with unnatural models CMSSM, split and
      minisplit SUSY.
      \label{fig:spectrum}}
\end{figure}

In Fig. \ref{fig:spectrum} we show sample spectra from the NUHM3 natural SUSY
model where first/second generation scalars have masses $\sim 30$ TeV but light higgsinos at $\sim 200$ GeV, highly mixed top-squarks with mass a few TeV and where $m_h=125$ GeV. We also show the CMSSM model
where the rather heavy higgsinos at the several TeV scale will require some other
contribution to balance out so as to maintain $m_{weak}\sim 100$ GeV (unnatural).
The original split-SUSY model conjectured scalars as high as $10^{9}$ GeV so that
$\Sigma_u^u(\tst_{1,2})$ and other contributions to Eq. \ref{eq:mzs} would be enormous, 
thus making it implausible that the weak scale could lie around 100 GeV. We also show 
minisplit SUSY wherein the scalar masses are dialed down so as to accommodate
$m_h=125$ GeV but where $\Sigma_u^u(\tst_{1,2})$ and other terms would still lie far beyond the measured weak scale, thus necessitating implausible cancellations in 
Eq. \ref{eq:mzs} to generate $m_{weak}\sim 100$ GeV. 

\begin{figure}[htb!]
\centering
    {\includegraphics[height=0.4\textheight]{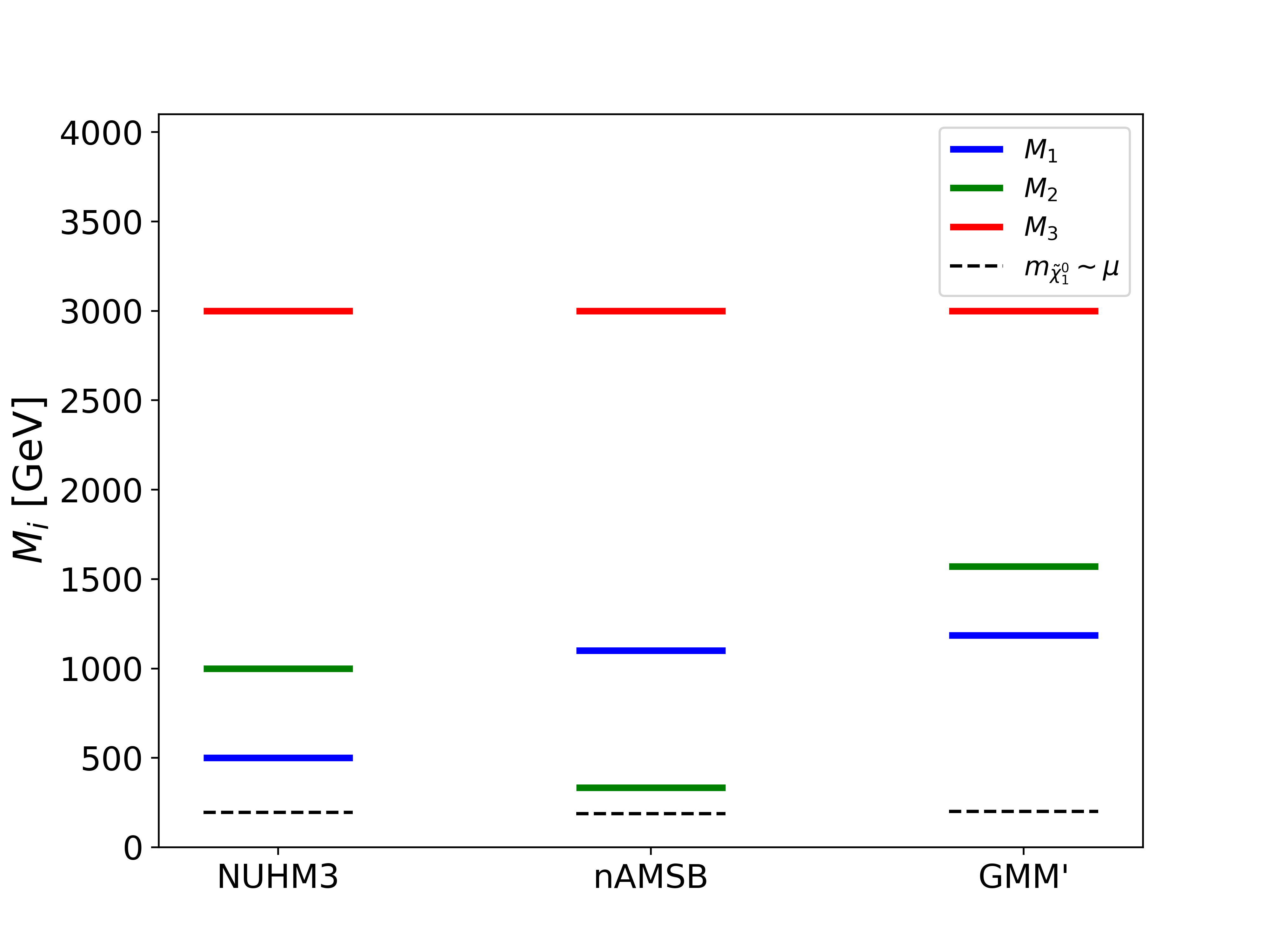}}
    \caption{Figurative plot of gaugino/higgsino masses in the
      three paradigm natural SUSY models with $m_{\tg}\simeq 3$ TeV
      and $\mu =200$ GeV (from Ref. \cite{Baer:2024tfo}).
      \label{fig:inomass}}
\end{figure}

Our goal in this review is to collect all results pertaining to the
SUSY discovery channels and reach of HL-LHC for {\it plausible} SUSY models.
For the three plausible SUSY models listed, we present in
Fig. \ref{fig:inomass} a figurative plot of the expected spectra from the
three surviving models. While all surviving plausible SUSY models have light
higgsinos with mass $\mu\sim 100-350$ GeV, the orientation of the associated
gaugino masses is quite different amongst the three cases.
Most of our results presented will focus on the NUHM2-4 models with
hidden sector singlet SUSY breaking, although we will also examine facets of
nAMSB and nGMM.

\section{Prospects for natural supersymmetry at   HL-LHC with
  $\sqrt{s}=14$ TeV and $3$ ab$^{-1}$ (NUHM)}
\label{sec:hllhc}

In this Section, we present a review of our main results: prospects for
discovery of natural supersymmetry in different discovery channels
at high luminosity LHC (which usually assumes LHC $pp$ collisions at
$\sqrt{s}=14$ TeV with 3 ab$^{-1}$ of integrated luminosity).

\subsection{Higgsino pair production}
\label{sec:higgsino}

Natural SUSY discovery via higgsino pair production at LHC was emphasized in Ref. \cite{Baer:2011ec}. The reaction 
$pp\to\tchi_1^+\tchi_1^-$ does not lead to a distinctive final state above SM backgrounds while the final state $\tchi_1^0\tchi_1^0$ is
(essentially) invisible. A more promising approach is to 
search for the reaction
\be
pp\to \tchi_1^0\tchi_2^0\ \ {\rm followed\ by}\ \tchi_2^0\to\tchi_1^0\ell^+\ell^-
\ee
where $\ell = e$ or $\mu$.
A characteristic feature of this reaction is that the invariant mass of the
dilepton pair is kinematically bounded:
$m(\ell^+\ell^-)<m_{\tchi_2^0}-m_{\tchi_1^0}$.
For natural supersymmetric (natSUSY) models with $\mu\ll m_{soft}$, then the mass gap
$m_{\tchi_2^0}-m_{\tchi_1^0}$ is of order $\sim 5-20$ GeV and the visible
opposite-sign/same-flavor dilepton pair is energetically rather soft
while the missing-$E_T$ ($\eslt$) can be soft as well.\footnote{Smaller
  dilepton mass gaps $\alt 5$ GeV are naturalness disfavored because
  then the wino mass $M_2$ must be very large which results in large
  $\Sigma_u^u(\tw )$ in Eq. \ref{eq:mzs}.}
A soft dimuon trigger may be in order if possible.

\begin{figure}[tbp]
\begin{center}
\includegraphics[height=0.4\textheight]{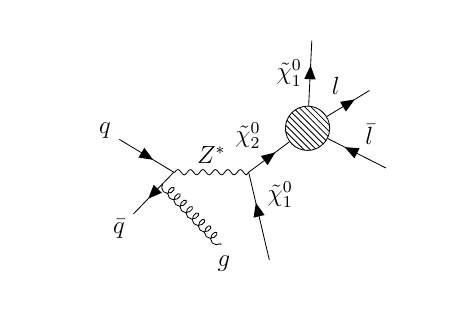}
\caption{Diagram for neutral higgsino pair production reaction at LHC leading to the OSDLJMET signature.
  \label{fig:ppZZll}}
\end{center}
\end{figure}

Han {\it et al.}\cite{Han:2014kaa} suggested to search for these events
where the final state recoils off a hard initial state jet emission which
could serve as a trigger. The hard jet emission can also 
serve to boost the dilepton pair and $\eslt$
to higher values.
The diagram is depicted in Fig. \ref{fig:ppZZll}.
This is the opposite-sign dilepton + jet + missing-$E_T$ signature (OSDLJMET).
Ref. \cite{Baer:2014kya} showed the importance of properly implementing the
ditau invariant mass cut to reject backgrounds from $\tau^+\tau^-j$
production: $m_{\tau\tau}^2<0$. 

The ATLAS\cite{ATLAS:2019lng} and CMS\cite{CMS:2021edw} collaborations
have both presented search results for
$pp\to\tchi_2^0\tchi_1^0 j$ production with $\tchi_2^0\to\tchi_1^0\ell^+\ell^-$.
Both collaborations have some small excesses around the $2\sigma$ level in
this channel. Analysis of Run 3 data is eagerly awaited to see if these
excesses accumulate or dissipate.

\begin{figure}[tbp]
\begin{center}
\includegraphics[height=0.4\textheight]{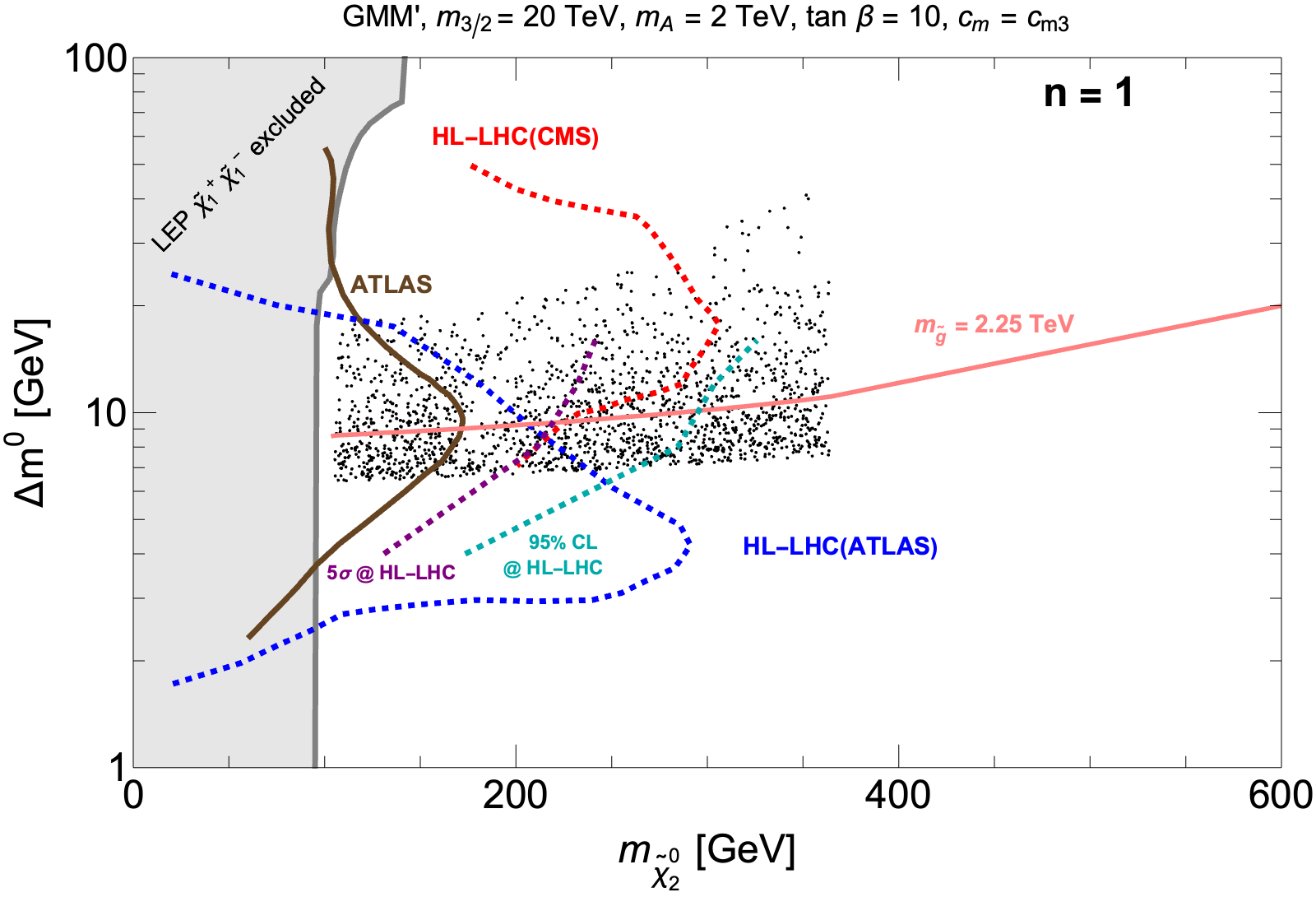}
\caption{Plot of $m_{\tchi_2^0}$ vs. $\Delta m^0$ plane including recent
  ATLAS excluded region (brown curve) along with projected
  HL-LHC reach from ATLAS, CMS and Ref. \cite{Baer:2020sgm}.
  We also show locus of points from the stringy natural GMM model.
  The region above the pink contour
  is excluded by LHC limits $m_{\tg}\agt 2.2$ TeV.
  \label{fig:hplane1}}
\end{center}
\end{figure}

For the 2019 European Strategy and US Snowmass 2021 meetings, some effort has been
applied to ascertain the reach of HL-LHC with 3 ab$^{-1}$ in the OSDLJMET channel. 
In Fig. ~\ref{fig:hplane1} the ATLAS and CMS HL-LHC reach projections are shown
as the blue and red dashed curves in the
higgsino discovery plane of $m_{\tchi_2^0}$ vs.
$\Delta m^0\equiv m_{\tchi_2^0}-m_{\tchi_1^0}$.
In Ref. \cite{Baer:2021srt}, further angular and other cuts were advocated to
reject SM backgrounds, and those HL-LHC reach results are shown by the purple and light blue dashed curves as $5\sigma$ and $95\%$CL reaches.
The region above the pink contour is excluded by 
$m_{\tg}\alt 2.2$ TeV.
The dots show the stringy natural favored region in the GMM model.
While it seems HL-LHC can cover much of the higgsino discovery plane,
some portions with $m_{\tchi_2^0}\agt 300$ GeV apparently remain beyond reach. 
Discovery of higgsino pair production appears difficult via monojets
only due to the huge $Zj$ backgrounds\cite{Baer:2014cua}, but mono-$Z/W$ signals appear more encouraging\cite{Carpenter:2021jbd}.

\subsection{Top squark pair production}
\label{ssec:stops}

\begin{figure}[htb!]
\begin{center}
  \includegraphics[height=0.3\textheight]{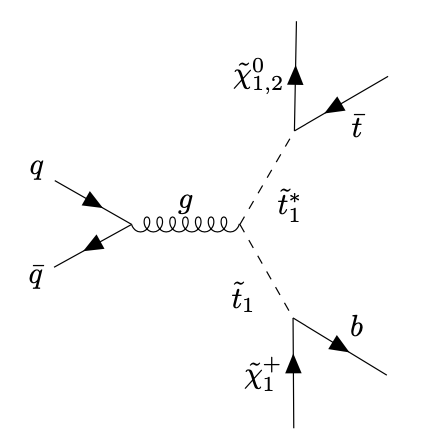}
  \caption{Representative diagram for top squark pair production and decay
    at LHC in natural SUSY.
\label{fig:stoppair}}
\end{center}
\end{figure}

As shown in Fig. \ref{fig:mass}{\it c}), one expects light top squarks of mass
$m_{\tst_1}\sim 1-2.5$ TeV from string landscape scans.
With the added naturalness requirement that $\mu\sim 100-350$ GeV,
this means the following decay modes are open and should dominate
the top squark branching fractions\cite{Baer:2016bwh}:
$\tst_1\to b\tchi_1^+$ and $\tst_1\to t\tchi_{1,2}^0$ where the former
occurs at $\sim 50\%$ and each of the latter at around $\sim 20-25\%$.
Putting top-squark pair production together with decays, and noting the visible decay products of heavier higgsinos are only quasi-visible,
we expect the following final states:
$t\bar{t}+\eslt$, $tb+\eslt$ and $b\bar{b}+\eslt$ as depicted in
Fig. \ref{fig:stoppair}.

\begin{figure}[htb!]
\begin{center}
  \includegraphics[height=0.4\textheight]{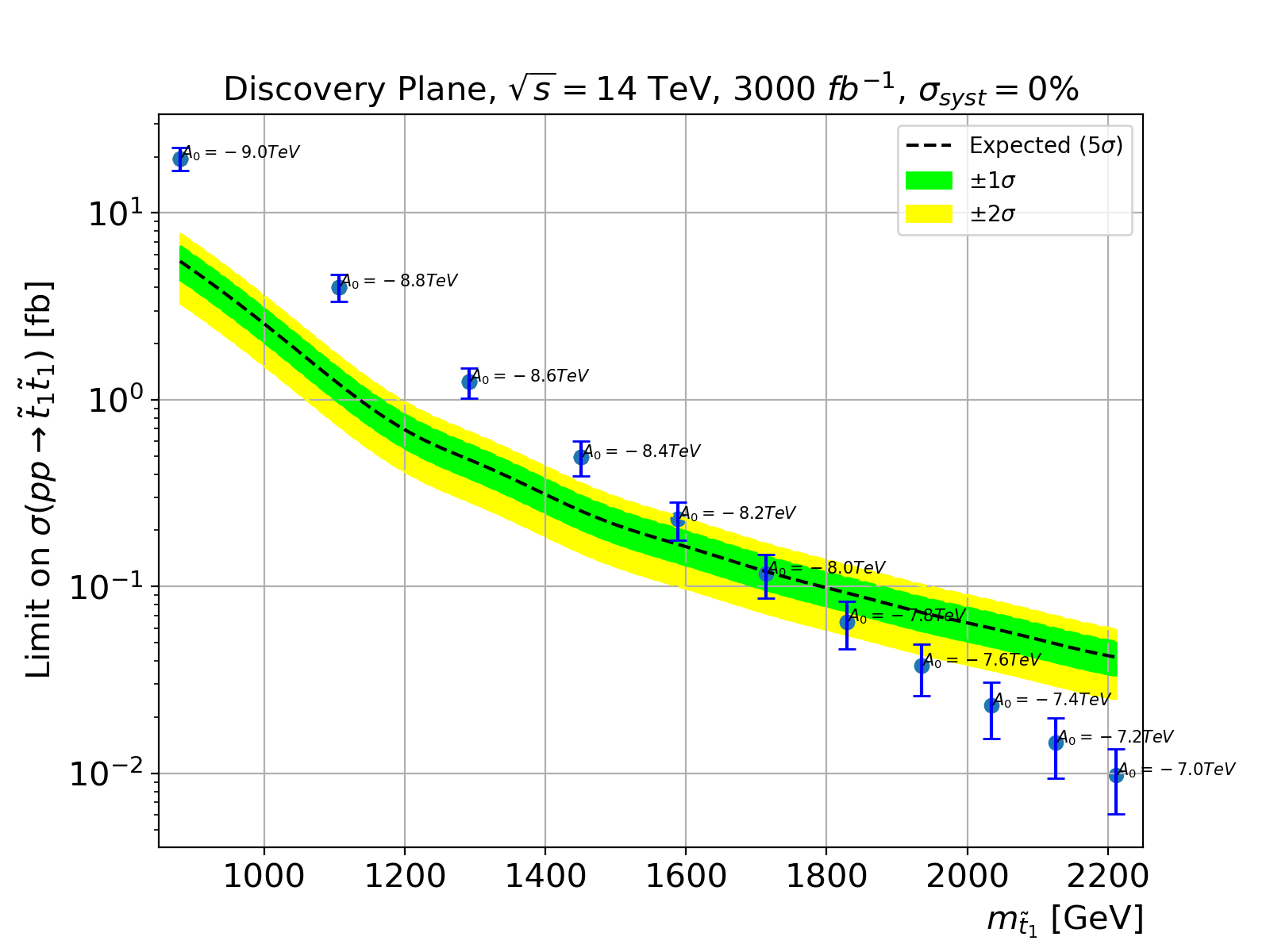}\\
    \includegraphics[height=0.4\textheight]{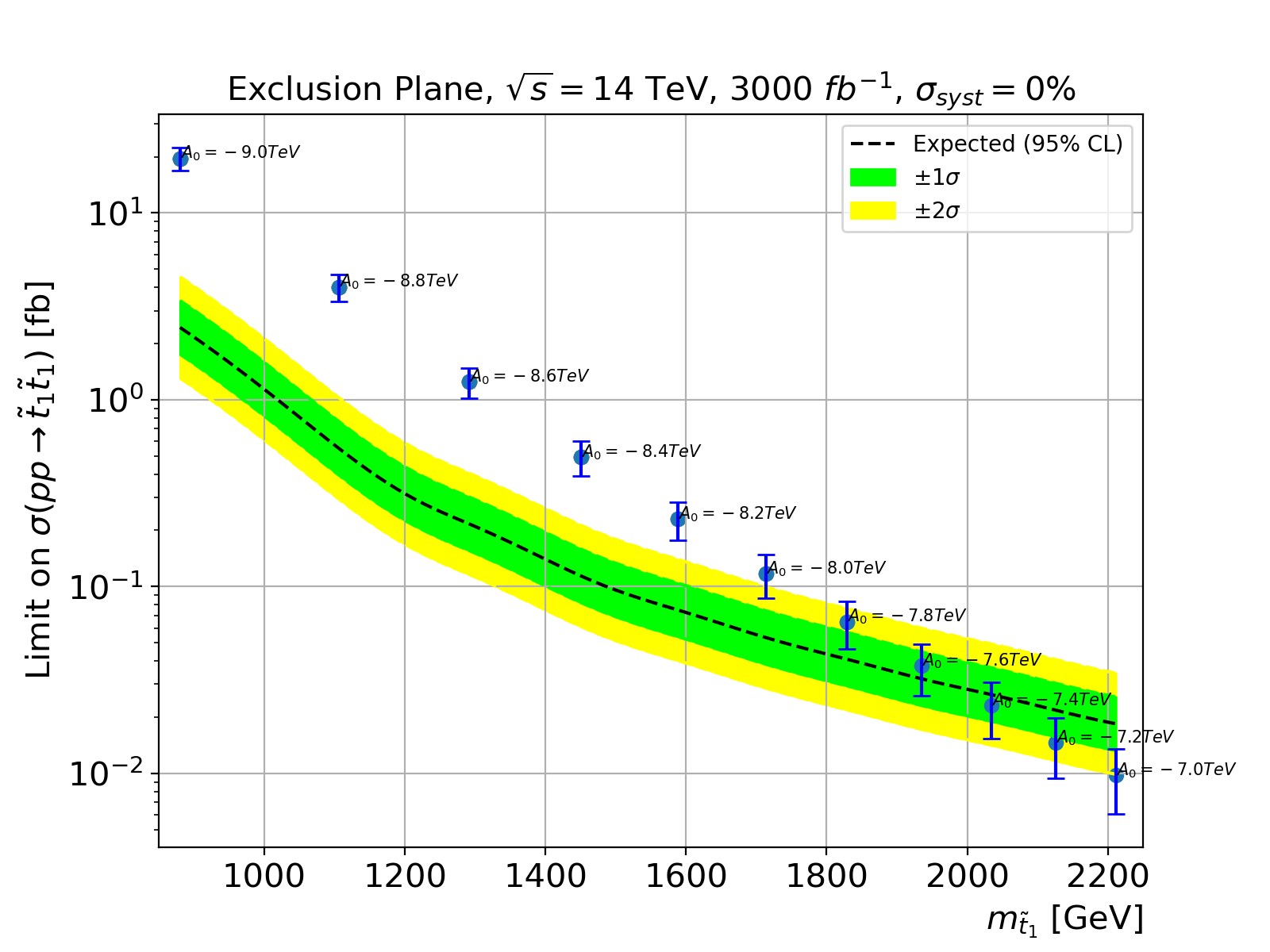}\\
  \caption{Expected 5$\sigma$ discovery limit and expected 95$\%$ CL exclusion limit on top-squark pair production cross section 
    vs. $m_{\tst_1}$ from a natural SUSY model line at HL-LHC with
    $\sqrt{s}=14$ TeV and 3 ab$^{-1}$ of integrated luminosity. Figures are updated from Ref.~\cite{Baer:2023uwo}
\label{fig:discexcl}}
\end{center}
\end{figure}

The reach of HL-LHC for these top-squark pair production final states
has been examined in Ref.~\cite{Baer:2023uwo}.
By requiring events with high $H_T$ and $\eslt$ and the presence of two
top-like or bottom-like objects, then the $m_{T2}$ variable~\cite{Barr:2003rg} can be reconstructed
which wants to be bounded by $m_{\tst_1}$.
The $m_{T2}$ distribution typically exhibit a bulge above SM backgrounds for
$m_{T2}\sim 1-2$ TeV and $m_{\tst_1}\alt 2$ TeV.
By combining all three top-squark discovery channels, then the HL-LHC
top squark mass reach for natural SUSY models can be extracted from
Fig. \ref{fig:discexcl}. It is found that
HL-LHC with 3 ab$^{-1}$ has a $5\sigma$ reach to $m_{\tst_1}\sim 1.7$ TeV
and a 95\%CL reach to $m_{\tst_1}\sim 2.05$ TeV when systematic uncertainty is treated as negligible.
These HL-LHC reach limits will cover much (but not all) of the
expected stringy natural parameter space from SUSY on the landscape!

\subsection{Wino pair production}

In the string landscape, gaugino mass soft terms should have the same pull to
large values as other soft terms, and so we expect the $U(1)_Y$ and $SU(2)_L$ 
bino and wino masses $M_1,\ M_2\gg \mu$, but where $\Sigma_u^u(wino,bino)$
must not contribute too much to the weak scale via Eq. \ref{eq:mzs}.
In NUHM2-4 models with gaugino mass unification, we expect
$M_1(weak)\sim M_2(weak)/2$ with $\tchi_2^\pm$ and $\tchi_4^0$
as the charged and neutral winos,
with $m_{\tchi_2^\pm}\simeq m_{\tchi_4^0}$, while for other models such as nAMSB or
GMM this arrangement is predictably different.
For typical NUHM2-4 models, it is typically found that
$M_2(weak)\sim 1-2$ TeV and $M_1\sim 0.5-1$ TeV.
Since there are several winos and $g_2\gg g_1$, we expect wino pair production
to be more important than bino production.
Thus, here we focus on wino pair production signatures in NUHM2-4:
$pp\to \tchi_2^\pm\tchi_2^\mp$ and $pp\to \tchi_2^\pm \tchi_4^0$ production. 

The various electroweakino (EWino) branching fractions
can be obtained from ISAJET\cite{Paige:2003mg}.
For models with gaugino mass unification,
and where phase space effects are unimportant, one typically
finds for charged wino decays that
$B(\tchi_2^+\to W\tchi_{1,2}^0): B(\tchi_2^+ \to h\tchi_1^+):B(\tchi_2^+\to Z\tchi_1^+)\simeq 2:1:1$,
with a very small fraction of the $\tchi_2^\pm$
decaying via the dynamically and kinematically suppressed decay to the bino. 
Here, we sum over the decays to the neutral higginos since
it is hard to identify the soft visible decay products of the higgsinos.
Likewise, for neutral winos,
$B(\tchi_4^0\to W^\mp\tchi_1^\pm):B(\tchi_4^0\to h\tchi_{1,2}^0):B(\tchi_4^0\to Z\tchi_{1,2}^0) =2:1:1$
while decay to the bino is again strongly suppressed.
Thus, production and decay of wino-like neutralinos in natSUSY leads to final
states of $VV$, $Vh$ and $hh+\eslt$ (where $V=W$ or $Z$).

\subsubsection{Same-sign diboson$+\eslt$ signature from wino pair production}
\label{sssec:ssdb}

\begin{figure}[htb!]
\begin{center}
  \includegraphics[height=0.3\textheight]{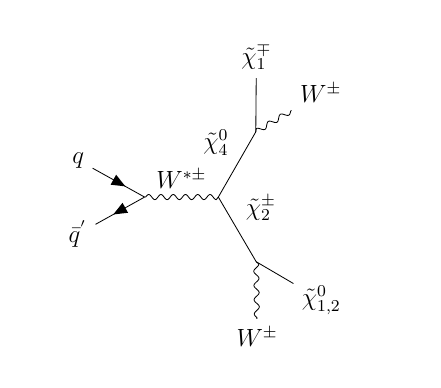}
  \caption{Feynman diagram for the SSdB signature from wino pair production and decay
    at LHC in natural SUSY models.
\label{fig:diagram_ssdb}}
\end{center}
\end{figure}

A very intriguing signature which only occurs in models with $M_2\gg \mu$
is the (relatively) jet-free same sign diboson (SSdB) final state\cite{Baer:2013yha}
\be
pp\to\tchi_2^\pm\tchi_4^0\to W^\pm W^\pm +\eslt\to \ell^\pm\ell^\pm +\eslt ,
\ee
see Fig. \ref{fig:diagram_ssdb}.
While the same-sign dilepton signature is suppressed by two leptonic
branching fractions, the SM backgrounds are expected to be tiny and thus this
distinctive signature may lead to an avenue for discovery of natSUSY
for a sufficiently large integrated luminosity as could be had at HL-LHC.

By requiring exactly two isolated hard same-sign leptons plus zero
$b$-jets plus additional cuts on $\eslt$ and transverse mass,
signal and SM background have been evaluated along a natural SUSY NUHM2
model line\cite{Baer:2017gzf}. For HL-LHC with 3 ab$^{-1}$,
the $5\sigma$ reach extends out to $M_2\sim 860$ GeV while the
95\%CL reach extends to 1080 GeV. Some distinctive features of the SSdB
signature include the same-sign dilepton (SSDL)
asymmetry ($\sigma (++)\gg\sigma (--)$
where the asymmetry depends on the wino mass, and the cleanliness
of the signature. Since the SSDL signature arises from the
well-predicted wino pair production cross section and relatively
well-understood wino branching fractions, the total signal rate might be used
to extract a measurement of the weak scale wino mass $M_2$
up to some error bars.
\begin{figure}[htbp]
\begin{center}
\includegraphics[width=15cm,clip]{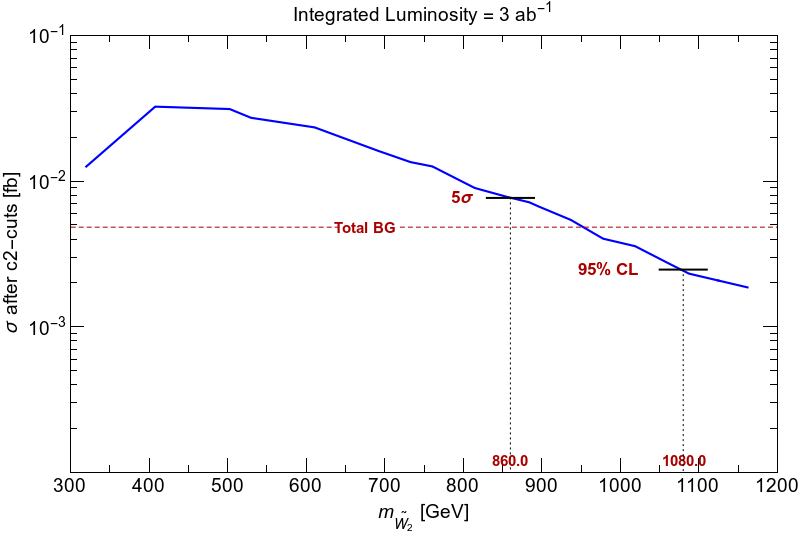}
\end{center}
\caption{Cross section for SSdB production 
after {\bf C2} cuts\cite{Baer:2017gzf} versus $m(wino)$ at the LHC with 
$\sqrt{s}=14$ TeV.  We show the $5\sigma$ and 95\% CL reach 
assuming a HL-LHC integrated luminosity of 3 ab$^{-1}$.
\label{fig:ssdb_reach}}
\end{figure}

\subsubsection{Eight channel combined analysis from wino pair production}
\label{sssec:winos}

\begin{figure}[htb!]
\begin{center}
  \includegraphics[height=0.35\textheight]{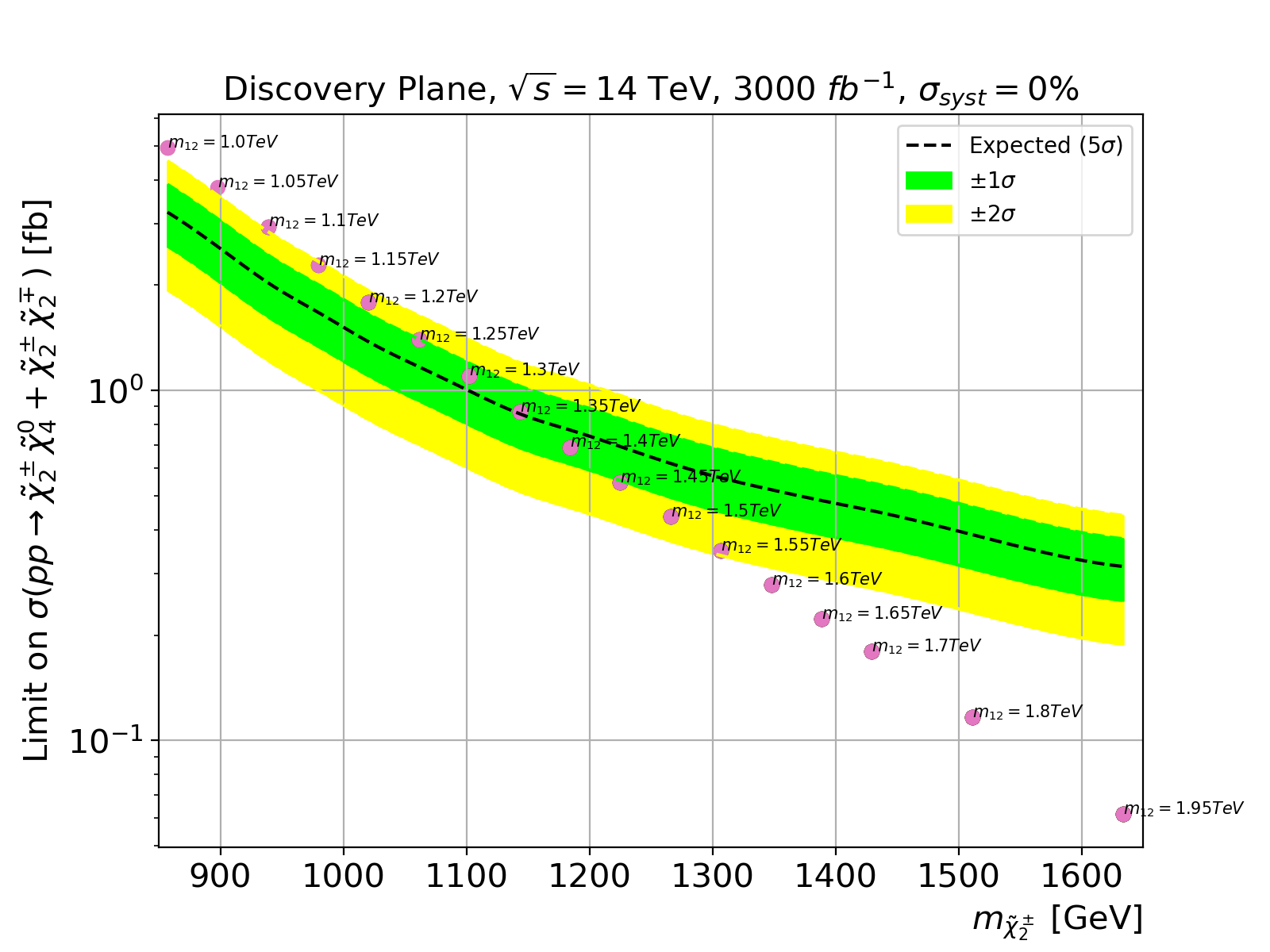}\\
  \includegraphics[height=0.35\textheight]{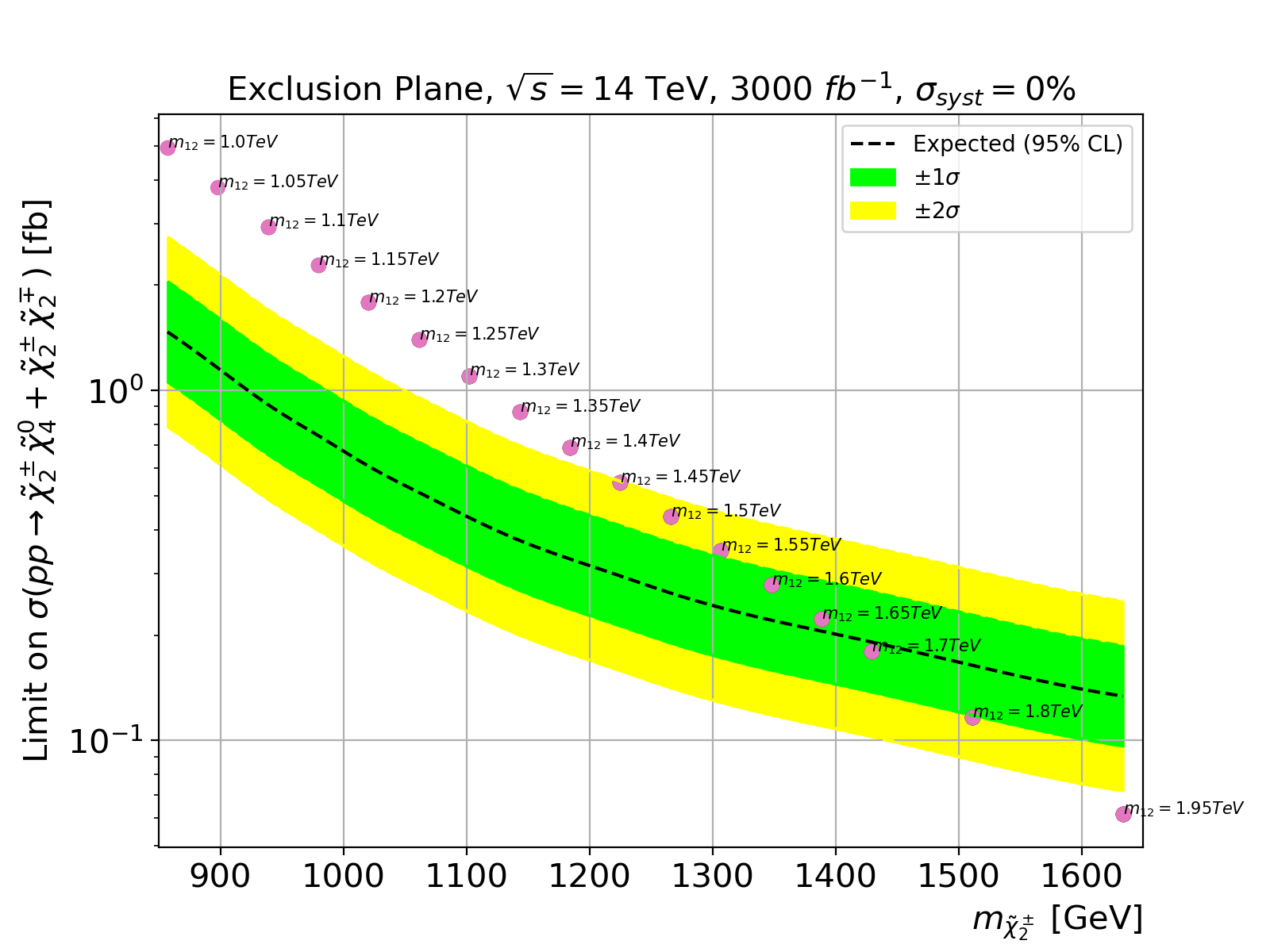}
  \caption{$5\sigma$ discovery reach and 95\%CL exclusion reach
    of HL-LHC from wino pair production
    as a function of $m_{1/2}$ along a NUHM2 model line,
    after combining the eight discovery channels
    detailed in Ref.~\cite{Baer:2023olq}. Both frame shows the reach with
    statistical errors alone, not including possible systematic errors. Figures are updated from Ref.~\cite{Baer:2023olq}
    \label{fig:EWino_reach}}
\end{center}
\end{figure}

Along with the SSdB signature, a variety of other intriguing signatures
arise from wino pair production including the following
classification into eight channels: 
\begin{enumerate}
\item $Z(\to \ell^+\ell^-) B+\eslt$, 

\item $h/Z(\to bb)B+\eslt$, 

\item $BB+\eslt$, 

\item $\ell h+\eslt$, 

\item $\ell B_{W/Z}+\eslt$, 

\item $Z(\to \ell^+\ell^-) +\eslt$,  

\item $h/Z(\to bb)+\eslt$, and

\item $\ell^\pm\ell^\pm +\eslt$ events from $q\bar{q'} \to
  \widetilde{W}^\pm(\to W^{\pm}\tilde{h}^0)\widetilde{W}^0(\to
  W^{\pm}\tilde{h}^{\mp})$, where the $W$ bosons decay leptonically and
  the decay products of higgsinos are soft so that these events have
  hadronic activity only from QCD radiation
  \cite{Baer:2013yha,Baer:2017gzf}.
\end{enumerate}
Here, $B$ (for boson) means any hadronically decayed $W$, $Z$ or $h$
boson, while $B_{W/Z}$ refers to hadronically
decaying $W$ and $Z$ bosons identified as large-radius (fat) jets (J) with
60~GeV~$<m_J<$~100~GeV.
In Ref. \cite{Baer:2023olq}, some optimized cuts for each channel were found and
SM backgrounds were evaluated. The above signal channels along with the
SSdB channel were combined to produce the wino pair production
reach plots for a NUHM2 model line and shown in Fig. \ref{fig:EWino_reach}. 

From Fig. \ref{fig:EWino_reach}, one can extract that the combined wino pair production reach of HL-LHC with $\sqrt{s}=14$ TeV and 3 ab$^{-1}$ extend to $M_2\sim 1.1$ TeV at $5\sigma$ level and $1.4$ TeV at 95\%CL. Thus, the wino pair production reach covers about half the expected natural SUSY parameter space.

\subsection{Gluino pair production}
\label{ssec:gluino}

Traditionally, gluino pair production has been one of the most important
avenues for SUSY searches, at least in the 20th century.
However, for natural SUSY, as emergent from the string landscape,
we expect a pull on $m_{\tg}$ to large values until the gluinos
contribution (at 2-loops) to the weak scale is too  large.
Thus, from the landscape, one expects $m_{\tg}\sim 2-6$ TeV.
From this perspective,
LHC with present mass limits $m_{\tg}\agt 2.2$ TeV is just beginning to explore the expected parameter space.

\begin{figure}[htb!]
\begin{center}
  \includegraphics[height=0.3\textheight]{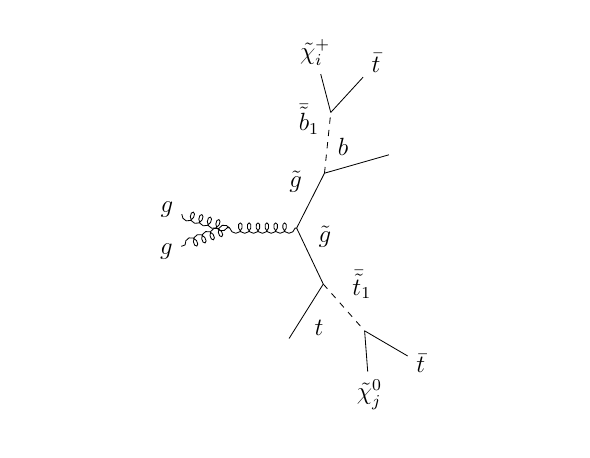}
  \caption{Feynman diagram for gluino pair production and decay
    at LHC in natural SUSY.
\label{fig:diagram_glgl}}
\end{center}
\end{figure}

In natural SUSY, with $m_{\tg}\sim 2-6$ TeV and $m_{\tst_1}\sim 1-2.5$ TeV, then
gluino two-body decays to top-squarks are (almost) always open, and so
we expect $\tg\to t\tst_1$ at a large branching fraction. This is followed
typically by $\tst_1\to b\tchi_1^+$ or $\tst_1\to t\tchi_{1,2}^0$
as mentioned in Subsec. \ref{ssec:stops}. Thus, we expect gluino pair
production to give rise to final states with multiple $t$- and $b$-jets
plus $\eslt$. After top decay, then we expect
\be
pp\to\tg\tg\to 4b's+\eslt +{\rm other\ stuff} .
\ee
A typical gluino pair production followed by cascade decay via third generation
squarks is depicted in Fig. \ref{fig:diagram_glgl}.
The light higgsinos in the final state typically yield soft decay debris.

\begin{figure}[htb!]
  \includegraphics[height=0.3\textheight]{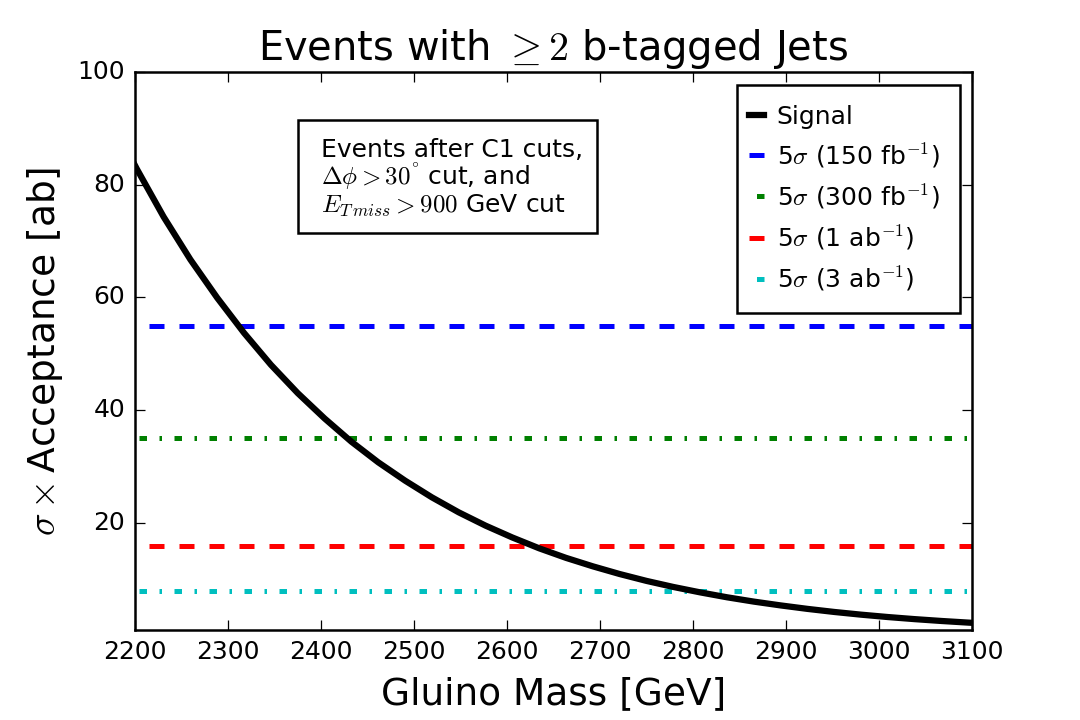}\\
  \includegraphics[height=0.3\textheight]{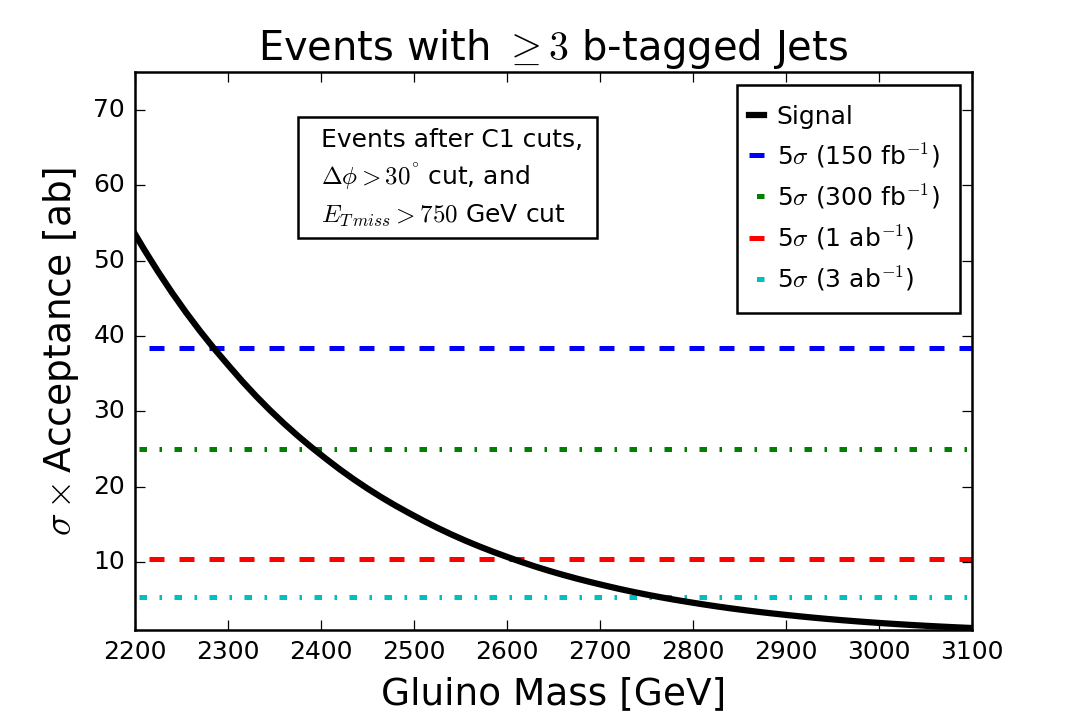}\\
\caption{ The gluino signal cross section for the $\ge 2$ tagged
  $b$-jet (left) and the $\ge 3$ tagged $b$-jet channels (right) after
  all the analysis cuts described Ref. \cite{Baer:2016wkz}.
  The horizontal lines
  show the minimum cross section for which the Poisson fluctuation of
  the corresponding SM background levels, 5.02 ab for $2b$ events and
  1.65~ab for $3b$ events, occurs with a Gaussian probability
  corresponding to $5\sigma$ for integrated luminosities for several
  values of integrated luminosities at LHC14.
\label{fig:fig_glreach}}
\end{figure}

In Ref. \cite{Baer:2016wkz}, a study was made of the reach of HL-LHC for
gluino pair production. In that study, events with 2 or 3 very high
$p_T$ $b$-jets were required along with high $\eslt\agt 750$ GeV or more.
The remaining signal after cuts along with $5\sigma$ reach for
various integrated luminosity options
is shown in Fig. \ref{fig:fig_glreach}. From the figure, it is seen that
HL-LHC with 3 ab$^{-1}$ has a $5\sigma$ reach out to $m_{\tg}\sim 2.8$ TeV
in each channel. The combined reach will be somewhat higher.

For heavy first/second generation squarks, as allowed by naturalness and
expected from the landscape, then the total $\tg\tg$ production
cross section is rather well-determined by just the gluino mass $m_{\tg}$.
For well-defined gluino branching fractions, as expected in natSUSY, then
in the event of a gluino pair production signal, the total signal rate
might be used to extract an estimate of the gluino mass.
Details can be found in Ref. \cite{Baer:2016wkz}.

\subsection{Slepton (right-stau) pair production}

Current LHC search limits for right (R) and left (L) slepton
pair production\cite{Baer:1993ew} followed by $\tell\to\ell\tchi_1^0$ 
($\tell = \te$ or $\tmu$) require
slepton masses to be $m_{\tell}\agt 700$ GeV for the lower
range of $m_{\tchi_1^0}$ and within the framework of the
assumed simplified model\cite{ATLAS:2019lff,CMS:2024gyw}.
While these limits are impressive, we can see from our discussion of
natural SUSY emergent from the landscape that first/second generation
slepton masses would likely inhabit the tens-of-TeV range. 

\begin{figure}[htb!]
\begin{center}
  \includegraphics[height=0.3\textheight]{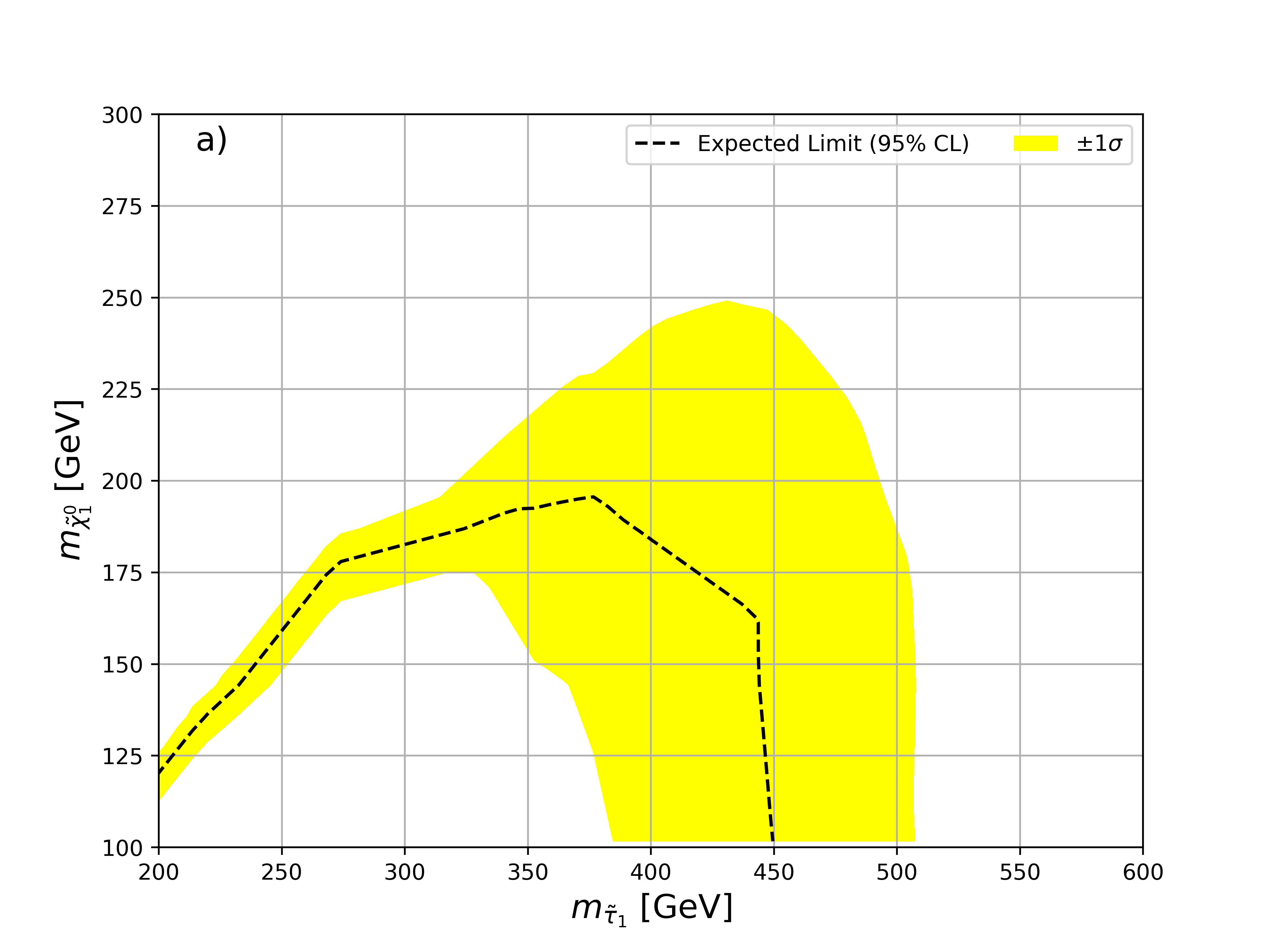}
  \caption{Region of the $m_{\ttau_1}$ vs. $m_{\tchi_1^0}$ plane which
    can be probed by HL-LHC in a natural SUSY model from
    Ref. \cite{Baer:2024hgq}.
    The region below he dashed contour can be excluded at 95\%CL
    but no discovery reach was found to be possible. The yellow region
    shows the $\pm 1\sigma$ uncertainty of the exclusion contour.
\label{fig:stauplane}}
\end{center}
\end{figure}

Alternatively, third generation sfermions have a large contribution
to the weak scale (owing to the large top-quark Yukawa coupling) and hence are
expected to lie within the few TeV range. Thus, a more likely target for slepton pair production would focus on stau pair production.
Of the staus, the mainly right stau is expected to be lighter than the
mainly left-stau due to RG running effects.
Thus, Ref. \cite{Baer:2024hgq} examined a natural SUSY model line but with
variable right-stau soft mass $m_{E_3}$. For low values of $m_{E_3}$,
then $m_{\ttau_R}$ values of a few hundred GeV are possible and are a
worthy target for collider searches. In this case, the reaction
$pp\to\ttau_R^+\ttau_R^-$ was examined, followed by $\ttau_R\to \tau+\tchi_1^0$
leading to a $\tau^+\tau^- +\eslt$ final state.
Both the double hadron final state $\tau_h\tau_h +\eslt$ and mixed
final state $\tau_h\ell+\eslt$ were examined, with hadronic tau decays $\tau_h$
identified by their low multiplicity 1- or 3- charged prong decays.
After a variety of cuts were imposed, the ditau distribution in $m_{T2}$
was examined. For HL-LHC with 3 ab$^{-1}$, no discovery reach was
found in the $m_{\ttau_1}$ vs. $m_{\tchi_1^0}$ plane for a natural SUSY model line
with variable $m_{\ttau_1}$. However, a 95\%CL exclusion limit
could be found, and is depicted in Fig. \ref{fig:stauplane}, along with
its $\pm 1\sigma$ uncertainty. The exclusion limit diminished after
imposing some assumed systematic uncertainty.
%

\subsection{Heavy Higgs production}

\begin{figure}[htb!]
\begin{center}
  \includegraphics[height=0.25\textheight]{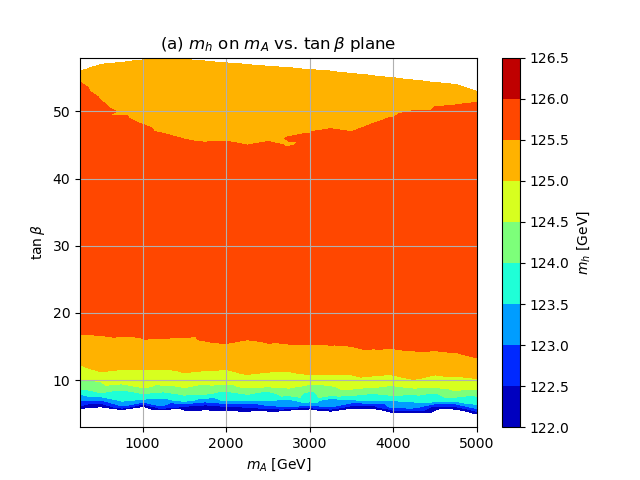}
  \includegraphics[height=0.25\textheight]{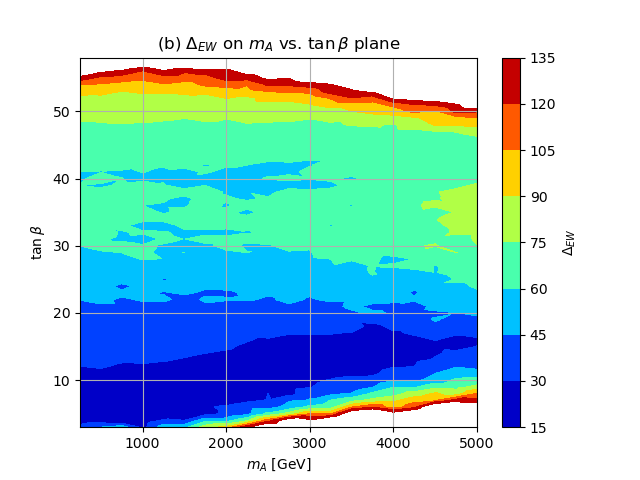}
\caption{{\it a}) Contours of $m_h$ in the $m_A$ vs. $\tan\beta$ plane using the
  $m_h^{125}({\rm nat})$ scenario from the NUHM2 model with $m_0=5$ TeV, $m_{1/2}=1.2$ TeV,
  $A_0=-8$ TeV and $\mu=250$ GeV. {\it b}) Regions of electroweak naturalness
  measure $\Delta_{EW}$ in the same plane as {\it a}).
  \label{fig:mhdew}}
\end{center}
\end{figure}

Along with LHC searches for sparticle pair production,
it has long been appreciated that SUSY may be searched for via resonant
production of the heavy neutral Higgs bosons\cite{Kunszt:1991qe} $H$ and $A$,
and also by searches for charged Higgs bosons $H^\pm$. Heavy Higgs boson
search results are usually displayed in the $m_A$ vs. $\tan\beta$
Higgs discovery plane.
For instance, a recent ATLAS search\cite{ATLAS:2020zms} for
$pp\to A,\ H\to\tau\bar{\tau}$ events from LHC Run 2 with 139 fb$^{-1}$
of data concluded that $m_{A}\agt 1$ TeV for $\tan\beta =10$ and
$m_A\agt 1.7$ TeV for $\tan\beta =30$.
(See Ref. \cite{CMS:2022rbd} for the corresponding CMS study.)
In such analyses, one must assume something about the remainder of the SUSY
spectrum; here, the ATLAS analysis assumed a $m_h^{125}$ SUSY benchmark
point\cite{Bagnaschi:2018ofa} where $m_h=125$ GeV and all sparticles were
assumed heavy with $m_{SUSY}=1.5$ TeV, $\mu =1$ TeV and $m_{\tg}=2.5$ TeV.
These parameter choices are rather unnatural at least due to the large
$\mu$ parameter and so would be hard-pressed to explain why
$m_{weak}\sim 100$ GeV. An alternative Higgs search scenario is the hMSSM
model\cite{Djouadi:2013uqa}, where $m_h$ is used as an input parameter
(guaranteeing $m_h=125$ GeV) while SUSY particles are all assumed heavy
and beyond LHC reach (unnatural).
In Ref. \cite{Baer:2022qqr}, it was noted that the heavy Higgs contribution to the naturalness
measure $\Delta_{EW}$ goes like $m_A/\tan\beta$, so while $\mu$ is constrained
to be less than $\sim 350$ GeV, $m_A$ would correspondingly be required to be
$m_A\alt 3.5$ TeV for $\tan\beta =10$, with higher upper bounds for larger
$\tan\beta$. A natural SUSY benchmark dubbed the $m_h^{125}(nat)$ scenario
was proposed where $m_h\sim 125$ GeV but where large regions of the
$m_A$ vs. $\tan\beta$ plane were natural with $\Delta_{EW}\alt 30$:
see Fig. \ref{fig:mhdew}.

\subsubsection{$A,\ H\to\tau\bar{\tau}$}

\begin{figure}[htb!]
\begin{center}
\includegraphics[height=0.25\textheight]{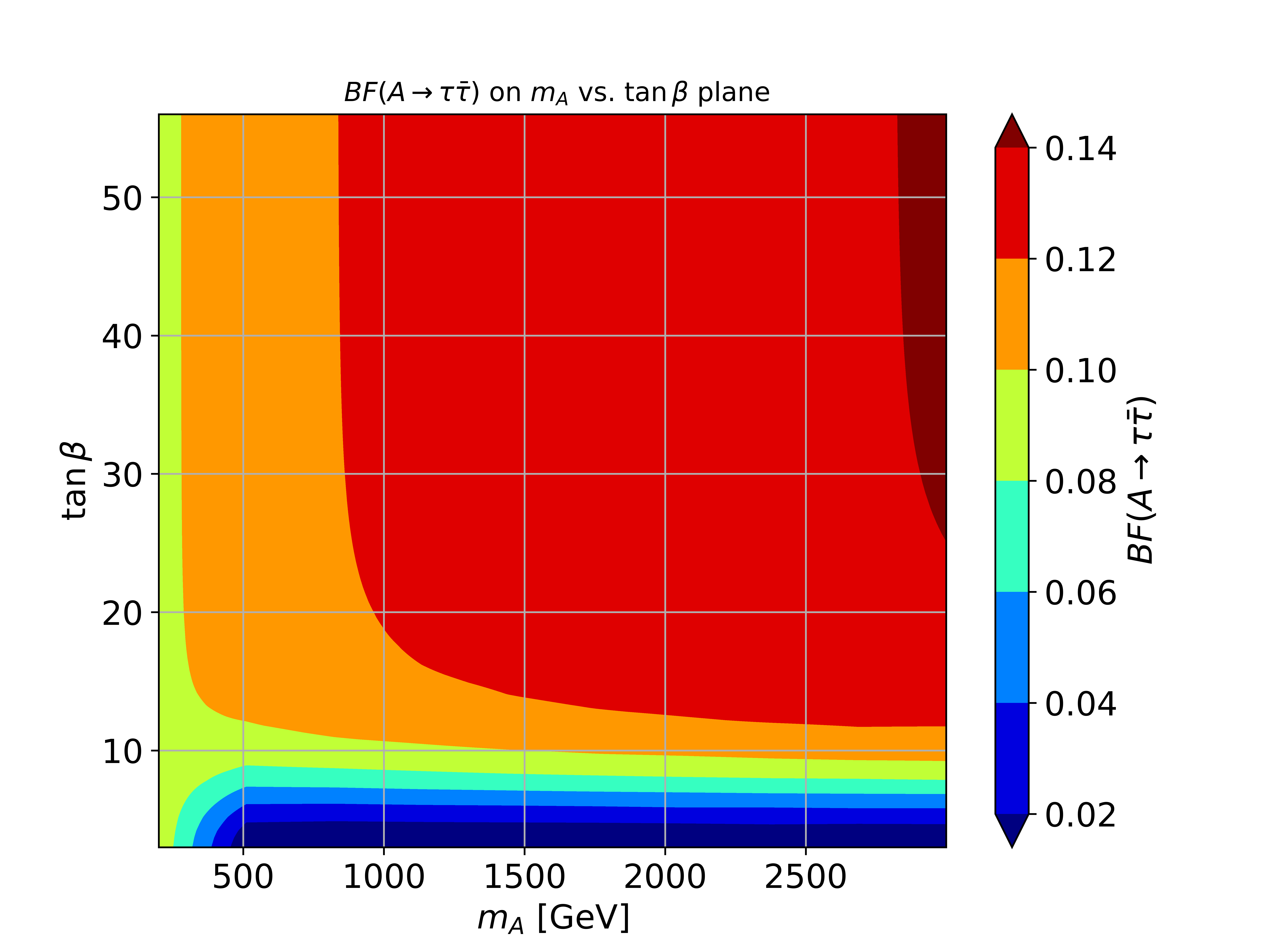}
\includegraphics[height=0.25\textheight]{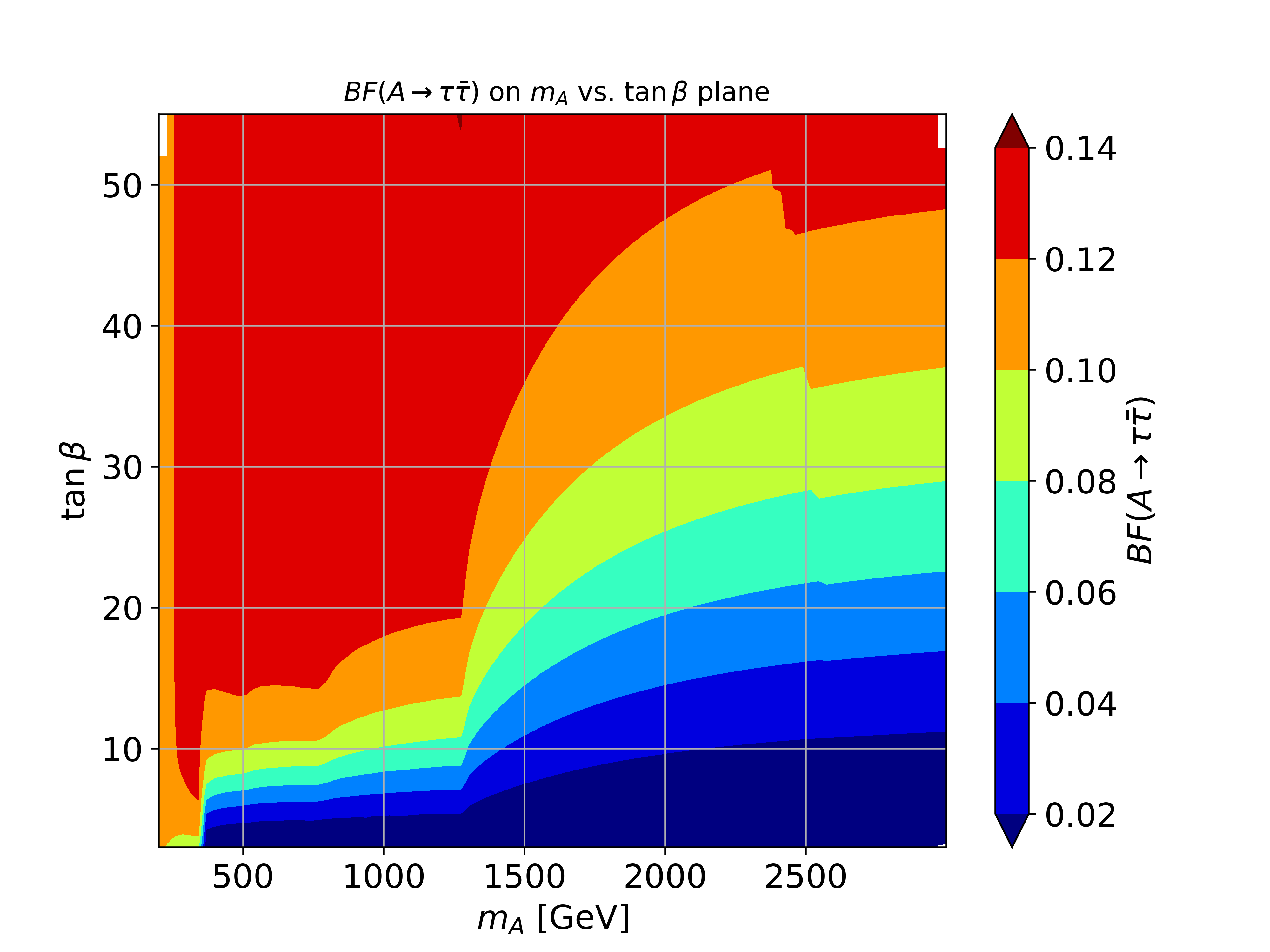}\\
\caption{Branching fraction of $A\to\tau\bar{\tau}$
  in the {\it a}) hMSSM and {\it b}) in the $m_h^{125}({\rm nat})$  benchmark case
  in the $m_A$ vs. $\tan\beta$ plane.
\label{fig:BFAtt1}}
\end{center}
\end{figure}

The most promising heavy Higgs $H,\ A$ search channel is
$pp\to H,\ A\to \tau\bar{\tau}$ due to 1. the large $H,\ A\to\tau\bar{\tau}$
branching fraction, 2. relatively low SM backgrounds mainly from $\gamma^*,\ Z^*\to\tau\bar{\tau}$ and 3. the possibility to reconstruct the
$m(\tau\bar{\tau})$ resonant bump using kinematic constraints from
$\tau$ decay and $\eslt$ in non-back-to-back (BtB) events.
However, in natural SUSY where $m(higgsinos)\sim 100-350$ GeV, then for
$m_A\agt m(higgsino)+m(gaugino)$, the heavy Higgs branching fraction to
gaugino plus higgsino opens up and quickly dominates the heavy Higgs decays.
The $BF(A\to\tau\bar{\tau})$ is shown in Fig. \ref{fig:BFAtt1}
for the hMSSM case and for the $m_h^{125}(nat)$ scenario.
From the Figure, it is clear that
when $m_A$ becomes large enough, then the $H$ and $A$ $\to SUSY$ 
decay modes
quickly dominate the SM modes and the projected heavy Higgs
reach via $\tau\bar{\tau}$ will be overestimated~\cite{Baer:2022qqr}.

\begin{figure}[htb!]
\begin{center}
  \includegraphics[height=0.25\textheight]{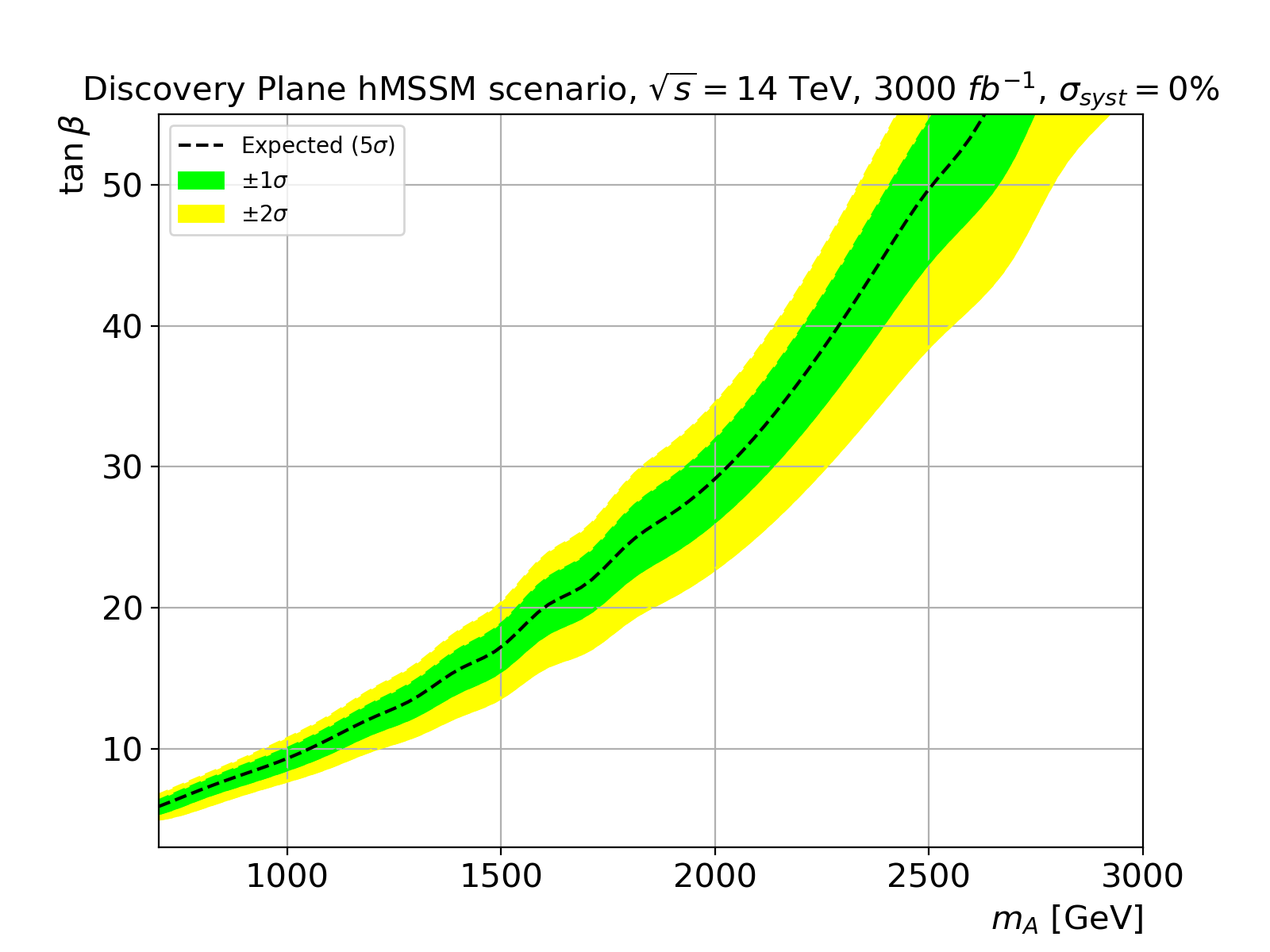}
  \includegraphics[height=0.25\textheight]{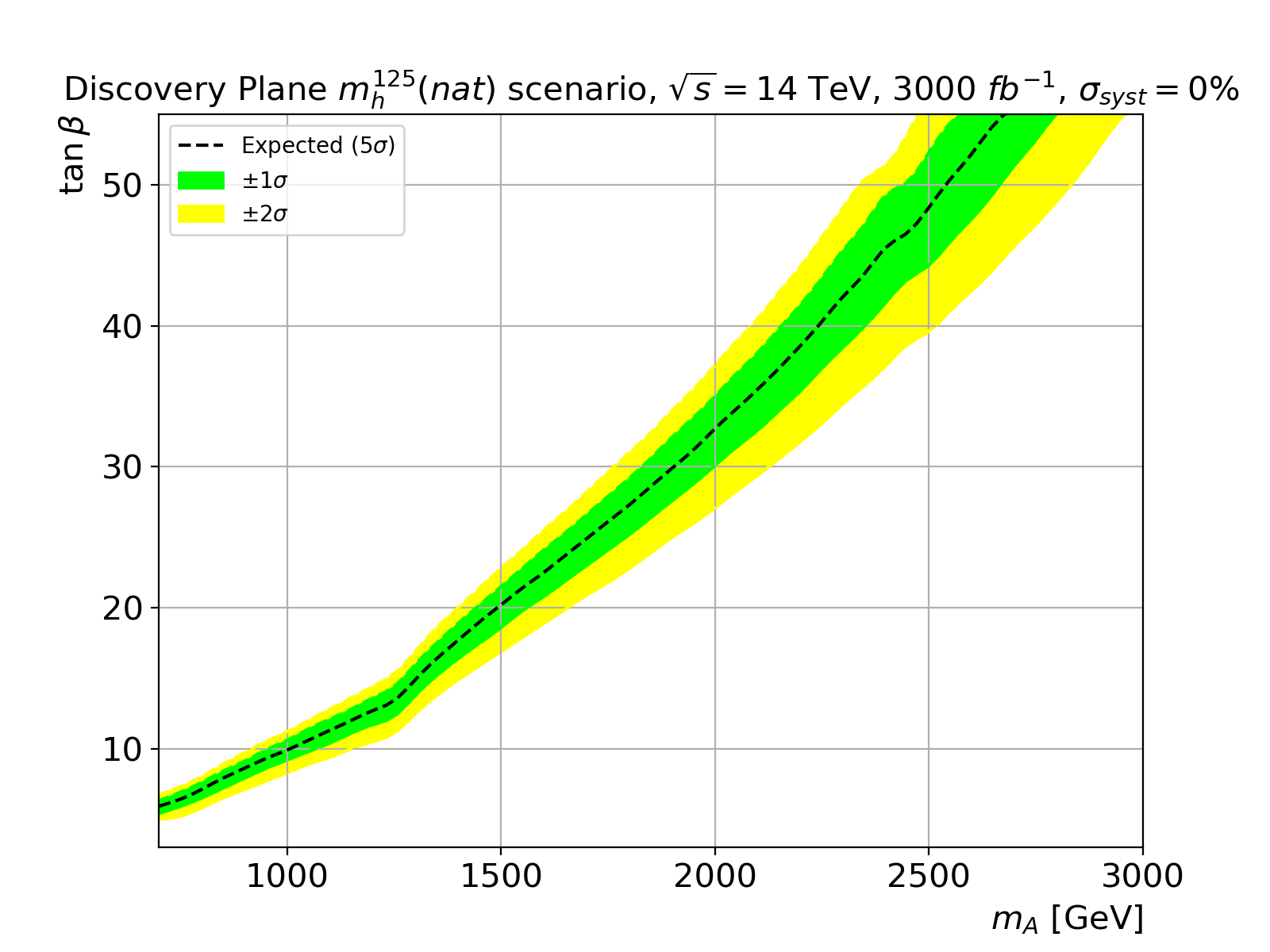}\\
\caption{The discovery sensitivity with $\sqrt{s}=14$ TeV and 3000 fb$^{-1}$
  for $H,\ A\to\tau\bar{\tau}$ in
  {\it a}) the hMSSM and {\it b}) the $m_h^{125}({\rm nat})$ scenario. Figures are updated from Ref.~\cite{Baer:2022qqr}.
  \label{fig:dis3000}}
\end{center}
\end{figure}

In Ref.~\cite{Baer:2022qqr}, both BtB and non-BtB ditau events
were examined and cuts were proposed to extract $H,\ A\to \tau\bar{\tau}$
signal from SM background. The HL-LHC $5\sigma$ discovery reach assuming
3 ab$^{-1}$ of integrated luminosity is shown in Fig. \ref{fig:dis3000}
for {\it a}) the hMSSM model and {\it b}) the $m_h^{125}(nat)$
scenario. For high $\tan\beta \sim 50$ where the
$H,\ A\to\tau\bar{\tau}$ branching fraction is hardly affected, the discovery reach is similar between the two cases. However, for, {\it e.g.}, $\tan\beta \sim 20$,
then the HL-LHC reach goes out to $m_A\sim 1650$ GeV for hMSSM but only
to $m_A\sim 1500$ GeV for the $m_h^{125}(nat)$ scenario.

\subsubsection{$A,\ H\to SUSY$}

\begin{figure}[htb!]
\begin{center}
\includegraphics[height=0.2\textheight]{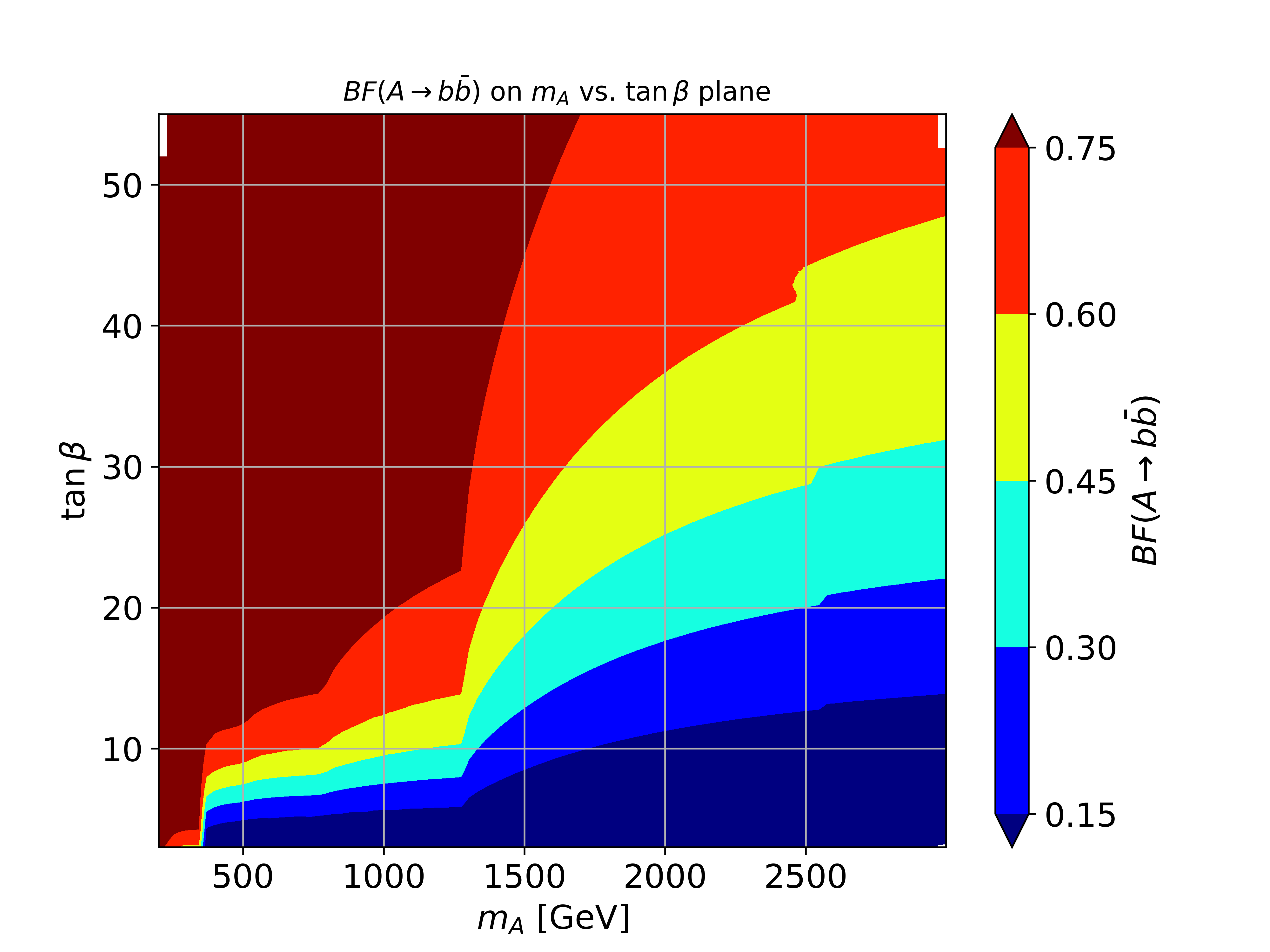}
\includegraphics[height=0.2\textheight]{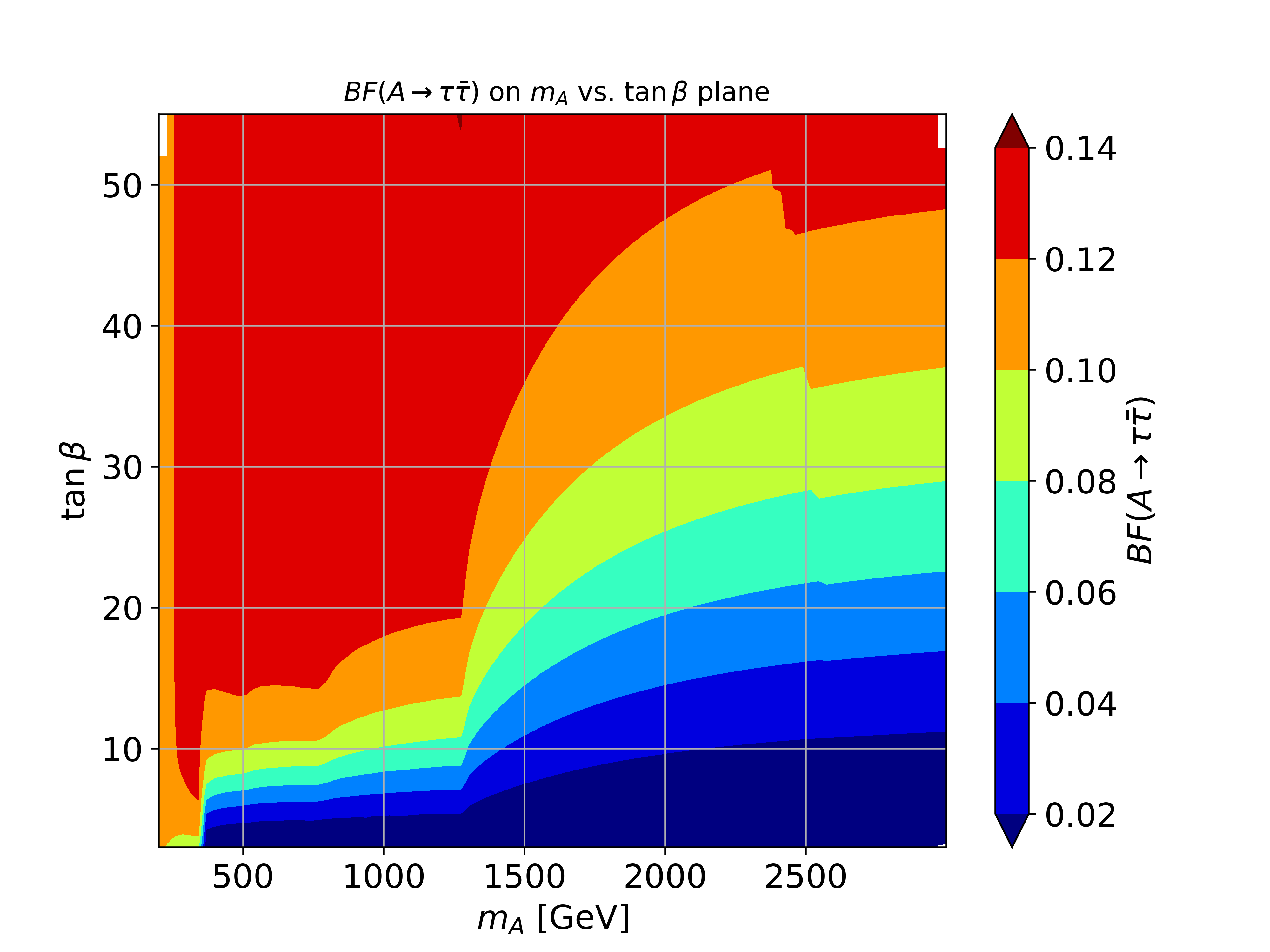}\\
\includegraphics[height=0.2\textheight]{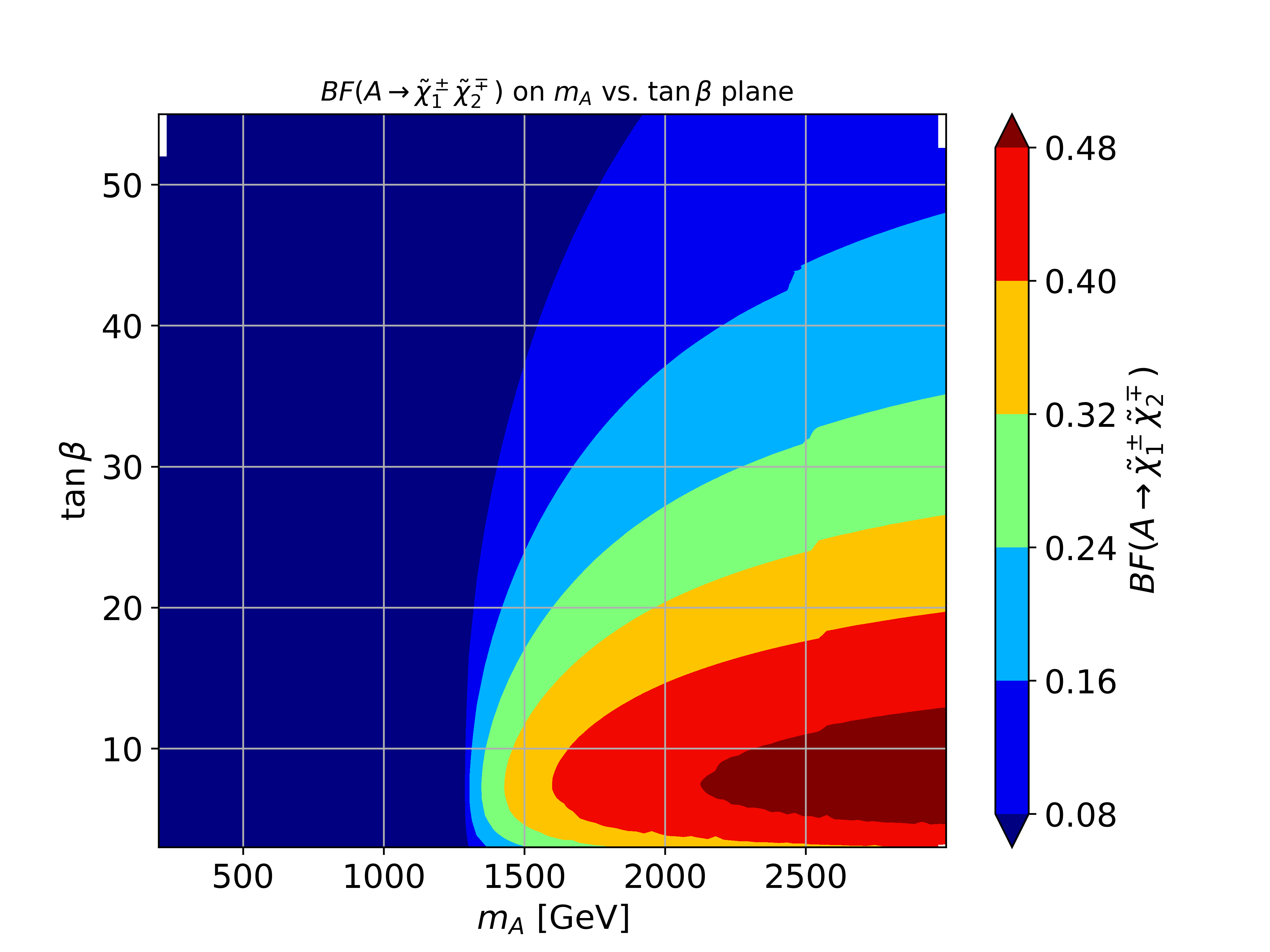}
\includegraphics[height=0.2\textheight]{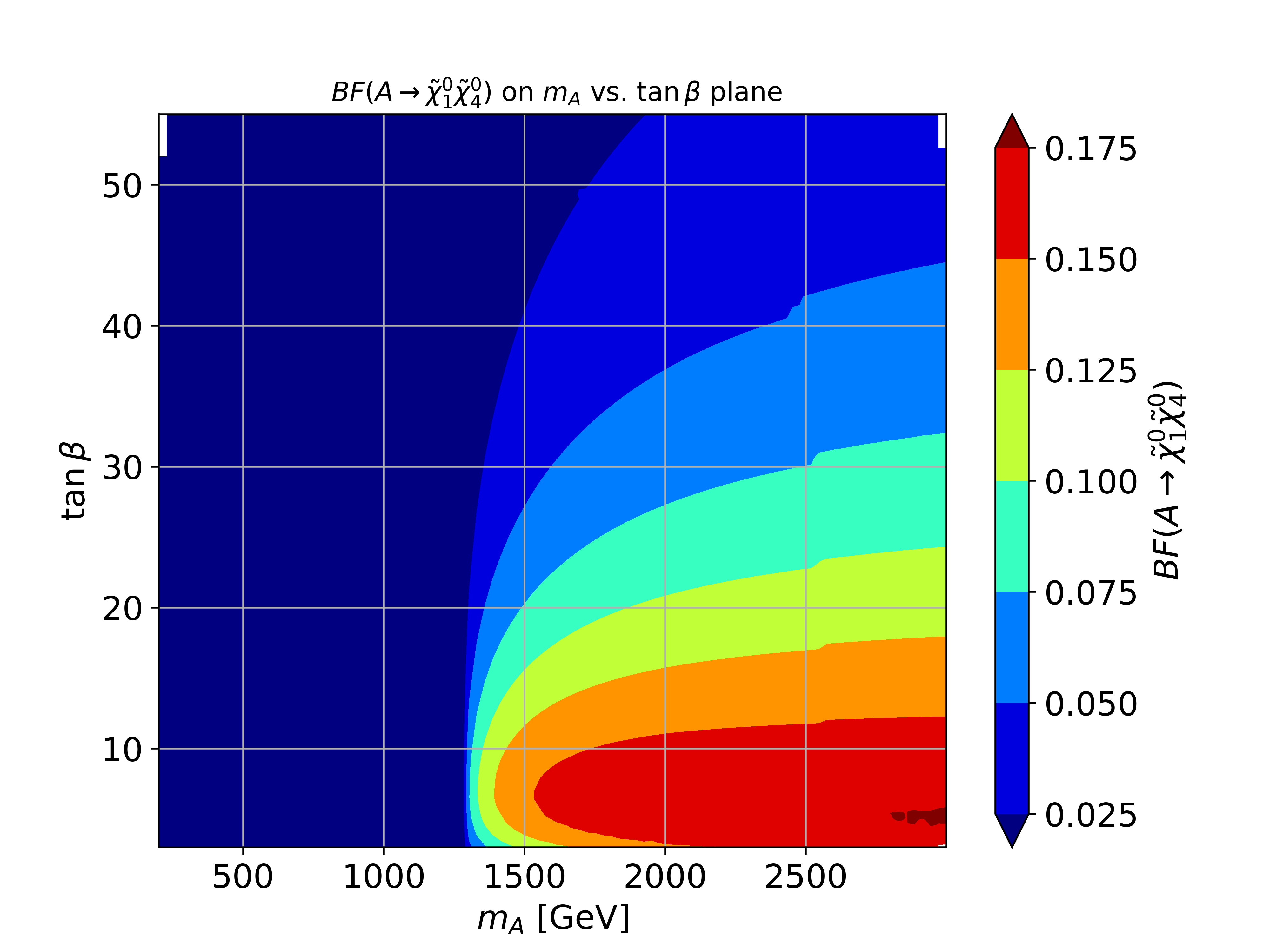}\\
\includegraphics[height=0.2\textheight]{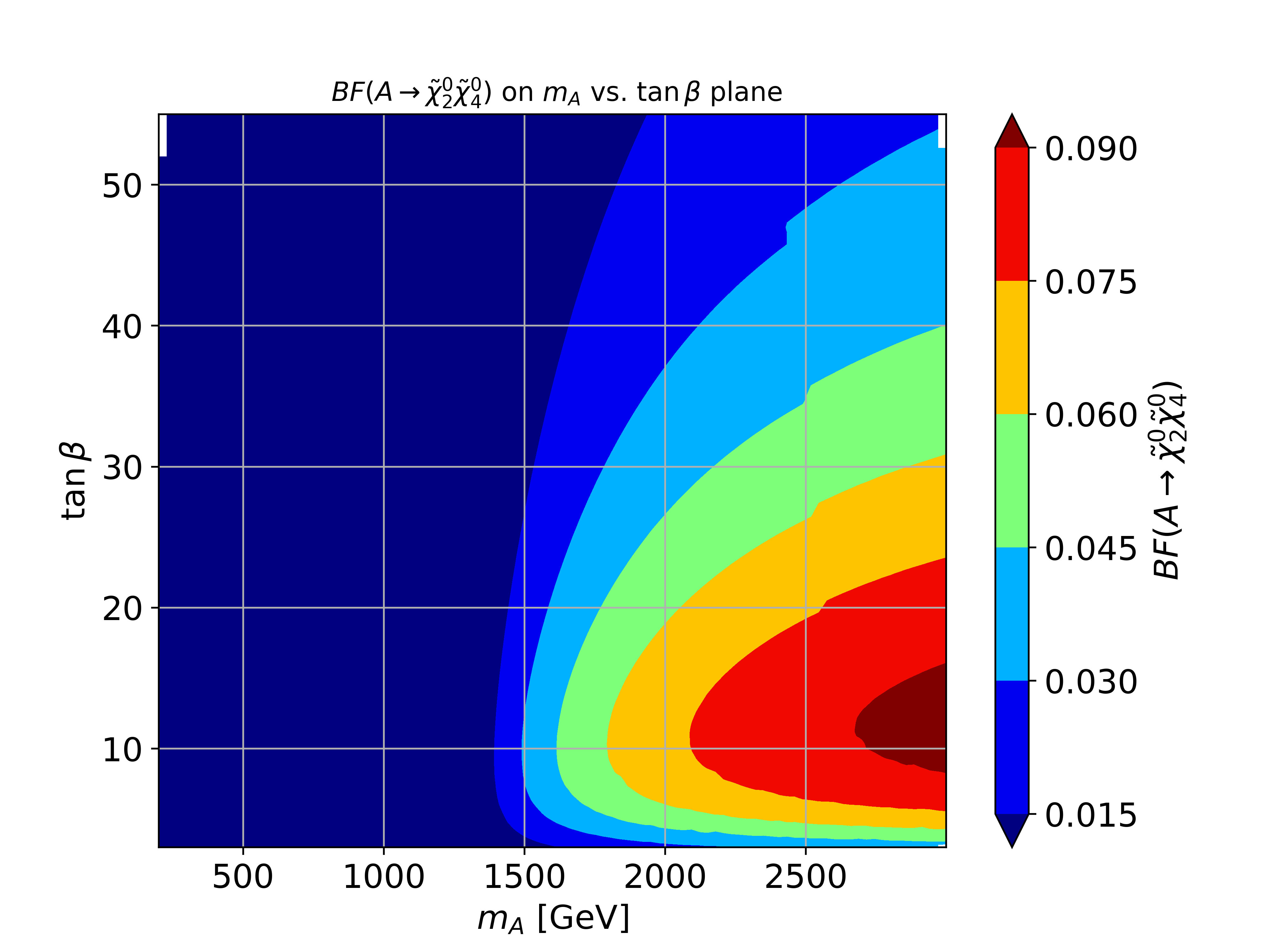}
\includegraphics[height=0.2\textheight]{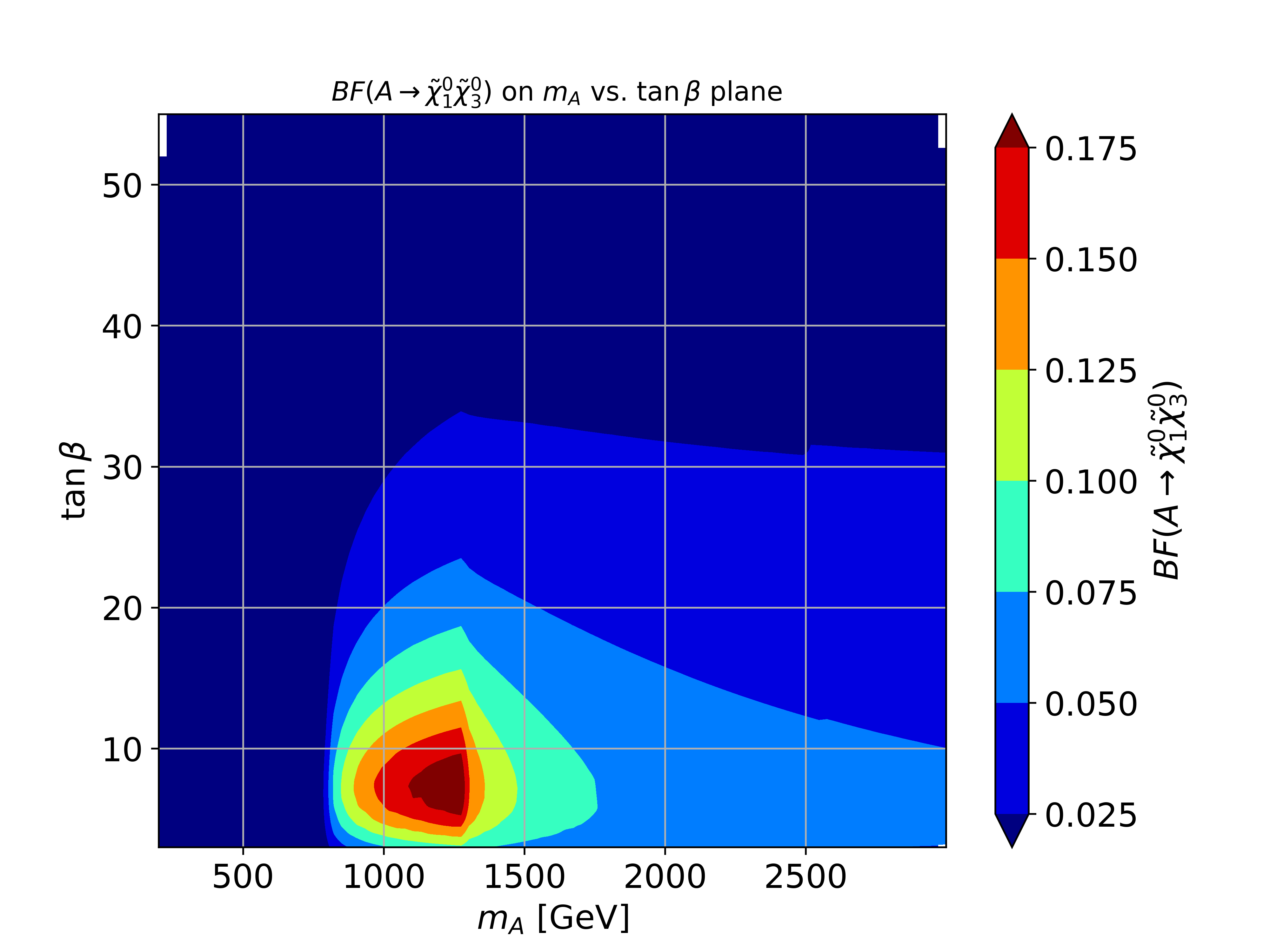}\\
\caption{Branching fractions for $A$ to {\it a}) $b\bar{b}$,
  {\it b}) $\tau\bar{\tau}$, {\it c}) $\tchi_1^\pm\tchi_2^\mp$,
  {\it d}) $\tchi_1^0\tchi_4^0$, {\t e}) $\tchi_2^0\tchi_4^0$
  and {\it f}) $\tchi_1^0\tchi_3^0$ from Isajet 7.88~\cite{Paige:2003mg}.
\label{fig:BFA}}
\end{center}
\end{figure}

In HL-LHC explorations for TeV-scale heavy Higgs bosons $H$ and $A$,
we expect the SM heavy Higgs boson branching fractions to diminish in
natural SUSY due to the opening up of SUSY decay modes, thus also
diminishing discovery prospects via decay modes such as $H,\ A\to\tau\bar{\tau}$.
However, the opening up of new SUSY decay modes for the heavy neutral Higgs
bosons also presents new opportunities for discovery~\cite{Baer:2022smj}.
In Fig.~\ref{fig:BFA}, we show the SUSY $A$ boson branching fractions in the
$m_A$ vs. $\tan\beta$ plane for the $m_h^{125}(nat)$ scenario,
where we show the  {\it a}) $b\bar{b}$,
  {\it b}) $\tau\bar{\tau}$, {\it c}) $\tchi_1^\pm\tchi_2^\mp$,
  {\it d}) $\tchi_1^0\tchi_4^0$, {\t e}) $\tchi_2^0\tchi_4^0$
  and {\it f}) $\tchi_1^0\tchi_3^0$ branching fractions as computed by
  Isajet 7.88\cite{Paige:2003mg}.
  Along with the diminished SM decay modes into $b\bar{b}$ and $\tau\bar{\tau}$
  shown in frames {\it a}) and {\it b}), we see the gaugino plus higgsino
  decay modes opening up for sufficiently heavy $m_A$ in
  {\it c}) $\tchi_1^\pm\tchi_2^\mp$,
  {\it d}) $\tchi_1^0\tchi_4^0$, {\t e}) $\tchi_2^0\tchi_4^0$
  and {\it f}) $\tchi_1^0\tchi_3^0$. The light higgsinos $\tchi_1^\pm$
  and $\tchi_{1,2}^0$ are either invisible or quasi-visible (in the case
  of very soft decay products), but the heavier EWinos, which are
  either bino- or wino-like, preferentially decay to $W$, $Z$ or $h$ plus
  light higgsino, thus yielding a final state of $A\to W,\ Z$ or $h$ plus
$\eslt$.  

\begin{figure}[htb!]
\begin{center}
  \includegraphics[height=0.25\textheight]{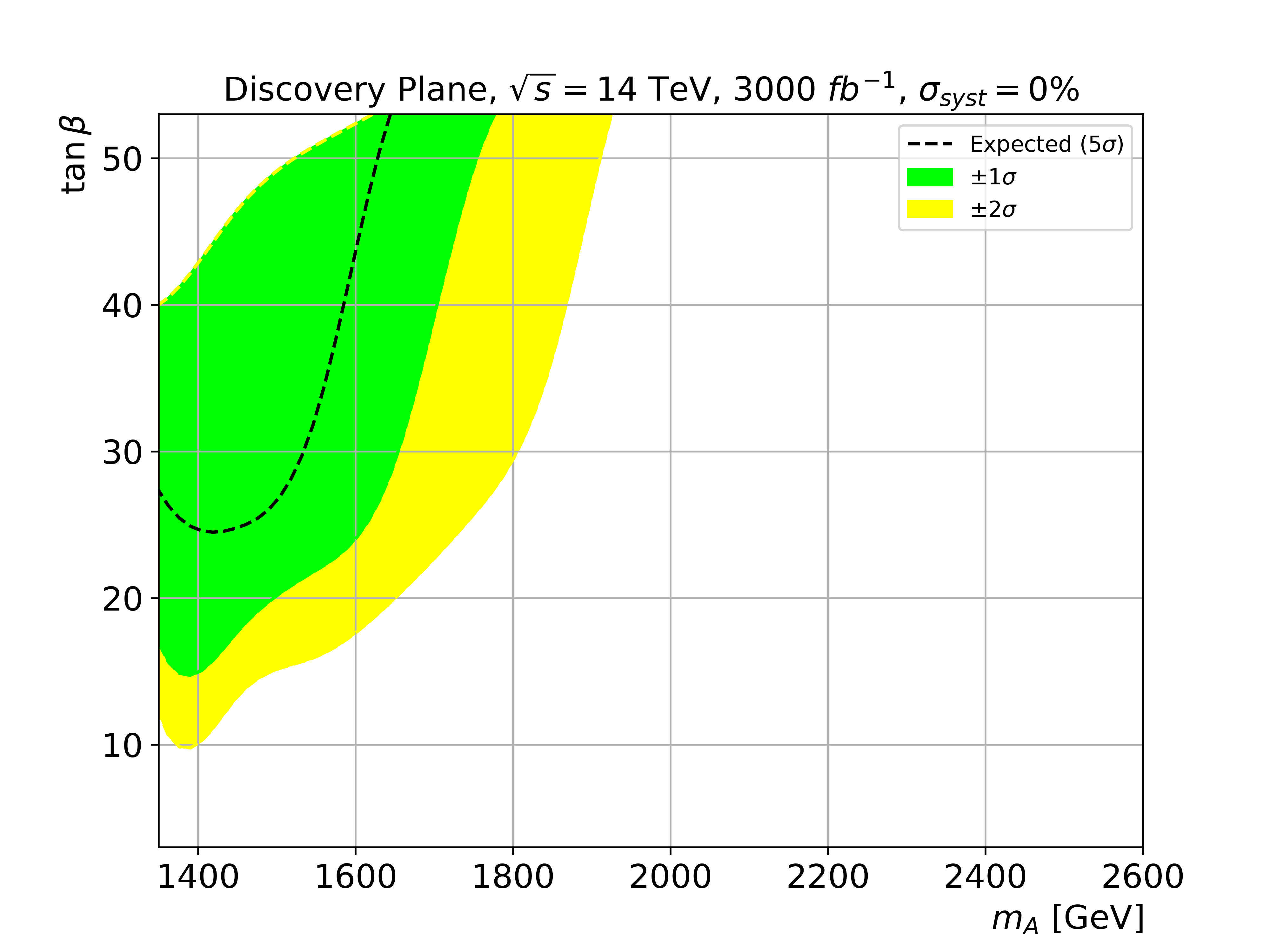}
  \includegraphics[height=0.25\textheight]{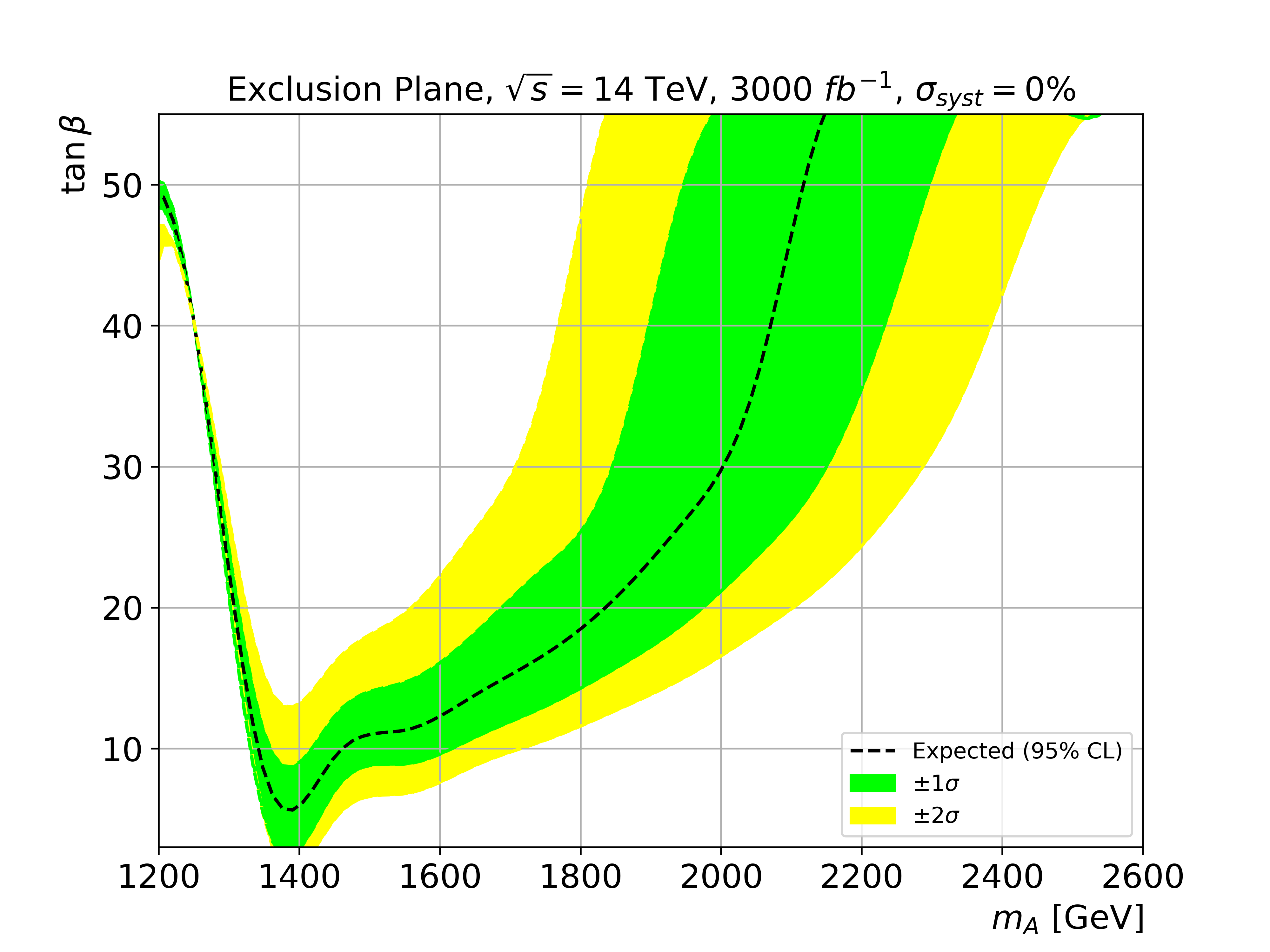}
  \caption{In {\it a}), we show the $5\sigma$
    contour combining all six new $H,\ A\to SUSY$ discovery channels,
    while in {\it b}) we plot the 95\% CL exclusion limits from the all six channels combined. 
    Figures are updated from Ref.~\cite{Baer:2022smj}.
    \label{fig:disc_excl}}
\end{center}
\end{figure}

In Ref.~\cite{Baer:2022smj}, these new channels for heavy neutral Higgs
bosons were examined:
\bi
\item $H,\ A\to W(\to\ell\nu)+\eslt$,
\item $H,\ A\to Z(\to \ell\bar{\ell})+\eslt$,
\item $H,\ A\to h(\to b\bar{b})+\eslt$,
\item $H,\ A\to 1LRj+ \ell +\eslt$,
\item $H,\ A\to 3\ell +\eslt$ and 
\item $H,\ A(\to bb)+\ell +\eslt$ .
\ei
Each channel was examined and cuts were proposed to help maximize signal
over SM background. A combined reach plot was developed using all six
search channels. The results are shown in Fig.~\ref{fig:disc_excl} for
HL-LHC with 3 ab$^{-1}$ at the {\it a}) $5\sigma$ level and {\it b}) the
95\%CL. From the plot, we see that new discovery modes are possible
mainly in the region where $m_A\agt 1.3$ TeV (so the decays to gaugino+higgsino turn on) and $m_A\alt 2.2$ TeV where 
for higher $m_{A,H}$ values the LHC $H$ and $A$
production cross sections diminish. 
In frame {\it b}), we show the projected exclusion limit of the HL-LHC.
While some of the new discovery region
is already ruled out by the ATLAS $H,\ A\to\tau\bar{\tau}$ analysis
(in the $m_h^{125}$ scenario), there is substantial parameter space beyond
this bound which is open to $H,\ A$ discovery via the new SUSY decay modes.

\subsubsection{Prospects for charged Higgs bosons at HL-LHC}
\label{sssec:HC}

\begin{figure}[htb!]
\begin{center}
\includegraphics[height=0.5\textheight]{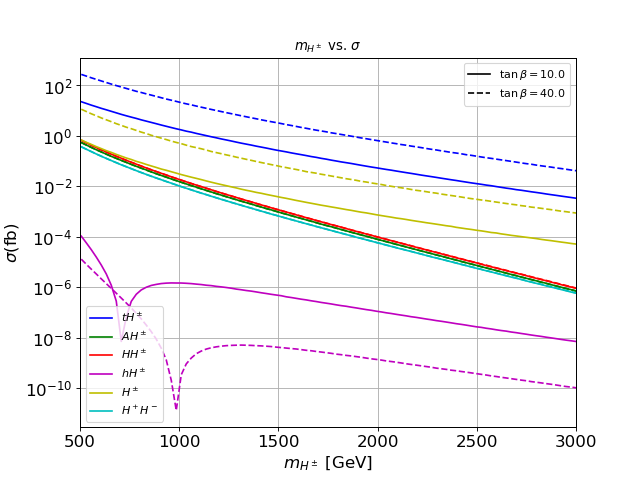}
\caption{The total cross section for $pp\to H^\pm +X$
via various production mechanisms at LHC14 for $\tan\beta =10$ (solid)
and $\tan\beta =40$ (dashed). 
\label{fig:sig1}}
\end{center}
\end{figure}

Detection of charged Higgs bosons in SUSY models has always been regarded
as very challenging. One reason is that $H^\pm$ is not produced resonantly
at a high rate, as are $H$ and $A$. In fact, as shown in Fig. \ref{fig:sig1}
(taken from Ref. \cite{Baer:2023yxk}),
single production of $H^\pm$ in association with a top quark is the largest
production reaction whereas resonant $H^\pm$ production is typically two orders
of magnitude smaller.
This is followed by the di-Higgs production reactions: $HH^\pm$, $AH^\pm$ and
$H^\pm H^\mp$ with $hH^\pm$ lowest of all.
For comparable heavy Higgs masses, $\sigma (pp\to tH^\pm )$ is in the vicinity
of an order of magnitude below the resonant $H$ and $A$ production
cross sections.

\begin{figure}[htb!]
\begin{center}
\includegraphics[height=0.28\textheight]{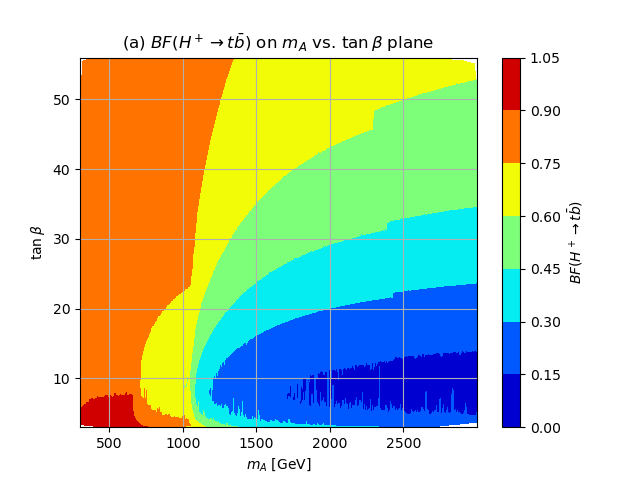 }
\includegraphics[height=0.28\textheight]{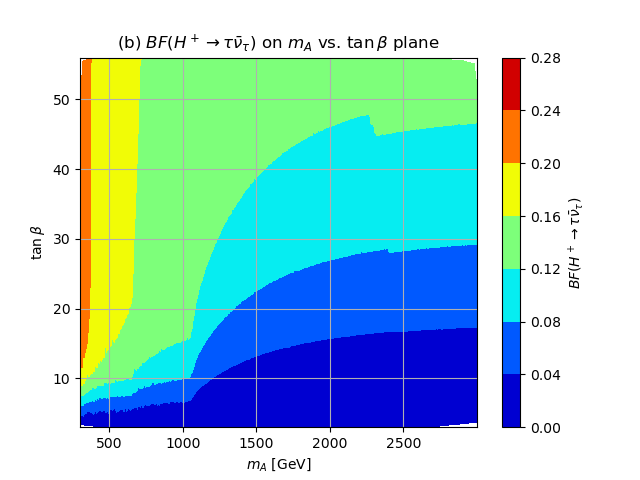}\\
\includegraphics[height=0.28\textheight]{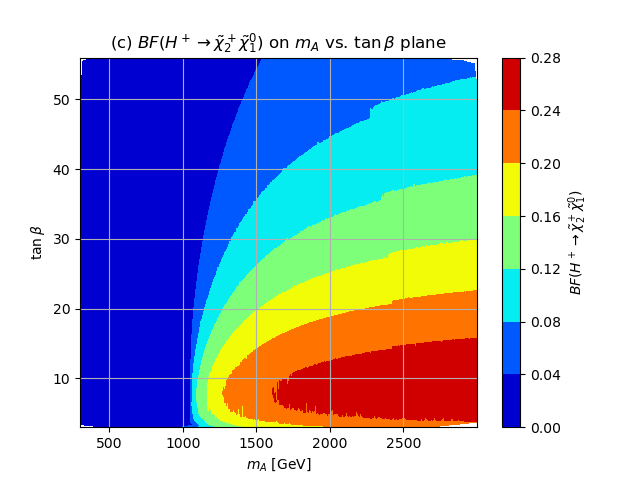}
\includegraphics[height=0.28\textheight]{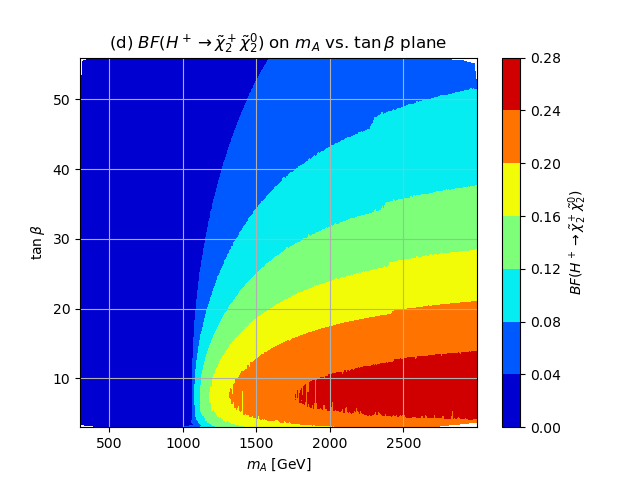}\\
\includegraphics[height=0.28\textheight]{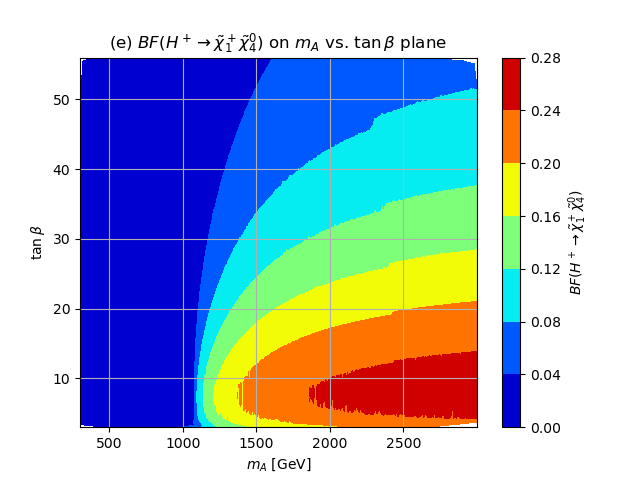}
\includegraphics[height=0.28\textheight]{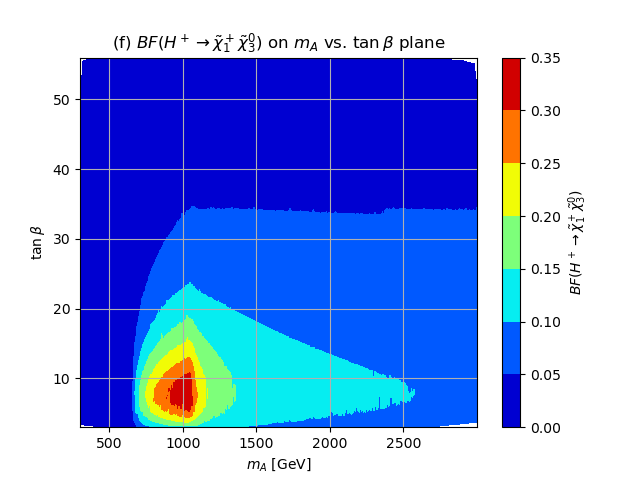}\\
\caption{Branching fractions in the $m_A$ vs. $\tan\beta$ plane for
  $H^+$ to {\it a}) $t\bar{b}$, {\it b}) $\tau^+\nu_{\tau}$, {\it c})
  $\tchi_2^+\tchi_1^0$, {\it d}) $\tchi_2^+\tchi_2^0$, {\t e})
  $\tchi_1^+\tchi_4^0$ and {\it f}) $\tchi_1^+\tchi_3^0$ from Isajet
  7.88~\cite{Paige:2003mg} for the model line introduced in the text.
\label{fig:BFHC}}
\end{center}
\end{figure}

For most charged Higgs analyses, the SM decay modes $H^+\to t\bar{b}$ and
$\nu_\tau\bar{\tau}$ are assumed at rates given by the two-Higgs doublet model
(2HDM). For natural SUSY with light higgsinos, then for TeV-scale charged
Higgs bosons, SUSY decay modes start opening up. Once $H^\pm$ decay
to gaugino+higgsino is open, these decay modes rapidly dominate the
charged Higgs branching fraction. This is shown in Fig. \ref{fig:BFHC},
where the first two frames show the SM-like charged Higgs decay modes
{\it a}) $t\bar{b}$ and {\it b}) $\tau^+\nu_\tau$, which dominate for
sub-TeV charged Higgs masses or at high $\tan\beta$.
For higher $m_{H^\pm}$
values (and lower $\tan\beta$), then the SUSY decay modes
{\it c}) $\tchi_2^+\tchi_1^0$, {\it d}) $\tchi_2^+\tchi_2^0$, {\t e})
$\tchi_1^+\tchi_4^0$ and {\it f}) $\tchi_1^+\tchi_3^0$ rapidly
dominate the charged Higgs decays.

\begin{figure}[htb!]
\begin{center}
  \includegraphics[height=0.5\textheight]{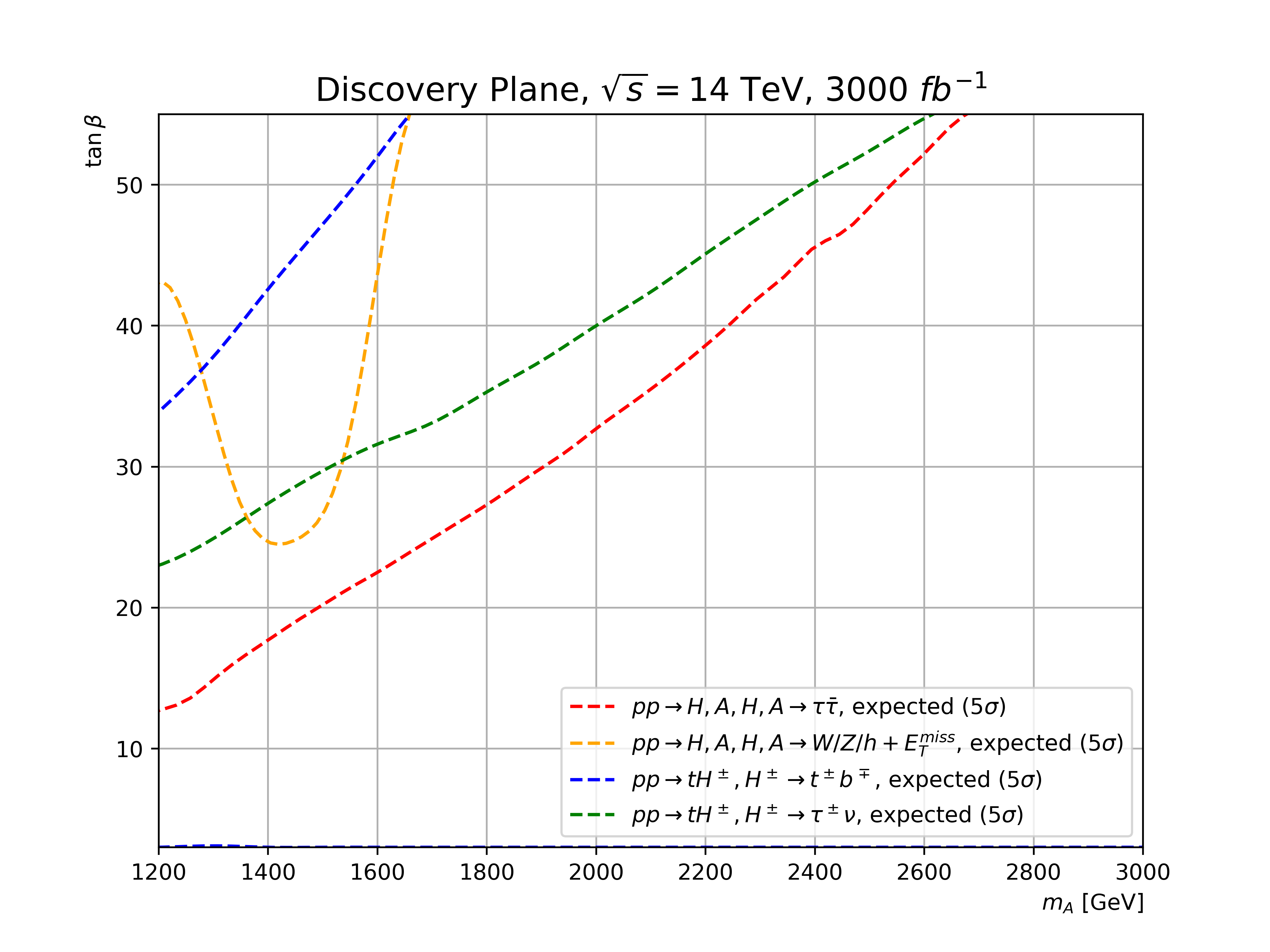}
  \caption{Plot of $5\sigma$ discovery projections for
    heavy SUSY Higgs boson searches at HL-LHC
    in the $m_A$ vs. $\tan\beta$ plane for the $m_h^{125}({\rm nat})$ scenario.}
\label{fig:disc}
\end{center}
\end{figure}

In Ref. \cite{Baer:2023yxk}, detection of charged Higgs bosons from natural
SUSY was investigated for HL-LHC with 3 ab$^{-1}$. Signals from
$tH^\pm$ were examined folllowed by the decays to $t\bar{b}$ and
$\tau^+\nu_\tau$. After judicious sets of cuts to optimize signal over
SM backgrounds, $5\sigma$ and $95\%CL$ reach contours were obtained.
The $5\sigma$ discovery contours for charged Higgs at HL-LHC are shown in
Fig. \ref{fig:disc} in the $t\bar{b}$ mode
(discovery above the blue-dashed contour) and the $\tau\nu_\tau$ mode
(above green-dashed contour).
No further discovery reach was found for $H^\pm \to SUSY$ decay modes.
For comparison, we show the $5\sigma$
discovery contours via $H,\ A\to\tau\bar{\tau}$ (above red-dashed contour)
and the $H,\ A\to SUSY$ modes (above gold-dashed contour).
While the charged Higgs contour does not blaze new parameter space in the
$m_A$ vs. $\tan\beta$ plane as compared to resonant $H,\ A$ production,
it would still be exciting to confirm the presence of signals arising from
charged Higgs production at HL-LHC.

\section{Prospects for natural anomaly-mediation at HL-LHC}

As shown in Fig. \ref{fig:m03m32}, the natural AMSB model is
different from NUHM and GMM models in that the gaugino masses
are expected to occur in the ratio $M_1:M_2:M_3\sim 3:1:8$ so that the wino
is the lightest gaugino even though the higgsinos are still expected to be
the lightest EWinos. Meanwhile, the upper limits on parameter space due to naturalness occur due
to an interplay of gluino and top-squark masses and in fact these high $m_{3/2}$
portions of nAMSB parameter space seem to be excluded by the ATLAS limits
on boosted hadronically-decaying gauge bosons produced from chargino and
neutralino decay\cite{ATLAS:2021yqv}. This leaves in Fig. \ref{fig:m03m32}
the upper and lower ranges of $m_{3/2}$ excluded by the boosted $W/Z$
searches and by gluino searches, respectively. An intermediate gap region of
nAMSB parameter space with $m_{3/2}\sim 90-200$ TeV is still LHC-allowed but with
$m_h\simeq 125$ GeV and $\Delta_{EW}\alt 30$\cite{Baer:2023ech}.

\begin{figure}[tbp]
\begin{center}
    \includegraphics[height=0.3\textheight]{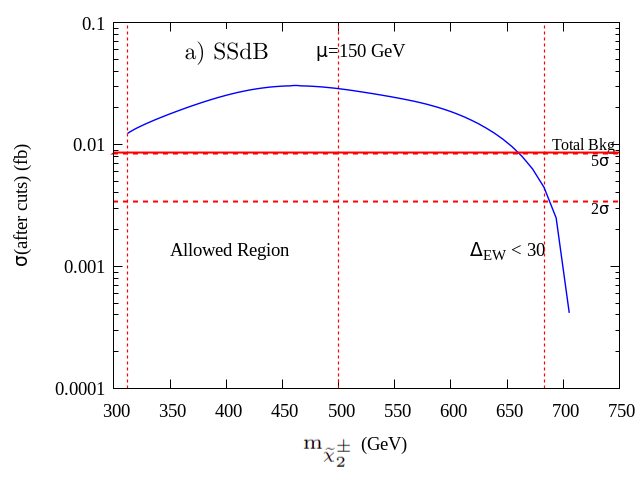}
    \includegraphics[height=0.3\textheight]{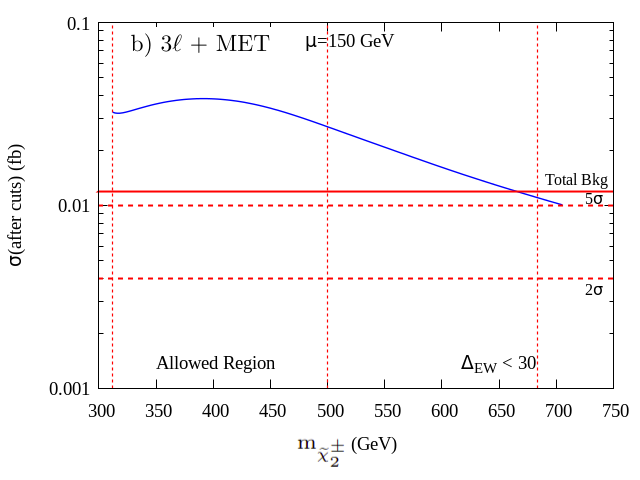}
    \caption{Cross section after cuts for the {\it a}) the SSdB signature
      and {\it b}) the clean $3\ell$ signature from wino pair production along a nAMSB model line
      for $\mu =150$ GeV.
      We also show the $5\sigma$ and $2\sigma$ reach levels for HL-LHC and the remaining total background rate.
      \label{fig:namsb3}}
\end{center}
\end{figure}

This nAMSB gap region typically contains $m(wino)\sim 275-500$ GeV
and so should be accessible to HL-LHC via wino pair production searches
$pp\to\tchi_2^\pm\tchi_3^0$ and $\tchi_2^\pm\tchi_2^\mp$. 
The charged and neutral winos typically decay to $W$, $Z$ or $h$ plus
higgsino where the higgsinos are (quasi)invisible.
In Ref. \cite{Baer:2024hqm},
the wino-pair signals in the nAMSB gap region were examined in the
context of HL-LHC. There it was found that the SSdB signal and a clean
trilepton signal should be detected throughout the nAMSB gap region 
(for $m_{\tchi_2^\pm}\sim 300-500$ GeV for $\mu =150$ GeV) by HL-LHC, 
as shown in Fig. \ref{fig:namsb3}.
These signals would augment the possible signals of OSDLJMET from higgsino
pair production and also top-squark pair production signatures.
Thus, the nAMSB model seems to be completely testable by HL-LHC
with 3 ab$^{-1}$ and the model cannot be saved by merely pushing the
sparticle masses to higher values.

\section{Putting it all together: prospects for SUSY discovery
  over the next decade}
\label{sec:conclude}

In this review we have attempted to present the current status of weak
scale supersymmetry along with prospects for discovery within the next decade,
which means mainly at high luminosity LHC with $\sqrt{s}=14$ TeV and
3 ab$^{-1}$ of integrated luminosity. At the present time, most LHC analyses are
performed within the context of simplified models which make little or
no contact with contemporary theoretical ideas. Many searches take place within
rather implausible scenarios or scenarios which connect to the wider world
of theoretical model building and thus may create a false impression that
SUSY parameter space has been largely explored, and within a plethora of
varied scenarios.

We attempt to remedy this situation by providing a
critical examination of the status of a variety of serious SUSY models
within the context of a {\it plausibility meter} that is largely based
on naturalness.
In this case, it is essential to define what one means by naturalness
and to provide a quantitative measure of when a model or spectra is
or is not natural.
Since several different naturalness measures exist, and their numerical
values which can provide upper limits on sparticle masses admittedly
have some arbitrariness, the impression has been given that naturalness is
somewhat or wholly subjective. We disagree with this viewpoint, and instead
subject contemporary measures to a critical evaluation.
We explain how and why measures such as $\Delta_{p_i}$ and $\Delta_{HS}$
overestimate finetuning and why the measure $\Delta_{EW}$ provides a
model-independent and conservative evaluation of the naturalness of a given
SUSY spectra (defined by the entire SUSY Les Houches Accord (SLHA) output
file from SUSY spectra generators).
$\Delta_{EW}$ is based on practical naturalness and compares the
independent derived contributions to the weak scale
$m_{weak}\simeq m_{W,Z,h}\sim 100$ GeV to its measured value. In fact, this
is where any finetuning actually occurs in SUSY spectra generator codes.
From practical naturalness, all independent contributions (in magnitude)
to the weak scale should be comparable to or less than  the weak scale.
This typically means to within a factor of several, so that
$\Delta_{EW}\alt 20-40$ should provide reliable upper bounds to
sparticle masses. The measure $\Delta_{EW}$ has a close link to the string theory landscape of vacua in that $\Delta_{EW}\alt 30$ corresponds to the
upper limit on the so-called ABDS window of anthropically-allowed
weak scale values. Rather general arguments from the landscape prefer a
statistical power-law draw to large soft terms which must then be tempered
by a derived pocket-universe weak scale within the ABDS window which
corresponds to the {\it atomic principle}: the value of the weak scale
in each allowable pocket universe must be such that complex nuclei,
and hence atoms and complex chemistry, can arise.
From a scan of soft SUSY breaking terms over a friendly landscape, it is
predicted that $m_h\simeq 125$ GeV is favored along with no signs yet of
sparticles: exactly what LHC has seen in Run 2.
SUSY models are also highly impacted by WIMP dark matter search
experiments. However, the results are usually interpreted within the context
of the likely over-simplified thermally produced SUSY WIMP scenario.
Even in the context of (not-so-well-well-motivated) R-parity conservation,
other dark matter candidates such as gravitinos or axinos or other hidden sector states may occur, along with a bevy of non-thermal processes which are to
be expected from SUSY axion theories (which solve the strong CP problem) and
string constructs which allow for light moduli fields.

\begin{figure}[htb!]
\centering
    {\includegraphics[height=0.4\textheight]{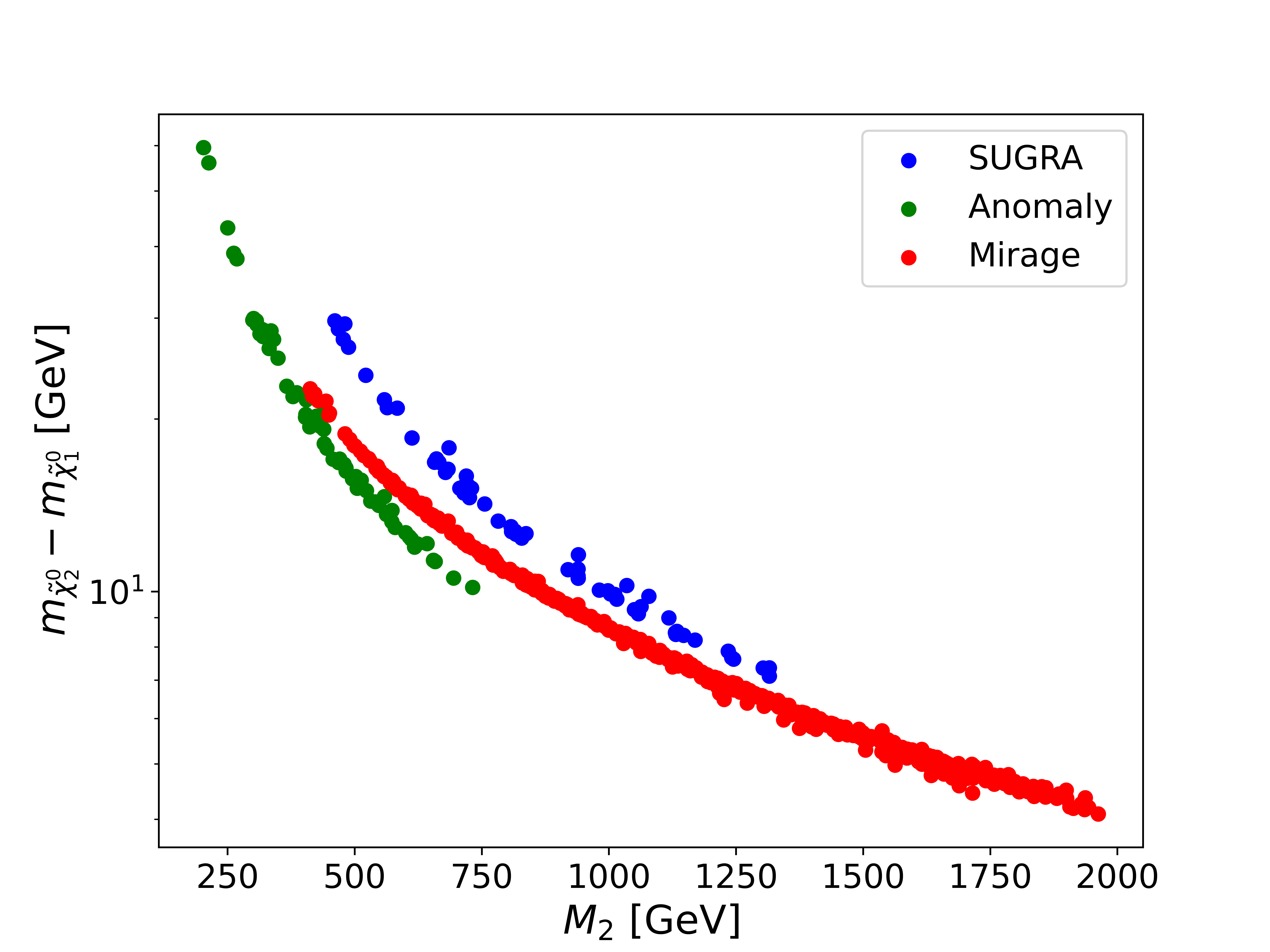}}
    \caption{Locus of landscape scan points from the NUHM3 model, the
      nAMSB model and GMM' model in the $m_{\tchi_2^0}-m_{\tchi_1^0}$
      vs. $m_{\tchi_2^+}\simeq M_2$ plane for fixed value $\mu =200$ GeV.
      \label{fig:plane1}}
\end{figure}

Putting the pieces together, we can now categorise many previously popular
SUSY models on the plausibility meter and find that models such as
CMSSM, GMSB, mAMSB, inoMSB and WTN are now rather implausible if not excluded.
On the other hand, models such as NUHM2-4, nAMSB and nGMM still contain
large portions of natural parameter space even in light of the measured
Higgs mass and lack of LHC sparticle signals.
These models point the way to new avenues for SUSY discovery at HL-LHC.
The expected light higgsinos with masses $\alt 350$ GeV give rise to
soft dilepton plus jet plus MET (SOSDLJMET) signatures, and indeed
LHC Run 2 has yielded slight excesses in this channel for both ATLAS and CMS.
It will be exciting to see whether these signals are confirmed or not in
LHC Run 3 and high-lumi data sets. If confirmed, then LHC experiments may be
able to distinguish amongst these natural models by measuring the 
kinematic limit on soft dilepton inariant mass 
$m(\ell^+\ell^-)< m_{\tchi_2^0}-m_{\chi_1^0}$ along with $M_2$, 
the wino mass,
which can be extracted from the rate for SSdB events\cite{Baer:2017gzf}.
The locus of these measurements is plotted for the several natural SUSY
models in Fig. \ref{fig:plane1}, where a measurement of the $\mu$ parameter helps to sharpen the results\cite{Baer:2024tfo}. 
Distinguishing amongst these models is referred to as decoding 
the Nilles-Choi gaugino code\cite{Choi:2007ka}.

We reviewed prospects for SUSY discovery in plausible models, focusing
mainly on NUHM2-4 since it presents the low energy EFT that is expected
from supergravity which is what is expected from compactified string theory
where singlet SUSY breaking fields reside in the hidden sector. 
For hidden sectors with charged SUSY breaking fields, 
then instead one might expect natural AMSB or generalized
mirage mediation to ensue, and the gaugino spectra may be correspondingly
rearranged.

\begin{figure}[tbp]
\begin{center}
  \includegraphics[height=0.4\textheight]{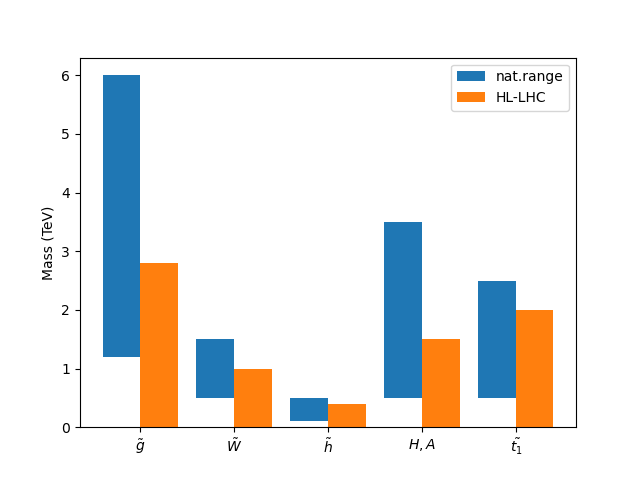}
\caption{We show the expected range of sparticle masses from natural SUSY as emergent from the string landscape in the NUHM3 model with $\tan\beta =10$ (blue histograms) and compare against the computed $5\sigma$ reach of HL-LHC (orange histograms) for various
sparticle and Higgs boson species.
    \label{fig:summary3}}
\end{center}
\end{figure}

In Fig. \ref{fig:summary3}, we show the expected range of sparticle and
Higgs masses from the NUHM3 model as realized from the string landscape,
where radiatively-driven natural SUSY is expected to emerge.
The expected range of masses is shown by the blue histogram, while the orange
histogram shows the derived reach of HL-LHC.
From the plot, we see that while gluinos may range up to 6 TeV in NUHM3, the
corresponding HL-LHC reach is only to about $m_{\tg}\sim 2.8$ TeV (at $5\sigma$ level). Winos may range up to $\sim 1.5$ TeV whilst HL-LHC reach extends to $\sim 1$ TeV. Top squarks can rage up to $2.5-3$ TeV compared to the HL-LHC reach
to $\sim 1.7-1.9$ TeV. Light higgsinos can range up to $\sim 350$ GeV
compared to a HL-LHC reach from Fig. \ref{fig:hplane1} of $\sim 250-300$ GeV.
Heavy Higgs $H$ and $A$ can range up to $\sim 3.5$ TeV although the HL-LHC reach is to about 1.5 TeV (for $\tan\beta \sim 10$).

From the big picture, HL-LHC should be sensitive to much, but not all, of
plausible SUSY model parameter space.
To definitively test plausible SUSY models,
a $pp$ collider with $\sqrt{s}\agt 30$ TeV seems required\cite{Baer:2017yqq,Baer:2017pba,Baer:2018hpb}.
A linear $e^+e^-$ collider with $\sqrt{s}>2m(higgsino)\sim 700$ GeV
would also work\cite{Baer:2014yta,Baer:2019gvu},
or a muon collider with $\sqrt{s}\agt 1$ TeV. 
We hope the ruminations and calculations summarized in this review
will help to guide the eventual discovery of weak scale supersymmetry.

\section*{Acknowledgments}

We thank Xerxes Tata for comments on the manuscript.
VB gratefully acknowledges support from the William F. Vilas Estate.
HB gratefully acknowledges support from the Avenir Foundation.

\bibliography{hllhc}
\bibliographystyle{elsarticle-num}

\end{document}